\documentclass[fleqn,usenatbib]{mnras}

\usepackage{newtxtext,newtxmath}

\usepackage[T1]{fontenc}
\usepackage{ae,aecompl}

\usepackage{threeparttable}

\usepackage{graphicx}	
\usepackage{subfigure}
\usepackage{amsmath}	
\usepackage{amssymb}	
\usepackage{color}
\usepackage[normalem]{ulem}

\definecolor{cacolor}{RGB}{51,0,102}

\title[How Biased Are Halo Properties in Cosmological Simulations?]{How Biased Are Halo Properties in Cosmological Simulations?}

\author[Mansfield \& Avestruz]{
Philip Mansfield,$^{1,2}$\thanks{E-mail: mansfield.astro@gmail.com}
Camille Avestruz$^{3,4}$
\\
$^{1}$Department of Astronomy \& Astrophysics, The University of Chicago, Chicago, IL 60637 USA \\
$^{2}$Kavli Institute for Cosmological Physics, The University of Chicago, Chicago, IL 60637, USA \\
$^{3}$Department of Physics, University of Michigan, Ann Arbor, MI 48109, USA
\\
$^{4}$Leinweber Center for Theoretical Physics, University of Michigan, Ann Arbor, MI 48109, USA
}

\date{Accepted XXX. Received YYY; in original form ZZZ}

\pubyear{2019}

\begin{document}
\label{firstpage}
\pagerange{\pageref{firstpage}--\pageref{lastpage}}
\maketitle

\begin{abstract}

Cosmological N-body simulations have been a major tool of theorists for decades, yet many of the numerical issues that these simulations face are still unexplored. This paper measures numerical biases in these large, dark matter-only simulations that affect the properties of their dark matter haloes. We compare many simulation suites in order to provide several tools for simulators and analysts which help mitigate these biases. We summarise our comparisons with practical `convergence limits' that can be applied to a wide range of halo properties, including halo properties which are traditionally overlooked by the testing literature. We also find that the halo properties predicted by different simulations can diverge from one another at unexpectedly high resolutions. We demonstrate that many halo properties depend strongly on force softening scale and that this dependence leads to much of the measured divergence between simulations. We offer an empirical model to estimate the impact of such effects on the rotation curves of a halo population. This model can serve as a template for future empirical models of the biases in other halo properties.
\end{abstract}




\section{Introduction}

\label{sec:intro}

Understanding the non-linear predictions of the $\Lambda$ Cold Dark Matter ($\Lambda$CDM) model requires the use of simulations. Simulations are required to understand the behaviour of almost every system smaller than the Lagrangian footprint of a large dark matter halo, whether it be the structure of dark matter haloes \citep[e.g.][]{de_Blok_2010}, the abundances of galaxies \citep[e.g.][]{Klypin_et_al_2015_2} and satellites \citep[e.g.][]{Moore_et_al_1999,Klypin_et_al_1999_missing}, or the properties of local dark matter streams \citep[e.g.][]{Vogelsberger_et_al_2009}.

The most common class of $\Lambda$CDM simulation is the N-body simulation. $N$-body simulations have been used to model both individual collapsed structures \citep[see review in][]{Griffen_et_al_2016} and large cosmological volumes (see review in Section \ref{sec:simulations}). While the predictions of $\Lambda$CDM include the behaviour of baryons, many simulators and analysts focus on `dark matter only' (DMO) simulations. Beyond the relative computational efficiency of DMO simulations, the fundamental reason for the popularity of DMO simulations lies in their parametrization. Baryonic simulations have a wide range of parameters, many of which have true physical meaning \citep[e.g. table 2 in][]{Hopkins_et_al_2018}. On the other hand, once a cosmology is specified, a DMO simulation has a much smaller set of parameters and all these parameters are purely numerical. This leads to the core fact that underpins all tests of DMO simulations: {\em any dependence on parametrization is evidence for numerical bias or error}.

Our study -- similar to most other DMO convergence studies -- focuses on the three most important parameters of DMO simulations: particle mass, $m_p$ (or mean interparticle spacing, $l=L/N$), timestepping, and the distribution of mass around each particle (`force softening').\footnote{The results of DMO simulations may depend on a number of other numerical parameters -- like starting redshift, box size, or force accuracy -- but the safe ranges for these parameters are generally better constrained. We have some discussion on the impact of these types of parameters in Section \ref{sec:multidark_illustris}.} Historically, statements about the correctness and biases associated with these three primary parameters come from a combination of four classes of analyses:

{\bf A priori models of numerical effects}: (e.g. \citealp{Knebe_et_al_2000}; \citealp{Dehnen_2001}; \citealp{Power_et_al_2003}; \citealp{Ludlow_et_al_2019}) In these studies, simulators create a model of how the numerical components of a simulation behave, often validating the predictions of this model with appropriate test simulations, and use that model to infer the correctness of other simulations.

{\bf Simulations of idealised systems}: \citep[e.g. ][]{Klypin_et_al_2015,van_den_Bosch_Ogiya_2018,Joyce_et_al_2020} In these studies, simulators either have a priori knowledge of the exact solution the simulation is expected to produce (such as the simulations of NFW haloes in \citealp{van_den_Bosch_Ogiya_2018}), or a priori knowledge of some invariant property of the system (such as the self-similar power spectra analysed in \citealp{Joyce_et_al_2020}). Measured deviations from these expectations are unambiguous numerical biases.

{\bf Resimulations of realistic systems}: \citep[e.g.][]{Power_et_al_2003,Navarro_et_al_2010,Ludlow_et_al_2019} In these studies, simulators resimulate a $\Lambda$CDM system with a variety of numerical parameters.  Systems are typically either a single halo \citep[e.g.][]{Power_et_al_2003,Navarro_et_al_2010}, or a small cosmological box \citep{Ludlow_et_al_2019}. This class of tests can also encompass comparisons between different simulation codes \citep[e.g.][and references therein]{Kim_et_al_2014}. Since there is no a priori expectation for these simulations, simulators will identify a region of numerical parameter space where results are locally independent of numerical parameters and measure deviations relative to this `converged' region.

{\bf Comparison of independent simulations}: (near-ubiquitous; e.g. \citealp{Klypin_et_al_2015}; \citealp{Villearreal_et_al_2017}; \citealp{Child_et_al_2018}) In these studies, simulators compare independently run simulations which inhomogogenously sample numerical parameter space, with the goal of identifying converged parameter ranges. While this type of test is particularly vulnerable to `false' convergence, it is substantially less labour-intensive than the previous classes of studies, and is often the only test available for assessing the correctness of expensive simulations which were not performed as part of a multi-resolution suite.

A simulator interested in assessing the biases of large cosmological DMO simulations -- the class of simulations targeted by this paper -- must rely on tests of all four types of studies. Although the last class of tests mentioned above will always be a necessary component of such assessments, simply comparing the results of cosmological simulations cannot establish that the `converged' solutions which these tests identify are correct. Such an inference must come from detailed comparison with the other classes of tests.

Despite the vast literature on convergence testing in cosmological DMO simulations, there are still unknowns, disparities, and limitations to the tests performed. Tests of the first three types mentioned above focus almost exclusively on radial density profiles at fixed radii. However, dark matter haloes are complex objects with a myriad of scientifically useful properties.  To the best of our knowledge, there are {\em no} published reliability requirements for many commonly used halo properties, such as the offset between a halo's centre of mass and its most bound particle, $X_{\rm off}.$  Even for the most well-tested halo properties, there is no clear consensus on what is required for reliability; examples including the peak of the rotation curve, $V_{\rm max},$ or the radius at which the logarithmic slope of the denisty profile is -2, $r_{-2}$. We surveyed twelve studies on the concentration mass-relation, all of which measure some form of $r_{-2}.$  From this survey, we found that the minimum particle counts ($N_{\rm vir}$) which different studies analysed ranged from from 500 to 10,000 particles, with the 1$\sigma$ scatter spanning more than a decade \citep{Neto_et_al_2007,Duffy_et_al_2008,Gao_et_al_2008,Zhao_et_al_2009,Prada_et_al_2012,Bhattacharya_et_al_2013,Ludlow_et_al_2013,Dutton_Maccio_2014,Diemer_Kravtsov_2015,Klypin_et_al_2016,Poveda_Ruiz_et_al_2016,Child_et_al_2018}. Lastly, tests focused solely on how many particles haloes are resolved with dominate much of the literature, despite demonstrations that force softening and timestepping have large effects on halo properties (see Sections \ref{sec:debias} and \ref{sec:timestepping}).

In this work, we aim to complete several components of the analysis needed to rectify these issues, incorporating components of all four classes of tests discussed above. We perform convergence tests using a large inhomogenous suite of publicly available cosmological simulations. These tests are performed over a wide range of halo properties, including halo properties which are traditionally overlooked by the testing literature. We also analyse the impact of timestepping and force softening parameters on halo properties.

We organise the paper as follows. In Section \ref{sec:methods}, we outline our methods for comparing cosmological simulations and extracting empirical convergence limits. In Section \ref{sec:results}, we report these empirical limits and consider the variation in limits between simulations. In Section \ref{sec:eps_dependence_mass_trend}, we study the dependence of various halo properties on the force softening scale. In Section \ref{sec:debias}, we outline a model for estimating the impact of large force softening scales on halo profiles and apply this model to our simulation suites. Lastly, in Section \ref{sec:discussion} we discuss our results (particularly the impact of timestepping), and in Section \ref{sec:conclusion} we summarise and conclude our analysis.

\section{Methods}
\label{sec:methods}

\subsection{Simulations}
\label{sec:simulations}

In this paper, we use eight widely-used simulations suites: Erebos\_CBol \citep{Diemer_Kravtsov_2014,Diemer_Kravtsov_2015}, Erebos\_CPla \citep{Diemer_Kravtsov_2015}, Multidark-Planck \citep{Klypin_et_al_2016}, Chinchilla \citep{Lehmann_et_al_2017}, Bolshoi \citep{Klypin_et_al_2011}, BolshoiP \citep{Klypin_et_al_2016}, $\nu^2$GC \citep{Ishiyama_et_al_2015}, and IllustrisTNG-Dark \citep{Naiman_et_al_2018,Pillepich_et_al_2018_b,Nelson_et_al_2018,Marinacci_et_al_2018,Volker_et_al_2018}. We list the cosmological and numerical parameters of these simulations in Table \ref{tab:simulations}. Access to the ESMDPL simulation from the Multidark-Planck suite has been generously provided by G. Yepes and S. Gottloeber.

Each simulation suite is the product of one of four simulation codes, each with varying gravity solvers and timestepping schemes. Bolshoi and BolshoiP were run using N-body ART \citep{Kravtsov_et_al_1997,Kravtsov_1999,Gottloeber_Klypin_2008}, the Multidark-Planck, Erebos\_CBol, Erebos\_CPla, and Chinchilla suites were run with Gadget-2 \citep{Springel_2005}, IllustrisTNG-Dark was run using \textsc{Arepo} \citep{Springel_2010,Weinberger_et_al_2019} which performs gravitational calculations using an updated version of the Gadget-2 gravity-solving algorithm. $\nu^2$GC was run with GreeM$^3$ \citep{Ishiyama_et_al_2012,Ishiyama_et_al_2015}.

An important aspect of these codes is the scheme they use for setting timestep sizes. Three of the four codes, Gadget-2, Arepo, and GreeM$^3$ use an adaptive timestepping scheme dependent on the local gravitational acceleration \citep[][T. Ishiyama, personal communication]{Springel_2005,Weinberger_et_al_2019}. The fourth code, ART, uses density-dependent timesteps \citep{Klypin_et_al_2011}.

In detail, Gadget-2, Arepo, and GreeM$^3$ calculate the timestep size, $\Delta t$, for each particle through
\begin{equation}
    \label{eq:gadget_timestep}
    \Delta t = \sqrt{2\eta\epsilon/|\vec{a}|}.
\end{equation}
Here, $\vec{a}$ is the local gravitational acceleration, $\epsilon$ is the `Plummer-equivalent' force softening scale which will be discussed below, and $\eta$ is a user-defined parameter (also referred to as \texttt{ErrTolIntAcc}) which is typically set to $\gtrsim 0.01.$ In practice, $\Delta t$ is evaluated for each particle, the values are used to place particles into the coarsest logarithmic timestepping bin, $\Delta t_i = t_0 2^{-i},$ such that $\Delta t_i \leq \Delta t.$ As such, the actual timestep size a particle experiences may be smaller than Eq.~\ref{eq:gadget_timestep} by as much as a factor of two. We note that while the initial GreeM$^3$ implementation used a different adaptive scheme \citep{Ishiyama_et_al_2009}, GreeM$^3$ used the adaptive scheme described above during the $\nu^2$GC runs (T. Ishiyama, personal communication).

ART timesteps vary at different depths of the refinement tree, meaning that they depend on density instead of acceleration. Both Bolshoi and BolshoiP use timesteps of $\Delta a \approx 2 - 3 \times 10^{-3}$ at the 0$^{\rm th}$ (coarsest) refinement level with time steps decreasing by a factor of two for each successive level of spatial refinement, leading to timesteps of $\Delta a \approx 2 - 3 \times 10^{-6}$ at the tenth level \citep{Klypin_et_al_2011}. The ART timestepping scheme leads to far finer timesteps than any of the other simulations considered in this paper.

\begin{table*}
  \centering
  \caption{A list of the simulations used in this work. The first six columns contain information common to all simulations in a given suite: the code used to run the suite, the suite name, the several important cosmological parameters ($\Omega_{M},$ $h_{100}=H_0/(100 {\rm\ km/s/Mpc)}$, and $\sigma_8$),
  and the Gadget-like timestepping parameter, $\eta.$ Note that the ART code does not use this timestepping scheme (see section \ref{sec:simulations} for details). The last four columns give information specific to each individual simulation: the simulation name, the box width, $L,$ the number of particles, $N^3,$ the particle mass, $m_p$, and the force softening scale at $z=0$ in units of the mean interparticle spacing, $\epsilon/l.$ We use Eq.~\ref{eq:plummer_equivalent} to convert from the formal resolution, $h,$ to $\epsilon$.}
  \label{tab:simulations}
  \begin{tabular}{lllllllllll}
  \hline\hline
    Code & Suite & $\Omega_{M}$ & $h_{100}$ & $\sigma_8$ & $\eta$ & Simulation &  $L$ & $N^3$ & $m_{p}$ & $\epsilon/l$ \\
     & & &  &  &  &  & ($h^{-1}$Mpc)  & & ($h^{-1}M_\odot$) &  \\
  \hline
GreeM$^3$ & $\nu^2$GC & 0.31 & 0.68 & 0.83 & 0.045 & $\nu^2$GC-L & 1120 & $8192^3$ & $2.27 \times 10^{8}$ & 0.04 \\
& & & & & & $\nu^2$GC-H1 & 140 & $2048^3$ & $2.75 \times 10^{7}$ & 0.04 \\
& & & & & & $\nu^2$GC-H2 & 70 & $2048^3$ & $3.44 \times 10^{6}$ & 0.04 \\
ART & Bolshoi & 0.27 & 0.7 & 0.82 & -- & Bolshoi & 250 & $2048^3$ & $1.36 \times 10^{8}$ & 0.0082\\
& BolshoiP & 0.307 & 0.678 & 0.823 & -- & BolshoiP & 250 & $2048^3$ & $1.55 \times 10^{8}$ & 0.0082\\
Gadget-2 & Chinchilla & 0.286 & 0.7 & 0.82 & 0.025 & L125 & 125 & $2048^3$ & $1.80 \times 10^{7}$ & 0.0082 \\
& & & & & & L250 & 250 & $2048^3$ & $1.44 \times 10^{8}$ & 0.0082 \\
& & & & & & L400 & 400 & $2048^3$ & $5.91 \times 10^{8}$ & 0.0082 \\
& Multidark & 0.307 & 0.678 & 0.823 & 0.01 & ESMDPL & 64 & $2048^3$ & $2.60 \times 10^{6}$ & 0.032 \\
& & & & & & VSMDPL & 160 & $3840^3$ & $6.16 \times 10^{6}$ & 0.024 \\
& & & & & & SMDPL & 400 & $3840^3$ & $9.63 \times 10^{7}$ & 0.014 \\
& & & & & & MDPL2 & 1000 & $3840^3$ & $1.50 \times 10^{9}$ & 0.019 \\
& & & & & & BMDPL & 2500 & $3840^3$ & $2.35 \times 10^{10}$ & 0.015 \\
& & & & & & HMDPL & 4000 & $4098^3$ & $7.92 \times 10^{10}$ & 0.026 \\
& Erebos\_CBol & 0.27 & 0.7 & 0.82 & 0.025 & CBol\_L63 & 62.5 & $1024^3$ & $1.70 \times 10^{7}$ & 0.016 \\
& & & & & & CBol\_L125 & 125 & $1024^3$ & $1.36 \times 10^{8}$ & 0.02 \\
& & & & & & CBol\_L250 & 250 & $1024^3$ & $1.09 \times 10^{9}$ & 0.024 \\
& & & & & & CBol\_L500 & 500 & $1024^3$ & $8.72 \times 10^{9}$ & 0.029 \\
& & & & & & CBol\_L1000 & 1000 & $1024^3$ & $6.98 \times 10^{10}$ & 0.034 \\
& & & & & & CBol\_L2000 & 2000 & $1024^3$ & $5.58 \times 10^{11}$ & 0.033 \\
& Erebos\_CPla & 0.32 & 0.67 & 0.82 & 0.025 & CPla\_L125 & 125 & $1024^3$ & $1.62 \times 10^{8}$ & 0.02 \\
& & & & & & CPla\_L250 & 250 & $1024^3$ & $1.29 \times 10^{9}$ & 0.024 \\
& & & & & & CPla\_L500 & 500 & $1024^3$ & $1.03 \times 10^{10}$ & 0.029 \\
Arepo & IllustrisTNG-Dark & 0.3089 & 0.6774 & 0.8159 & 0.012 & TNG100-1-Dark & 75 & $1820^3$ & $6.00 \times 10^{6}$ & 0.012 \\
& & & & & & TNG100-2-Dark & 75 & $910^3$ & $4.80 \times 10^{7}$ & 0.012 \\
& & & & & & TNG100-3-Dark & 75 & $455^3$ & $3.84 \times 10^{8}$ & 0.012 \\
  \hline
  \end{tabular}
\end{table*}

\subsection{Force Softening}
\label{sec:force_softening}
Cosmological simulations do not model particles as point masses. Infinitesimal point sources will scatter off one another during close encounters \citep[e.g. fig.~6 in][]{Knebe_et_al_2000}, which leads to aphysical energy exchange between particles and can potentially thermalise the inner regions of dark matter haloes \citep[see overview in][]{Ludlow_et_al_2019}. Additionally, these close encounters require much finer timesteps to resolve than typical orbits through a halo's potential, meaning that codes are forced to either spend large amounts of computation time resolving an aphysical process or risk conservation of energy errors (See Section \ref{sec:timestepping}). To minimise these effect, codes will `soften' forces to be weaker than $1/r^2$ below some resolution level, $h.$ The exact meaning of $h$ varies between codes.

The GreeM$^3$ code softens forces through a Plummer kernel \citep{Ishiyama_et_al_2012,Ishiyama_et_al_2015}, the simplest force softening scheme. In this scheme, the gravitational potential of a particle is given by
\begin{equation}
\label{eq:plummer_softening}
    \phi(r) = \frac{GM}{\sqrt{r^2 + h_{\rm Plummer}^2}}.
\end{equation}
Here, $\phi$ is the gravitational potential a distance $r$ away from a particle of mass $M.$   

In Gadget-based simulations \citep{Springel_et_al_2001,Springel_2005,Springel_2010}, the density distribution function of particles, $\delta(r),$ changes from a Dirac delta function to the SPH kernel of \citet{Monaghan_Lattanzio_1985}:
\begin{equation}
    \label{eq:gadget_softening}
    \delta(x) = \frac{8M}{\pi h^3}
  \begin{cases}
  1 - 6x^2 + 6x^3, & \text{if $x < \frac{1}{2}$}, \\
  2\,(1 - x)^3, & \text{if $\frac{1}{2} < x < 1$}, \\
  0, & \text{if $x > 1$},
  \end{cases}
\end{equation}
for $x=r/h_{\rm Gadget}.$ This leads to a perfectly Newtonian force beyond $r>h_{\rm Gadget}.$

In ART \citep{Kravtsov_et_al_1997,Kravtsov_1999,Gottloeber_Klypin_2008}, truncation errors in the underlying grid naturally soften gravitational forces according to the local grid cell width, $h_{\rm ART}.$ Because ART grids are adaptive, this means that the formal resolution is also adaptive. Typically, the finest resolution level used within a halo is cited as the formal resolution of that halo.

The analysis in this paper focuses on the impact of force softening at large scales.  We therefore adopt the following convention for converting between formal resolutions, which matches their impact on the halo rotation curves for $r \gtrsim \epsilon,$
\begin{align}
    \label{eq:plummer_equivalent}
    \epsilon=1.284\,h_{\rm Plummer}=h_{\rm ART}=0.357\,h_{\rm Gadget}.
\end{align}
The methodology behind this convention is laid out in Appendix \ref{sec:recalibrate_plummer}, along with the best-fitting impact of large-$\epsilon$ on circular velocity curves.

Note that our convention differs somewhat from those used in previous works in that $\epsilon \neq h_{\rm Plummer}.$ This is because our fits in Appendix \ref{sec:recalibrate_plummer} imply a `Plummer-equivalent' conversion for Gadget which is different from the commonly used $h_{\rm Plummer} = 0.357\,h_{\rm Gadget}$ \citep{Springel_et_al_2001}. Because simulations using Eq.~\ref{eq:gadget_softening} for force softening are far more common than those using Eq.~\ref{eq:plummer_softening}, we choose to use a non-Plummer-equivalent convention to maintain compatibility with as many studies as possible.

\subsection{Halo Finding}
\label{sec:halo_finding}

We use catalogues constructed by the \textsc{Rockstar} halo finder \citep{Behroozi_et_al_2013a}. When available, we also used merger trees constructed by \textsc{consistent-trees} \citep{Behroozi_et_al_2013b} to determine growth history-dependent halo properties. 

The halo catalogues we analyse were generated with a number of different versions of \textsc{Rockstar}. Because \textsc{Rockstar} has undergone many bug fixes since its release, this difference in versions could potentially lead to divergences in halo properties between simulations which are unrelated to numerical issues in the simulations themselves.
In Appendix \ref{sec:rockstar_versions} we analyse the impact of \textsc{Rockstar} versions and bugs on our analysis. We find that after applying a few previously established corrections, \textsc{Rockstar} bugs and versioning do not impact our results. Because correcting for these bugs is version dependent, we recommend that authors specify the \textsc{Rockstar} version they use when publishing catalogues.

Note that this result only establishes the consistency of the \textsc{Rockstar} finder, which is sufficient to establish numerical differences in the underlying simulations. We direct readers interested in assessing the robustness of \textsc{Rockstar}'s underlying algorithm to halo finder comparison projects such as \citet{Knebe_et_al_2011,Knebe_et_al_2013}.

We also generate \textsc{Rockstar} catalogues for the $z=0$ snapshots of the IllustrisTNG-Dark simulations listed in Table \ref{tab:simulations}, made available through the IllustrisTNG public data release \citep{Nelson_et_al_2019}. We used \textsc{Rockstar} as downloaded on June 10$^{\rm th}$ 2019.\footnote{\texttt{git} hash: 99d56672092e88dbed446f87f6eed87c48ff0e77.} We use $M_{\rm vir}$ as our primary mass definition, consistent with other catalogues. As with the other catalogues in this paper, we do not use strict spherical overdensity masses and remove `unbound' particles prior to analysis. We use a coarse-grained friends-of-friends linking length of $b=0.28\cdot l$ for load-balancing. Note that this last setting leads to inaccurate $M_{\rm 200m}$ masses \citep[see section 4.3 and appendix A of][]{mansfield_kravtsov_2019}, but we choose this setting for consistency with the other catalogues used in this study. Some analysis in this paper also uses \textsc{Rockstar} catalogues generated for the baryonic IllustisTNG simulations. In these cases, we use the same \textsc{Rockstar} parameters as we do with IllustrisTNG-Dark, but only consider dark matter particles when computing halo properties. We rescale the particle masses used in these catalogues by $\Omega_{\rm m}/(\Omega_{\rm m}-\Omega_{\rm b})$ to account for the removed baryons.

\subsection{Halo Properties} 
\label{sec:properties}

In this Section, we summarise the halo properties studied in this paper. We compute all properties with the \textsc{Rockstar} halo finder and the \textsc{consistent-trees} merger tree code.  Since the original code papers \citep{Behroozi_et_al_2013a, Behroozi_et_al_2013b}, \textsc{Rockstar} and \textsc{consistent-trees} have incorporated additional halo properties and modified some methodology for property calculations. While descriptions of all these halo properties can be found throughout the literature, we collect them here for pedagogical convenience.

{\bf Bound vs. Unbound Particles}: \textsc{Rockstar} separates particles into `bound' and `unbound' groups and primarily analyses bound particles. This is done because if particles were classified with a simple geometric cut, subhaloes would be contaminated with a large number of particles from their host haloes. There is no unambiguous way to perform this procedure due to the importance of tidal fields in true boundedness calculations, but \textsc{Rockstar} takes a reasonable approach and determines boundedness by performing pairwise potential calculations and comparing against the kinetic energy of particles in the rest frame of the halo centre.

{\bf Halo mass:} The most basic properties of a halo are its size and, equivalently, its mass. We adopt the near-ubiquitous `overdensity radius' definition of the halo boundary, i.e. that the halo is a sphere of radius $R_\Delta$ which encloses the bound mass $M_\Delta=M_{\rm bound}(<R_\Delta)$ such that
\begin{equation}
    M_\Delta = \frac{4\pi}{3}\Delta\rho_{\rm ref} R_\Delta^3.
\end{equation}
Here, $\Delta$ is some constant and $\rho_{\rm ref}$ is a cosmological reference density.  The reference density is typically either the background matter density, $\rho_{\rm m},$ or the critical density, $\rho_{\rm c}.$ 

Our primary radius definition is $R_{\rm vir},$ with $\Delta\rho_{\rm ref}$ given by the relation in \citet{Bryan_Norman_1998}. For completeness, we also consider the bound masses enclosed within $R_{\rm 200m}$ ($\Delta\rho_{\rm ref}=200\rho_{\rm m}$), $R_{\rm 200c}$ ($\Delta\rho_{\rm c}=200\rho_{\rm c}$), $R_{\rm 500c}$ ($\Delta\rho_{\rm ref}=500\rho_{\rm c}$), and $R_{\rm 2500c}$ ($\Delta\rho_{\rm ref}=2500\rho_{\rm c}$).

We note that \textsc{Rockstar} computes overdensity radii by constructing radial density profiles using only particles within the coarse-grained friends-of-friends (FOF) group that contains the halo centre. The linking length parameter used to identify the FOF group has a substantial effect on the convergence properties of $M_{\rm 200m}$ \citep{mansfield_kravtsov_2019} if high-precision measurements of halo masses are needed.

{\bf Virial Scaling:} We use $M_{\rm vir}$ and $R_{\rm vir}$ as characteristic scales to remove the dimensionality of halo properties and to reduce the dynamic ranges of fits and plots throughout this paper. To scale halo properties containing dimensions of time, we also use the virial velocity, 
\begin{equation}
    V_{\rm vir} = \sqrt{\frac{GM_{\rm vir}}{R_{\rm vir}}}.
\end{equation}

{\bf Maximum circular velocity:} We also consider velocity-based measurements of the halo's potential depth. We look at $V_{\rm max},$ the maximum circular velocity implied by the bound mass profile of the halo, and $V_{\rm rms},$ the 3D root mean square velocity of bound particles within $R_{\rm vir}.$ These properties are well-defined measurements of halo size in their own rights, but when scaled by $V_{\rm vir}$ they also give a measure of the `concentration' of the halo: the degree to which mass is concentrated in the core or the outskirts of the halo.

{\bf Halo concentration:} The canonical measurement of halo concentration comes from an analytic fit to the halo profile. \textsc{Rockstar} fits the bound radial density profile of every halo with the NFW \citep{Navarro_et_al_1997} form:
\begin{equation}
    \label{eq:nfw}
    \rho(r) = \frac{\rho_0}{r/R_s(1 + r/R_s)^2},
\end{equation}
where $\rho_0$ and $R_s$ are free parameters of the fit. $c_{\rm vir}\equiv R_{\rm vir}/R_s$ is then a measurement of the
concentration. This fit is delicate and different fitting strategies lead to different concentration statistics. \textsc{Rockstar} performs a $\chi^2$-minimisation of Eq.~\ref{eq:nfw} against binned density profiles, ignoring bins with fewer than 15 particles and heavily down-weighting bins with $r < 3\epsilon.$ We also investigate $R_{1/2},$ the radius which encloses half of the bound mass within $R_{\rm vir},$ but this quantity is a relatively recent addition to \textsc{Rockstar} and few of our catalogues contain it.

{\bf Halo shape: } \textsc{Rockstar} follows the recommendations of \citet{Zemp_et_al_2011}, and computes halo shapes using iterative, weighted mass distribution tensors. Specifically, \textsc{Rockstar} first computes the mass distribution tensor 
\begin{equation}
    M_{ij} = \frac{\sum_k^N (\vec{r}_k)_i(\vec{r}_k)_j|\vec{r}_k|^{-2}}{N \sum_k^N |\vec{r}_k|^{-2}}
\end{equation}
over all bound particles $k$ within $R_{\rm vir}$ and computes the eigenvalues, $\lambda_i,$ of $M_{ij}.$ Then, \textsc{Rockstar} estimates axis ratios as $\sqrt{\lambda_i/\lambda_j}$ for each pair of axes, $i$ and $j,$ repeating the process for all bound particles in an ellipsoid with the corresponding axis ratios and a minimum axis length of $R_{\rm vir}$. This process repeats until axis ratios converge to 1\%. Note that the axis ratio measurement is sensitive to the central mass distribution.

{\bf Halo spin:} To track halo spin, we use the dimensionless Peebles and Bullock spin parameters. The classical Peebles spin parameter \citep{Peebles_1969} is given by,
\begin{equation}
    \lambda_{\rm Peebles} = \frac{|\vec{J}|}{G |E_{\rm tot}| M_{\rm vir}^{5/2}}
\end{equation}
where $\vec{J}$ is the angular momentum vector of the halo and $E_{\rm tot}$ is the total energy of the bound particles. 
However, the normalisation by $E_{\rm tot}$ presents pragmatic difficulties (see the discussion of boundedness above) and makes $\lambda_{\rm Peebles}$ sensitive to recent merger history which is often undesirable. An alternate dimensionless parameter is the simpler Bullock spin parameter \citep{Bullock_et_al_2001} which normalises by virial properties:
\begin{equation}
    \lambda_{\rm Bullock} = \frac{|\vec{J}|}{\sqrt{2}M_{\rm vir}R_{\rm vir}V_{\rm vir}}.
\end{equation}

\indent {\bf Dynamical State Indicators:} We also consider single-epoch properties that indicate the dynamical state (the `dynamical relaxation') of a dark matter halo. We include the following properties: $T/|U|$, the ratio of kinetic to potential energy, $x_{\rm off}=X_{\rm off}/R_{\rm vir},$ the normalised offset between the density peak of the halo and its centre of mass, and $V_{\rm off},$ the offset between the velocity of the halo's density peak and the mean velocity of all its particles. The first two have been found to correlate with recent accretion activity \citep{power_et_al_2012} and age indicators, such as concentration \citep{Neto_et_al_2007}.

{\bf Mass Accretion History:} Beyond the single-epoch halo properties measured by \textsc{Rockstar}, we use the consistent-trees merger tree code to compute a number of properties along the mainline progenitor branch. The most fundamental such property is the mass accretion rate, 
\begin{equation}
    \Gamma(\Delta t) = \frac{M_{\rm vir}(t_0) - M_{\rm vir}(t_0 - \Delta t)}{\Delta t},
\end{equation}
where $t_0$ is the current age of the universe. We specifically focus on $\Gamma(t_{\rm dyn})$ measured over the halo's dynamical time,
\begin{equation}
    t_{\rm dyn} = \frac{1}{\sqrt{\frac{4}{3}\pi G(\Delta\rho_{\rm m})_{\rm vir}}}.
\end{equation}
Here, $(\Delta\rho_{\rm m})_{\rm vir}$ is the \citet{Bryan_Norman_1998} virial density contrast.

Accretion rates are most sensitive to recent mass growth. To trace older mass growth, we use the half-mass scale factor, $a_{0.5}$. This quantity corresponds to the earliest scale factor at which a mainline progenitor of the halo had half the mass of the present-day halo. We also consider major merger scale, $a_{\rm MM},$ the most recent scale factor at which consistent-trees detected a merger where the secondary-to-primary mass ratio was larger than 0.3.

Finally, we consider $M_{\rm peak}$ and $V_{\rm peak},$ the largest values that $M_{\rm vir}$ and $V_{\rm max}$ have taken on throughout the lifetime of the halo, respectively. These values are frequently used when analysing subhalos because a the dark matter halo of a satellite galaxy is disrupted long before the central stellar component is. `Peak' quantities allow modelling in which galaxies grow their stellar mass components in step with their dark matter haloes and maintain if after being captured by a host halo. Such modelling has been shown to be effective a predicting a wide range of observables \citep[e.g.][]{Reddick_2013}.

{\bf Mass and Velocity Functions} Using these halo properties, we measure differential mass and velocity functions, $\phi(X) = {\rm d}\, n(X)/{\rm d} \log_{10}(X)$. Here, $X$ is an arbitrary mass or velocity definition, and $n(X)$ is the number density at a given value of $X$. We consider the mass and velocity definitions of $X\in\left\{M_{\rm vir}, \,M_{\rm 2500c}, \,M_{\rm 500c}, \,M_{\rm 200c}, \, M_{\rm 200m}, \,M_{\rm peak}, \,V_{\rm max}, \,V_{\rm rms}, \,V_{\rm peak}\right\}$. 

{\bf Isolated Halo vs. Subhalo Classification}: We consider the distributions of all halo properties described above for both isolated halos and subhaloes. \textsc{ROCKSTAR} identifies isolated haloes as haloes whose centres are outside $R_{\rm vir}$ of all larger haloes in the simulations, while subhaloes have centres which lie within $R_{\rm vir}$ of a larger halo.

{\bf Other Properties}: There are a number of quantities in \textsc{Rockstar} and consistent-trees catalogues which we do not explicitly study in this paper. In most cases this is because the convergence behaviour of these properties is identical to that of another property: we find that the convergence limits for $\Gamma(t_{\rm dyn})$ are essentially the same as accretion rates defined over any other time scale tracked by any version of consistent-trees. We therefore only consider $\Gamma(t_{\rm dyn}).$ Similarly, we find that the convergence properties of $b/a$, $(c/a)(<R_{\rm 500c}),$ and $(b/a)(<R_{\rm 500c})$ are nearly identical to those of $c/a$ and thus only consider $c/a.$ Later versions of \textsc{Rockstar} track the maximum single-halo tidal force on each halo, but we do not track convergence behaviour for tidal force calculations. This is because too few of our catalogues contain this property to achieve meaningful statistics. We additionally note that computing the tidal force on haloes has subtleties that indicate that the approximation used by \textsc{Rockstar} may not be sufficiently physical (see section 2 of \citealp{van_den_bosch_et_al_2018} and section 2.5 and appendix C of \citealp{mansfield_kravtsov_2019}).%

\subsection{Finding Empirical Convergence Limits}
\label{sec:procedure}

\begin{figure*}
   \centering
   \subfigure{\includegraphics[width=0.49\textwidth]{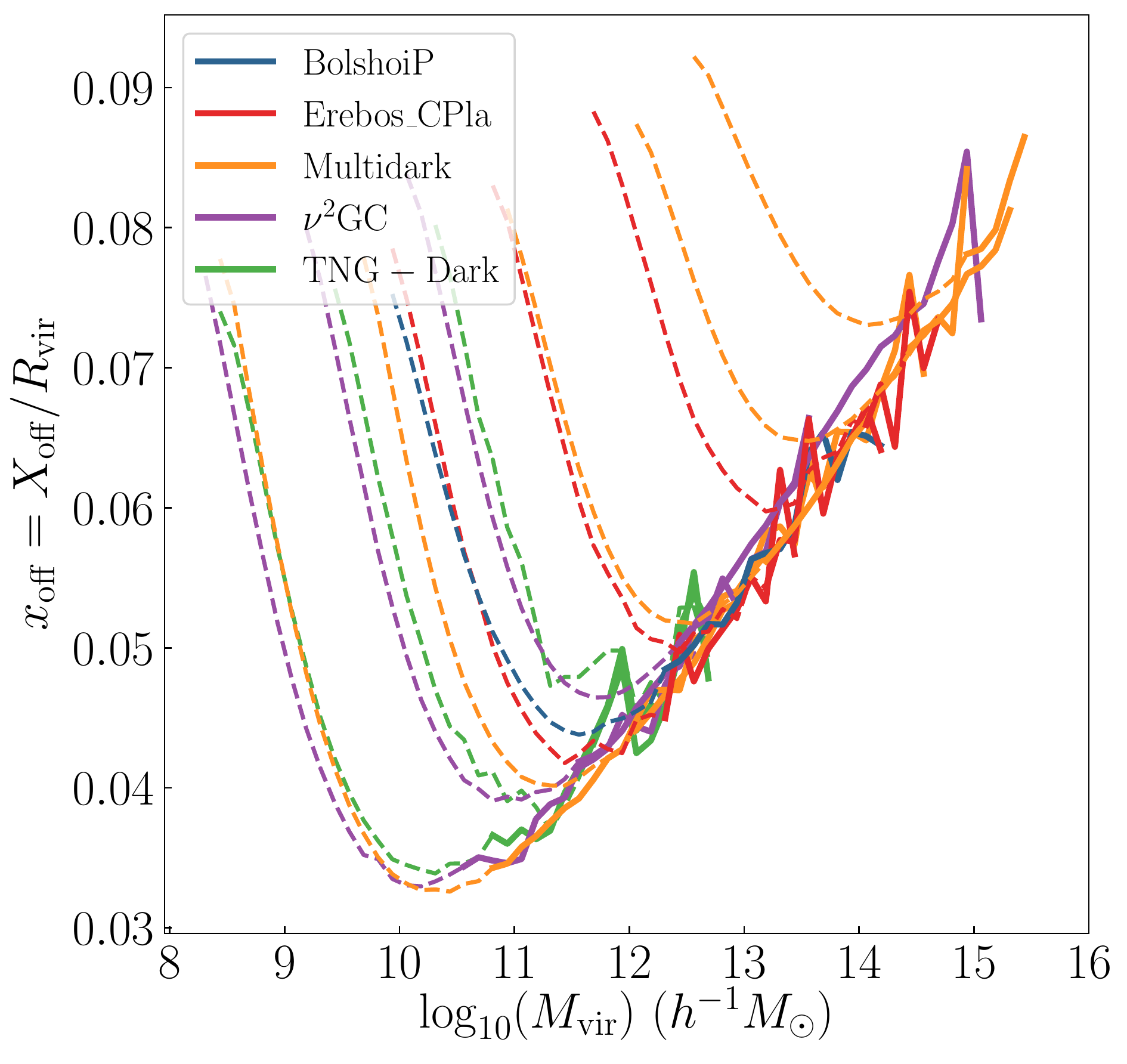}}
   \subfigure{\includegraphics[width=0.49\textwidth]{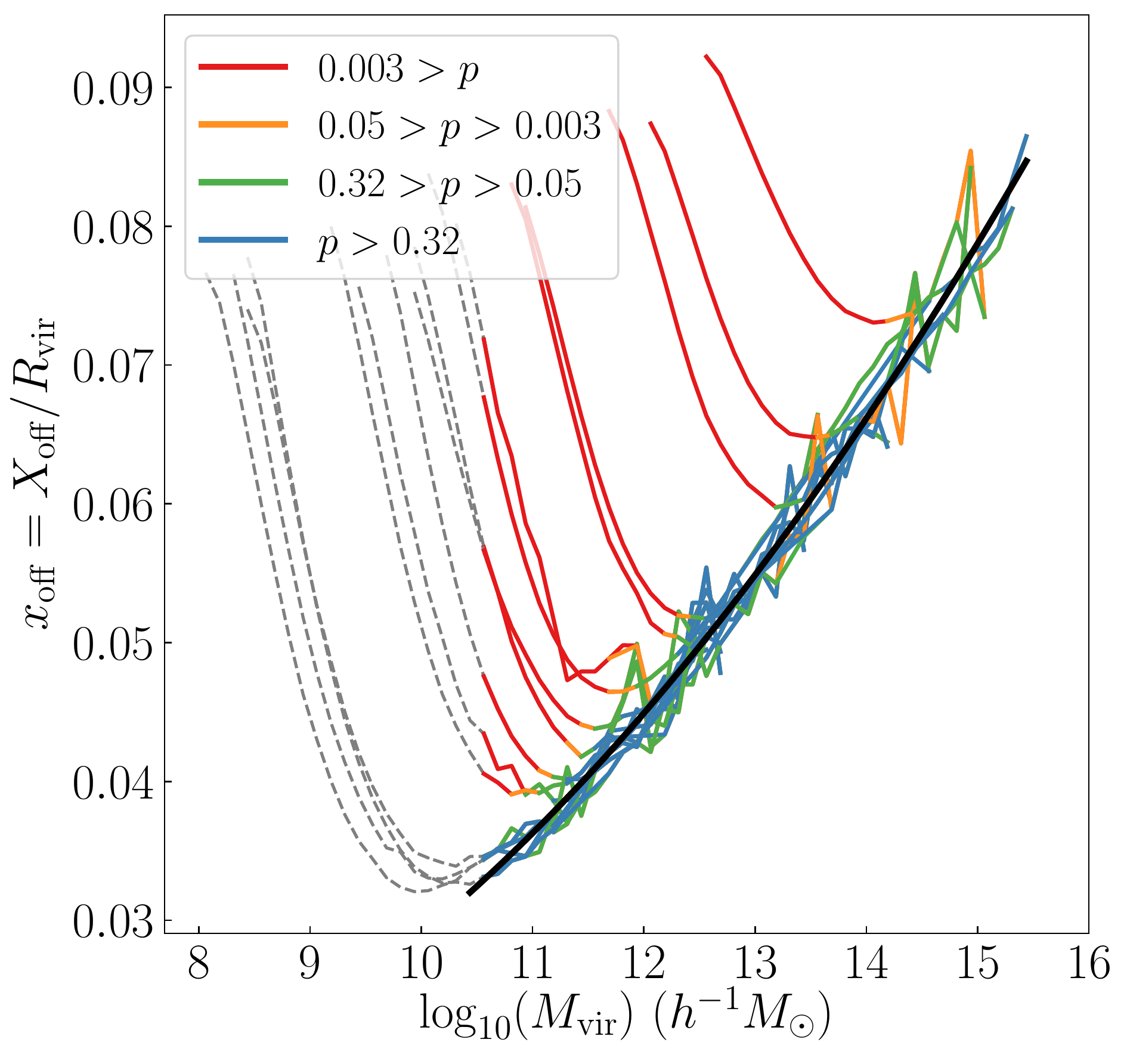}}
\caption{Illustration of the procedure used to determine the particle counts at which simulations diverge from the global, high-resolution mass trend. This is illustrated with the halo property $x_{\rm off}$ for isolated haloes in Planck-cosmology simulations. We fit subhaloes and WMAP-simulations separately. The left panel shows the steps prior to fitting. We collect all simulations with measured $x_{\rm off}$ values and visually identify a conservative $N_{\rm vir}$ cutoff such that all simulations fall along a single mass relation. Masses above this cutoff are shown as solid curves and masses below this cutoff are shown as dashed lines. Simulations are coloured by the suite which they belong to (Table \ref{tab:simulations}). In the right panel, we show the fitting and significance estimation steps. The fit is shown by the black line. After this, we estimate the significance of deviations from the fit using the procedure described in Appendix \ref{sec:significance}. Curves are shown as dashed grey lines when they are outside the range of the fit and are coloured by the significance of the deviation, $p,$ when they are within the range of the fit. Convergence limits are the lowest mass where curves transition from orange to green. The full procedure is outlined \ref{sec:procedure} and described in detail in Appendix \ref{sec:procedure_app}. Fits can be found in the online supplement.
}
\label{fig:procedure}
\end{figure*}

For each halo property, $X,$  we determine its `convergence limits:' the $N_{\rm vir}$ values at which $\langle X(M_{\rm vir})\rangle_s$ for each simulation, $s,$ deviates from the combined mass relation, $\langle X(M_{\rm vir})\rangle_{\rm HR},$ of higher resolution simulations.

Studies can accommodate varying levels of numerical bias in different halo properties. We therefore parametrize convergence limits by the desired fractional accuracy, $\delta.$ We take $\delta=0$ as our fiducial choice throughout this paper, in which case our tests reduce to detecting statistically significant differences between $\langle X(M_{\rm vir})\rangle_s$ and $\langle X(M_{\rm vir})\rangle_{\rm HR}$. However, this should not be taken as a normative statement that simulations should never be used below such limits.

We give the full details and additional discussion of our procedure in Appendix \ref{sec:procedure_app}. We summarise the key steps below and illustrate those steps in Fig.~\ref{fig:procedure}.
\begin{enumerate}
    \item Before analysis, we split simulations by cosmology (WMAP- and Planck-based parameters). Within each cosmology, we split halos by subhalo and isolated halo status. We analyse the four subgroups separately (Appendix \ref{sec:subgroups}).
    \item For each property, we identify conservative $N_{\rm vir}$ limits, $N_{\rm HR}$, which define the ``high resolution'' mass range of each simulation. We choose $N_{\rm HR}$ such that the $z=0$, $\langle X(M_{\rm vir}) \rangle_s$ relations agree for all $s$ for halos with $N_{\rm vir} > N_{\rm HR}$ (Appendix \ref{sec:hr_ranges}).
    \item For some halo properties, a handful of simulations deviate from other simulations at abnormally high $N_{\rm vir}.$ We do not consider these simulations when setting $N_{\rm HR}$ and do not include them in subsequent fits. These simulations are discussed in Appendix \ref{sec:fitting} and Section \ref{sec:results}. Their existence features heavily in Sections \ref{sec:results} - \ref{sec:debias}.
    \item Using mass ranges where $N_{\rm vir}>N_{\rm HR},$ we fit for a combined high-resolution mass relation, $\langle X(M_{\rm vir}) \rangle_{\rm HR}$ (Appendix \ref{sec:fitting}).
    \item We measure the deviation $\langle X(M_{\rm vir})\rangle_s - \langle X(M_{\rm vir})\rangle_{\rm HR}.$ To measure the significance of this deviation, we use the $z$-test to compute the probability that this deviation could be measured under the null hypothesis, $H_0(\delta),$ that the fractional difference between these two mass relations is smaller than some minimum tolerance, $\delta.$ For the $z$-test, we consider the mass-dependent sample variance in simulation $s$ and the dispersion of high-resolution simulations around $\langle X(M_{\rm vir})\rangle_{\rm HR}$. We take $p$-value less than 0.05 to indicate non-convergence (Appendix \ref{sec:significance}).
\end{enumerate}

We illustrate this procedure in Fig.~\ref{fig:procedure}. The left panel shows steps (i) and (ii). Shown are $\langle x_{\rm off}(M_{\rm vir})\rangle_s$ for isolated haloes in Planck-cosmology simulations. Simulations are coloured by the suite they belong to and transition from solid to dashed when $N_{\rm vir}$ drops below $N_{{\rm HR}, x{\rm off}}=10^4.$ A similar plot exists for WMAP-cosmology isolated haloes, Planck-cosmology subhaloes, and WMAP-cosmology subhaloes. Only the solid potions of curves are used when fitting $\langle x_{\rm off}(M_{\rm vir})\rangle_{\rm HR}$ in step (iv). As is the case for most halo properties, no simulations need to be removed from the sample, so step (iii) is skipped. 

The right panel shows steps (iv) and (v). The best-fitting $\langle x_{\rm off}(M_{\rm vir})\rangle_{\rm HR}$ is shown as a black curve and $\langle x_{\rm off}(M_{\rm vir})\rangle_s$ is coloured by the probability of observing $|\langle x_{\rm off}(M_{\rm vir})\rangle_s - \langle x_{\rm off}(M_{\rm vir})\rangle_{\rm HR}|$ under the null hypothesis, $H_0(\delta=0).$ Simulation-specific convergence limits are the $M_{\rm vir}$ where curves transition from orange to green. Curves are dashed and grey below the lowest mass where $\langle x_{\rm off}(M_{\rm vir})\rangle_{\rm HR}$ is valid.


\section{The Empirical $N_{\rm vir}$ Convergence Limits of Simulations}
\label{sec:results}

\begin{table}
  \centering
  \caption{An excerpt of the measured particle count cutoffs, $N_{\rm vir}$, associated with different halo properties, simulations, halo isolation classifications, and tolerance levels. $N_{\rm iso,\delta}$ indicates the number of particles where fractional deviations larger than a tolerance level of $\delta$ can be reliably measured from the mean value of the given halo property for isolated haloes. $N_{\rm sub,\delta}$ indicates the corresponding value for subhaloes.  An empty value indicates that we cannot make a reliable measurement, often due to the high resolution of the simulation. The full table is available at \texttt{https://github.com/phil-mansfield/halo\_convergence} with accuracies ranging from $\delta=0$ to $\delta=0.10.$}
  \label{tab:sample_table}
  \begin{tabular}{llccc}
        \hline\hline
        Property & Simulation & $N_{\rm iso,0.00}$ & $N_{\rm sub,0.00}$ & $N_{\rm iso,0.01}$ \\
        \hline
        $x_{\rm off}$ & Bolshoi & $2.8\times10^3$ & --- & $2.8\times10^3$ \\
        $x_{\rm off}$ & BolshoiP & $4.2\times10^3$ & $1.8\times10^3$ & $4.2\times10^3$ \\
        $x_{\rm off}$ & Chinchilla\_L125 & --- & --- & --- \\
        $x_{\rm off}$ & Chinchilla\_L250 & $1.4\times10^4$ & $6.0\times10^3$ & $1.4\times10^4$ \\
        & & ... & & \\
        \hline
    \end{tabular}
\end{table}

\label{sec:deviations}
\begin{table}
  \centering
  \caption{The particle count cutoffs at which 90\% of the simulations in our sample show no measurable deviation from fits to high-resolution halo samples for various halo properties. We show these cutoffs, $N_{\rm vir}$, for mean mass relations, $\langle X(M_{\rm vir})\rangle$, in the top block. The middle block shows the particle count cutoff, $N_X$, for each corresponding mass function $\phi(M_X)$. We provide cutoffs for both isolated and subhalo populations. Stars indicate limits that we cannot express with $N_{\rm vir}$ or $N_X$ alone because of a strong dependence on the force softening scale, $\epsilon$.  All numbers are accurate to 0.125 dex. The online supplement at \texttt{https://github.com/phil-mansfield/halo\_convergence} contains the cutoffs measured for individual simulations at varying degrees of accuracy. This includes starred halo properties.
  }
  \label{tab:iso_cut}
  \begin{tabular}{lll}
    \hline\hline
    Variable & $N_{\rm iso}$ &$N_{\rm sub}$ \\
  \hline
  $M_{\rm 2500c}/M_{\rm vir}$ & $\star$ & $\star$ \\
  $M_{\rm 500c}/M_{\rm vir}$ & $8.5 \times 10^{2}$ & $3.2 \times 10^{2}$ \\
  $M_{\rm 200c}/M_{\rm vir}$ & $1.3 \times 10^{2}$ & $1.4 \times 10^{2}$ \\
  $M_{\rm 200b}/M_{\rm vir}$ & $1.1 \times 10^{2}$ & $1.2 \times 10^{2}$ \\
              $V_{\rm max}$ & $\star$ & $\star$ \\
              $V_{\rm rms}$ & $\star$ & $\star$ \\
              $c_{\rm vir}$ & $\star$ & $\star$ \\
                  $R_{1/2}$ & $3.5 \times 10^{3}$ & $4.6 \times 10^{3}$ \\
                      $c/a$ & $\star$ & $\star$ \\
    $\lambda_{\rm Peebles}$ & $4.5 \times 10^{2}$ & $3.9 \times 10^{2}$ \\
    $\lambda_{\rm Bullock}$ & $1.1 \times 10^{2}$ & $4.9 \times 10^{2}$ \\
                    $T/|U|$ & $\star$ & $\star$ \\
              $x_{\rm off}$ & $2.9 \times 10^{3}$ & $1.2 \times 10^{3}$ \\
              $V_{\rm off}$ & $4.8 \times 10^{3}$ & $1.7 \times 10^{3}$ \\
      $\Gamma(t_{\rm dyn})$ & $1.1 \times 10^{2}$ & 83 \\
                  $a_{0.5}$ & $1.4 \times 10^{2}$ & 93 \\
               $a_{\rm MM}$ & $2.7 \times 10^{2}$ & $1.3 \times 10^{2}$ \\
  $M_{\rm peak}/M_{\rm vir}$ & $3.5 \times 10^{2}$ & $1.1 \times 10^{2}$ \\
             $V_{\rm peak}$ & $\star$ & $\star$ \\
  \hline
      $\phi(M_{\rm 2500c})$ & $\star$ & $\star$ \\
       $\phi(M_{\rm 500c})$ & $1.6 \times 10^{2}$ & $\star$ \\
       $\phi(M_{\rm 200c})$ & $1.6 \times 10^{2}$ & $\star$ \\
        $\phi(M_{\rm vir})$ & $1.5 \times 10^{2}$ & $\star$ \\
       $\phi(M_{\rm 200b})$ & $1.2 \times 10^{2}$ & $\star$ \\
       $\phi(M_{\rm peak})$ & $1.4 \times 10^{2}$ & $\star$ \\
  \hline
        $\phi(V_{\rm max})$ & $\star$ & $\star$ \\
        $\phi(V_{\rm rms})$ & $\star$ & $\star$ \\
       $\phi(V_{\rm peak})$ & $\star$ & $\star$ \\
  \hline
  \end{tabular}
\end{table}

\subsection{Typical Convergence Limits}
\label{sec:typical}

\begin{figure}
   \centering
   \subfigure{\includegraphics[width=0.49\textwidth]{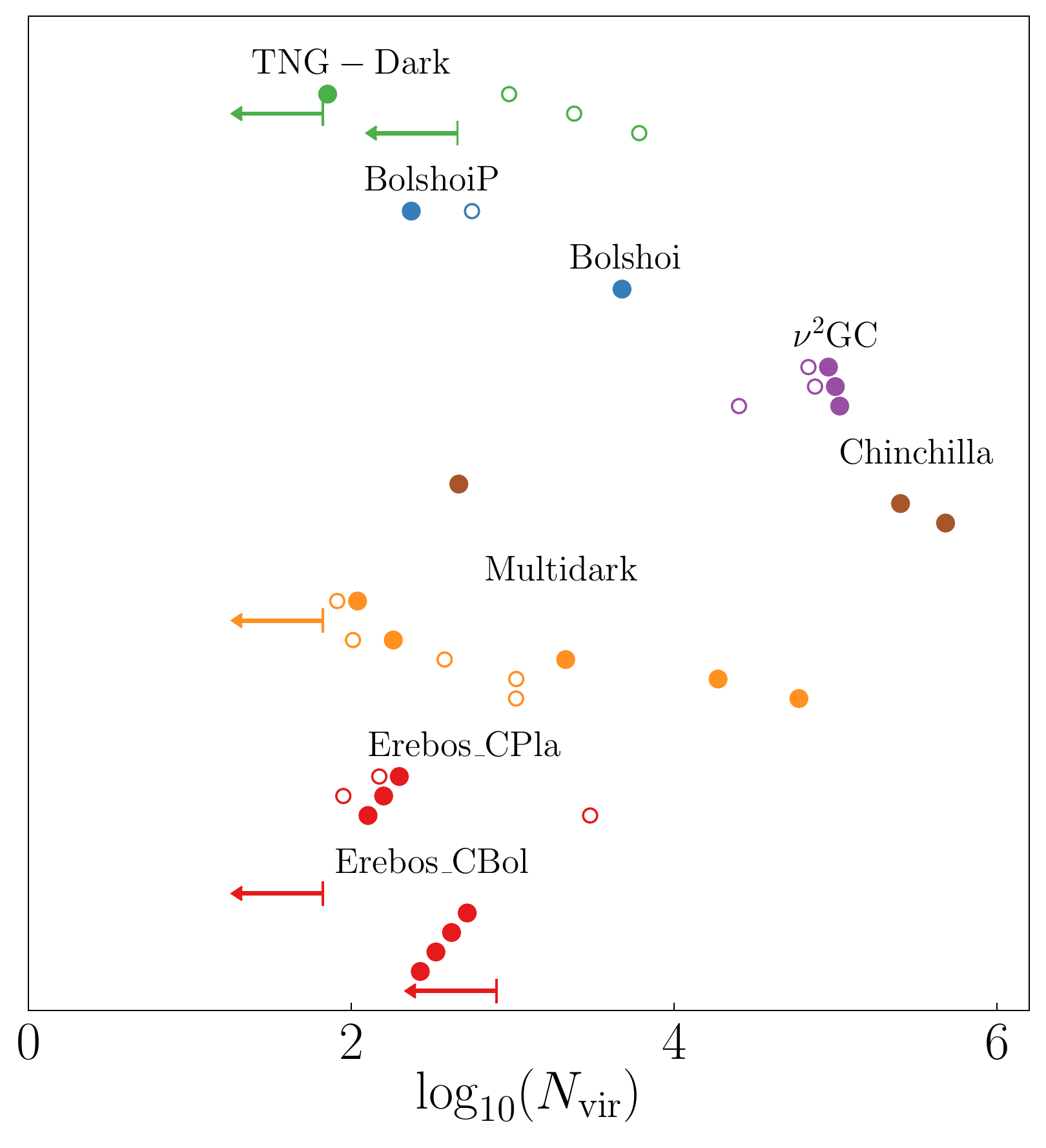}}
\caption{The $N_{\rm vir}$ values below which numerical effects measurably bias mean $V_{\rm max}$ value for isolated haloes in each simulation in Table \ref{tab:simulations}. These are conservative limits. There is significant variation in these limits from simulation to simulation. We first use colour to group simulations by suite, then vertically order the simulations by particle mass; the bottom dot in each suite corresponds to the highest resolution box of that suite. We use points to indicate simulations where we measure diverging behaviour and upper limits for simulations where we were not able to measure a divergence.  As discussed in Section \ref{sec:multidark_illustris}  and Appendix \ref{sec:fitting}, the two highest resolution Planck suites, TNG-Dark and Multidark, appear to converge to two different $\langle V_{\rm max}(M_{\rm vir})\rangle$ relations. The solid circles show cutoff values when the high-resolution fit does not include Multidark boxes and the open circles show cutoff values when the high-resolution fit does not include TNG-Dark.}
\label{fig:variation}
\end{figure}

\begin{figure*}
   \centering
   \subfigure{\includegraphics[width=0.48\textwidth]{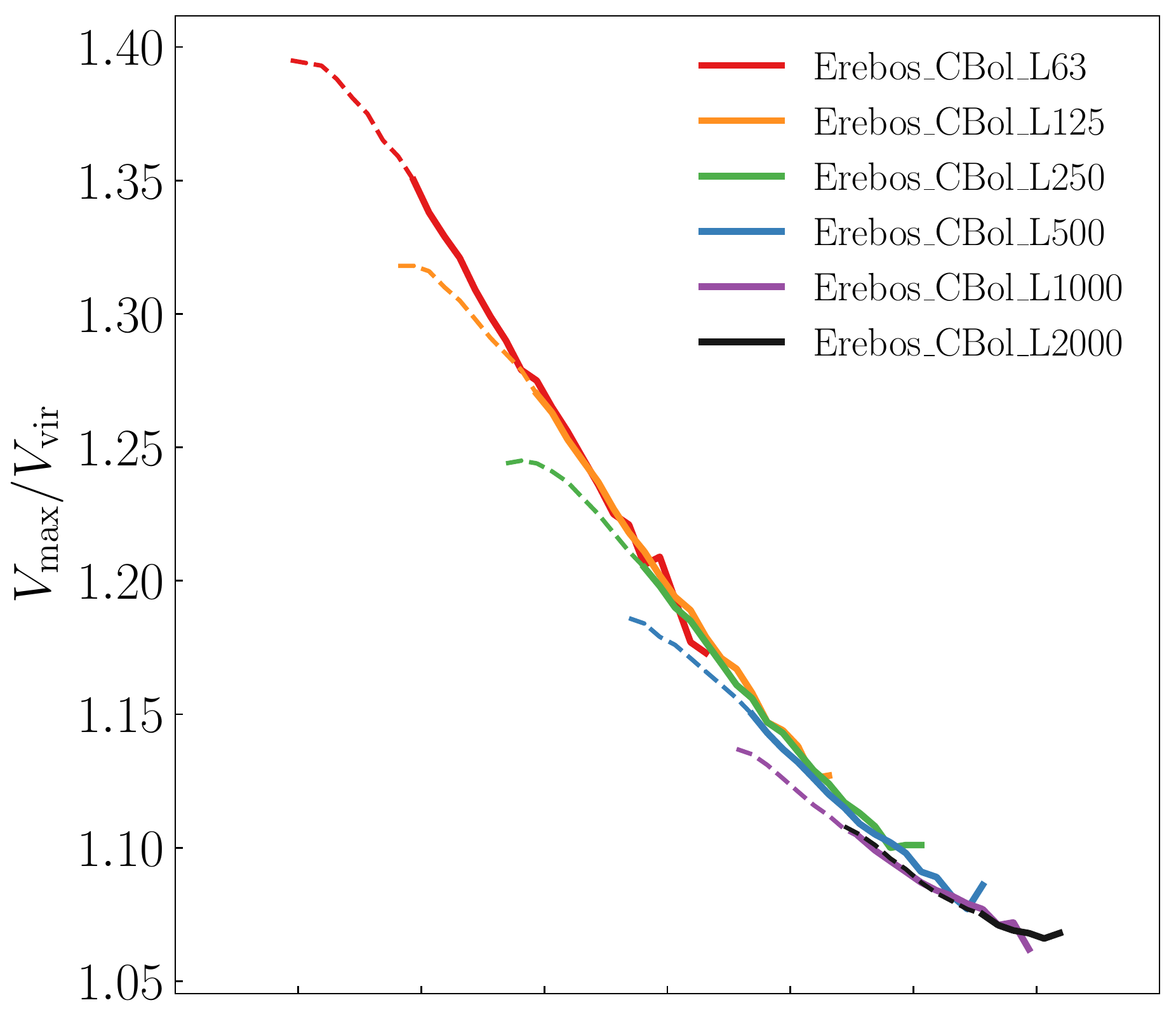}}\ \ \ \ 
   \subfigure{\includegraphics[width=0.475\textwidth]{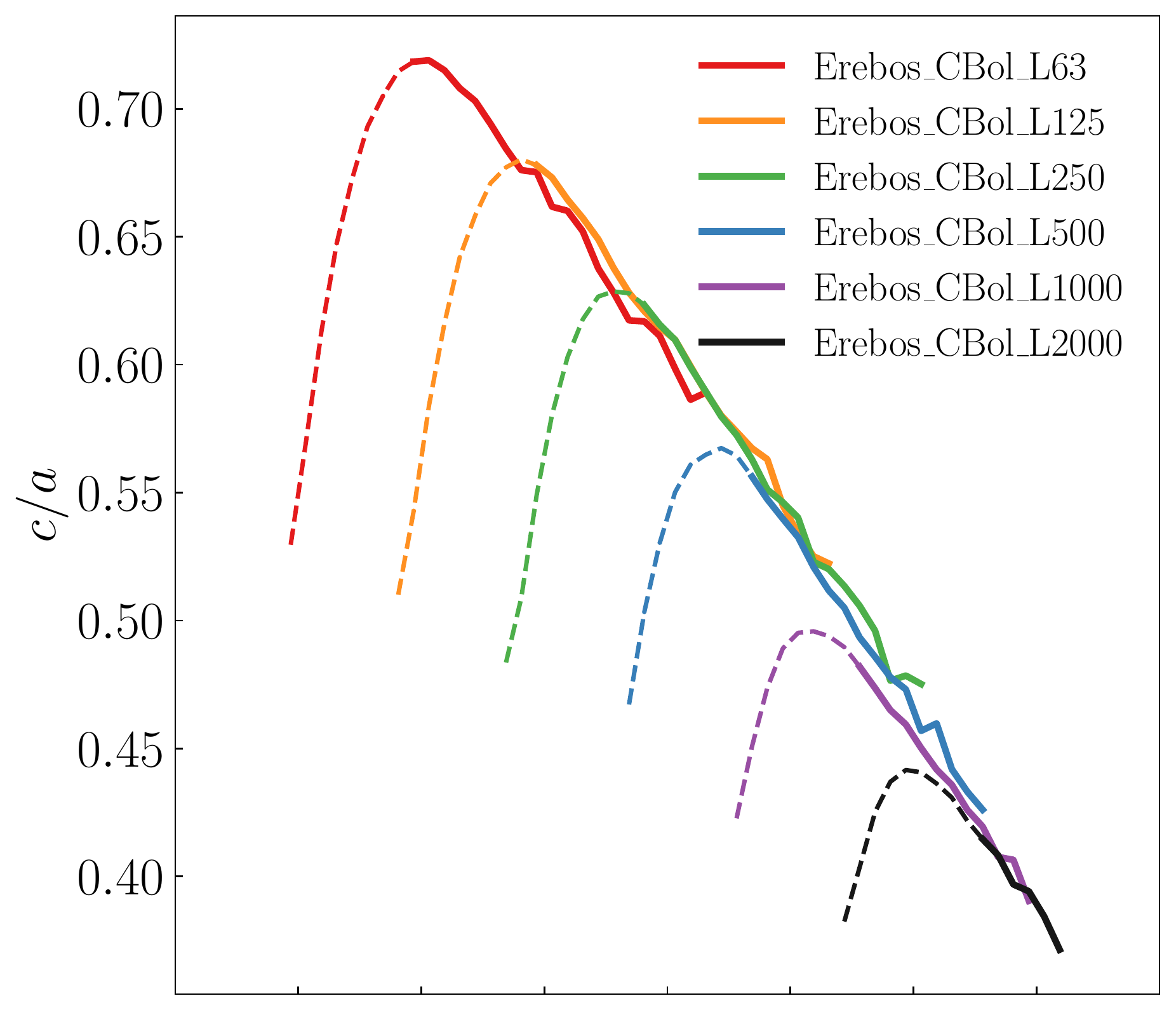}} \\[-2ex]
   \subfigure{\includegraphics[width=0.49\textwidth]{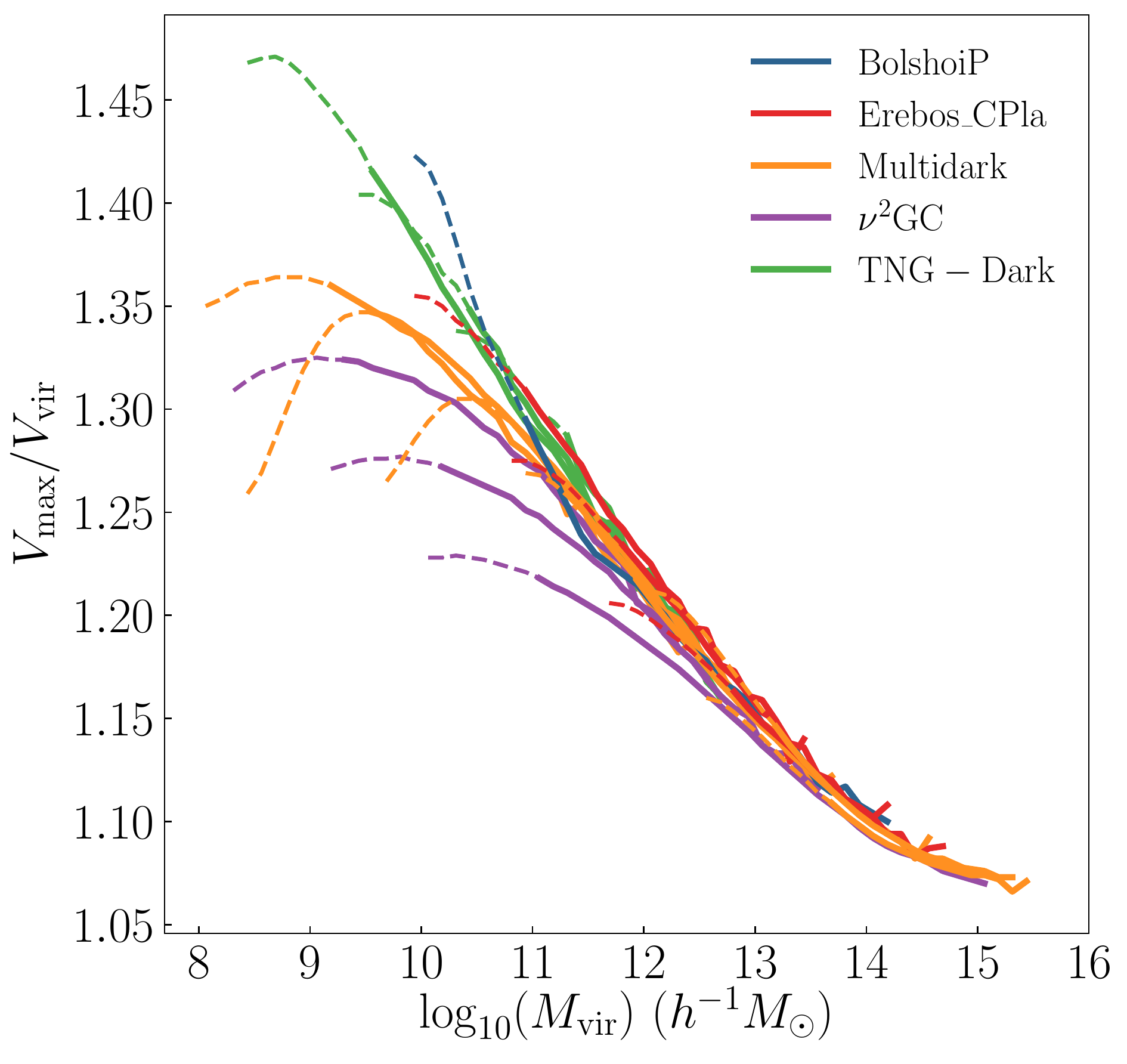}}
   \subfigure{\includegraphics[width=0.49\textwidth]{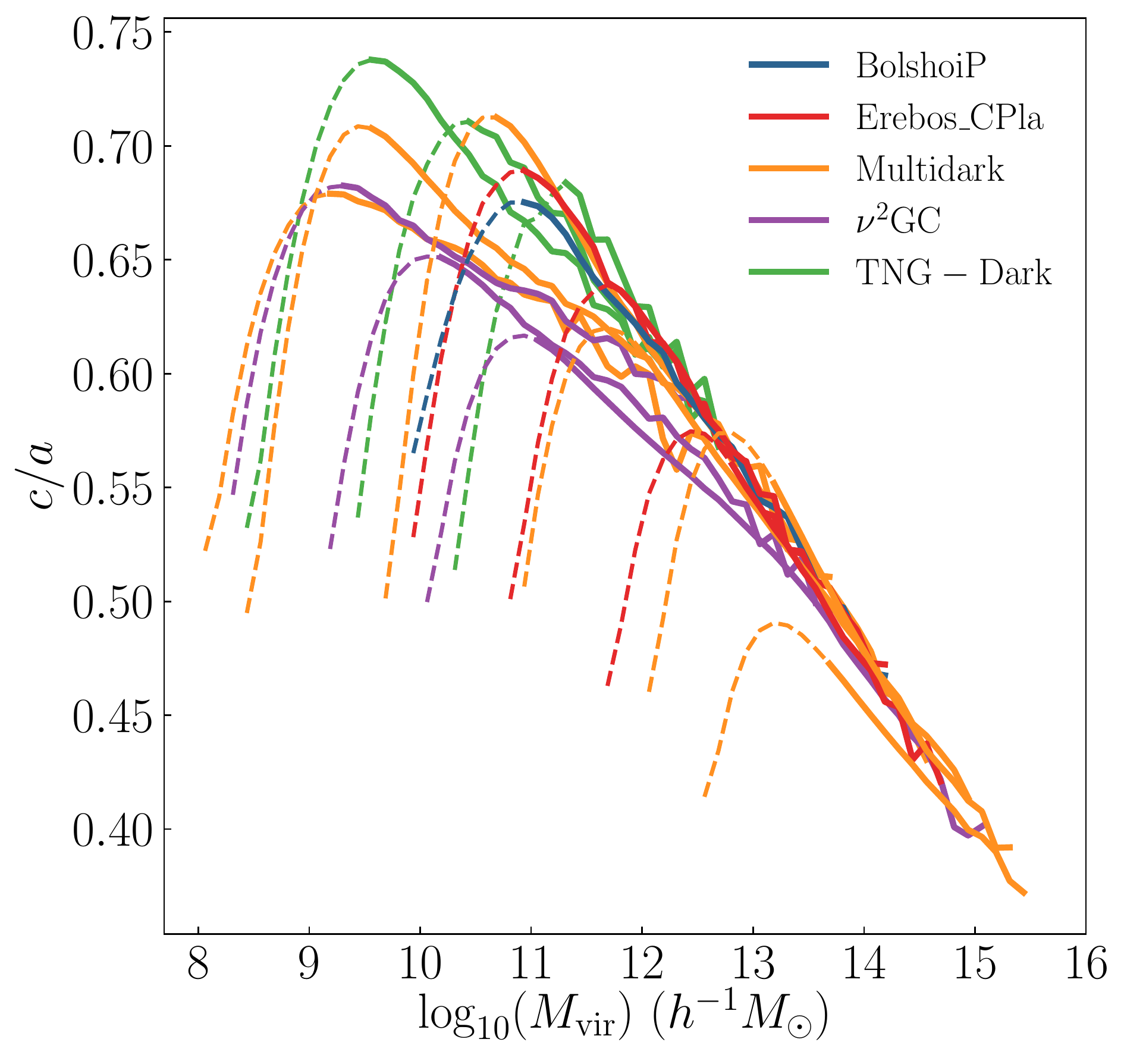}} \\
\caption{The convergence behaviour of $\langle V_{\rm max}/V_{\rm vir}(M_{\rm vir})\rangle$ and $\langle c/a(M_{\rm vir})\rangle$ for isolated haloes. Top: `Classical' single-suite convergence tests using the six boxes from the Erebos\_CBol suite. Each curve colour corresponds to a different box, and the linestyles transition from solid to dashed at $N_{\rm vir} \leq 500.$ In isolation, these plots imply that rotation curve peaks and halo shapes measured for halos above 500 particles are converged. Bottom: The $\langle V_{\rm max}/V_{\rm vir}(M_{\rm vir})\rangle$ and $\langle c/a(M_{\rm vir})\rangle$ relations for every Planck-cosmology simulation in Table \ref{tab:simulations}. Simulations are coloured by suite and the solid-to-dashed transition still occurs at $N_{\rm vir}\leq 500.$ There is disagreement between the mass relations well above the convergence limit implied by the top plots. The bottom two panels contain a number of noteworthy features which we highlight in section \ref{sec:limit_variation}.}
\label{fig:mass_trends}
\end{figure*}
We use the procedure described in Section \ref{sec:procedure} to find the particle counts, $N_{\rm vir}$, at which each simulation in Table \ref{tab:simulations} deviates from high resolution fits for a given mass relation, $\langle X(M_{\rm vir})\rangle$, of a halo property $X$. Table \ref{tab:sample_table} shows example particle count cutoffs for $X=x_{\rm off}$, with particle count cutoffs provided for both isolated haloes and subhaloes and example fractional accuracy tolerances of $\delta=0.00$ and $\delta=0.01$. The online supplement of this paper includes results for all properties listed in Section \ref{sec:properties} with accuracy tolerances ranging from $\delta=0.00$ to $\delta=0.10$, \footnote{\texttt{https://github.com/phil-mansfield/halo\_convergence}}. Blank table entries indicate that we were not able to make a reliable measurement of a deviation from the mean mass relation for that property at that accuracy tolerance.

In Table \ref{tab:iso_cut}, we show conservative `convergence limits' for many halo properties. These correspond to $N_{\rm vir}$ values at which 90\% of the simulations in our sample show no measurable deviation from high resolution fits ($\delta=0$). One can safely assume that haloes above these $N_{\rm vir}$ limits do not suffer significant numerical biases. Using halo properties from haloes below these limits may be acceptable for many types of analysis, but the prerogative falls upon the authors of such analyses to understand how numerical biases impact their results. We recommend that any analysis using haloes with $N_{\rm vir}$ below these limits either use the accuracy-dependent limits in Table \ref{tab:sample_table} or explicitly perform resolution tests. In either case, we recommend that numerical biases be explicitly included in such analyses' error estimates. 

For each halo property in Table \ref{tab:iso_cut}, we have performed detailed tests on how strongly this property depends on $\epsilon$  (see Section \ref{sec:eps_dependence_mass_trend}). Properties which strongly depend on $\epsilon$ have been marked by a $\star,$ as we cannot express convergence limits in terms of $N_{\rm vir}$ alone. The limits in Table \ref{tab:sample_table}, which correspond to individual simulation boxes are still valid for such properties

It is difficult to compare this table to previous tests in the literature. For most of the common properties with existing testing literature  (e.g., $V_{\rm rms}$; \citealp{Evrard_et_al_2008} or $T/|U|$; \citealp{power_et_al_2012}), we conclude that there is such a strong dependence on $\epsilon$ that we cannot endorse a single $N_{\rm vir}$ limit. For many of the remaining properties, such as, $x_{\rm off}$ or $a_{MM},$ we are not aware of any previous convergence tests. That said, we note that our cutoff for $\lambda_{\rm Peebles}$ is consistent with the results of \citet{Villearreal_et_al_2017}, and that our criteria for isolated halo abundances are consistent with existing literature on the topic \citep[e.g.][]{Aungulo_et_al_2012,Ishiyama_et_al_2015,Ludlow_et_al_2019}, although different authors adopt different target accuracies. Finally, we note that our input catalogues did not have subhaloes with fewer than 50 particles; we were not able to put competitive constraints on mass definitions with limits near or below this value.

\subsection{Variation in Limits Between Simulations}

\label{sec:limit_variation}

Fig.~\ref{fig:variation} shows the $N_{\rm vir}$ values at which $\langle V_{\rm max}(M_{\rm vir})\rangle$ for host haloes in every simulation in our suite measurably deviate from from high resolution fits to $\langle V_{\rm max}(M_{\rm vir})\rangle$. These values correspond to $N_{\rm iso,0.00}$ in Table \ref{tab:sample_table}. These $N_{\rm vir}$ limits are conservative ($\delta=0$), and applications which can accommodate modest biases in $\langle V_{\rm max} \rangle$ may be able to use haloes with smaller values of $N_{\rm vir}.$ Nevertheless, the limits shown in Fig.~\ref{fig:variation} show the resolution scales at which numerical effects begin to measurably influence the behaviour of the $V_{\rm max}$ distribution.

There is substantial variation in these convergence limits from simulation to simulation, with several simulations only reaching full statistical convergence at $10^5-10^6$ particles. Because TNG-Dark and Multidark appear to converge to different $V_{\rm max}$ relations (see Section \ref{sec:multidark_illustris}), we perform this analysis twice with separate fits to both suites. The filled-in circles correspond to the fit to TNG-Dark and the empty circles correspond to the fit to Multidark. Note that the limits for TNG-Dark become higher when Multidark is used to fit low-mass haloes, and the opposite is true when TNG-Dark is used. However, the overall scatter in the convergence limits does not depend on this choice.  Note that simulations with WMAP-like cosmologies (Chinchilla, Erebos\_CBol, Bolshoi) are unaffected by this fitting choice because they were fit separately.

The simulation-to-simulation variation in convergence limits is not an artefact of our convergence procedure. In Fig.~\ref{fig:mass_trends} we qualitatively demonstrate this effect for $V_{\rm max}$ and another commonly used halo property, $c/a$. 

The top panels of Fig.~\ref{fig:mass_trends} show a `classical,' single-suite, convergence test for $V_{\rm max}$ and $c/a$, using the seven boxes in the Erebos\_CBol suite. We show the mass relations, $\langle V_{\rm max}(M_{\rm vir})\rangle$ and $\langle c/a(M_{\rm vir})\rangle$, for isolated haloes using different colours for each box in the suite. The curves are solid for halo masses corresponding to $N_{\rm vir}>500$ and dashed for halo masses below this particle count. These simulations agree with one another above this visually-identified convergence limit.  There is some slight variation in the amplitude due to sample variance. This agreement {\em seems} to indicate that both quantities are converged above 500 particles.

However, we do not find such agreement when comparing across simulation suites. The bottom two panels of Fig.~\ref{fig:mass_trends} show the same mass relations for all of our Planck-cosmology simulations. Most of the simulations have many times more particles than boxes in the Erebos\_CBol suite and many go to far smaller $m_p$. As in the top panels, the curves are solid above 500 particles and dashed below. Unlike the top panels, there can be disagreement between the simulations at halo masses corresponding to approximately $10^5$ particles, {\em even for simulations in the same suite}.

We have ruled out many factors other than of numerical non-convergence that could potentially cause a difference in these mass relations. We address these factors in other sections of this paper, but we collect them here for convenience.
\begin{itemize}
    \item As discussed in Appendix \ref{sec:rockstar_versions}, we cross-matched catalogues to demonstrate that varying versions, bugs, and parametrizations of the \textsc{Rockstar} halo finder cannot cause this disagreement.
    \item The statistically estimated cutoffs shown in Fig.~\ref{fig:variation} are consistent in detail with the qualitative disagreement shown in Fig.~\ref{fig:mass_trends}. As described in Section \ref{sec:procedure} and Appendix \ref{sec:significance}, our statistical cutoffs explicitly account for sample variance. Additionally, the disagreement extends to some very large boxes, such as $\nu^2$GC-L. This means that the disagreement is not caused by sample variance.
    \item Fig.~\ref{fig:mass_trends} only contains isolated haloes, so this disagreement cannot be due to the stricter convergence criteria on subhalo resolution. We have also inspected the distribution of halo properties at a constant mass and determined that a small population of outliers is not driving the differences.
    \item In some cases, simulations which diverge from the typical mass relation will also diverge from other simulations in the same suite. Simulations within the same suite use identical codes, identical cosmologies, and nearly identical initial conditions setups. This means that differences of this type cannot be the sole cause of the disagreement.
\end{itemize}

\subsection{Differences Between Multidark and Illustris-TNG}
\label{sec:multidark_illustris}

\begin{figure}
   \centering
    \subfigure{\includegraphics[width=0.49\textwidth]{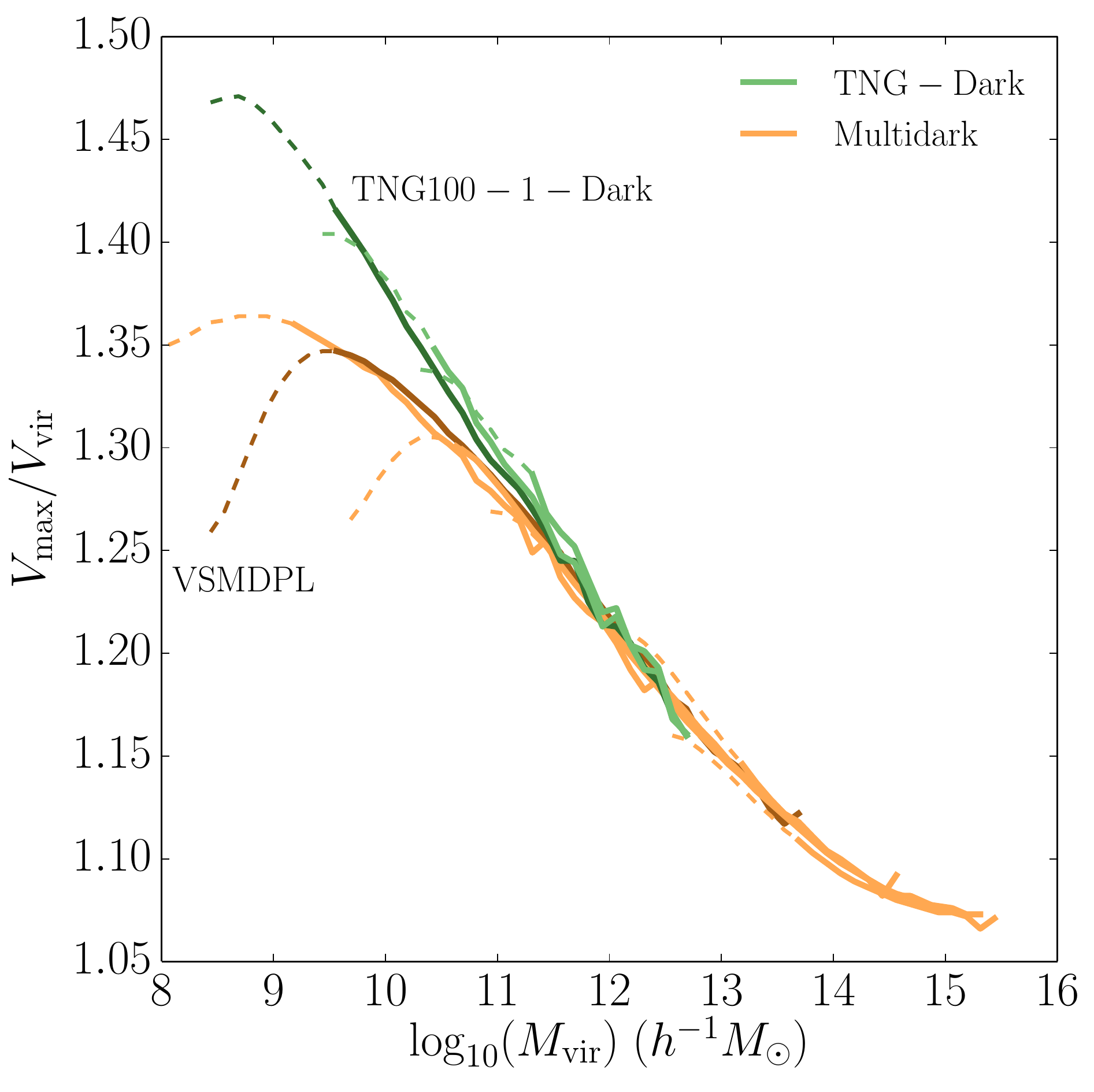}}
\caption{The same as the lower left panel of Fig.~\ref{fig:mass_trends}, except restricted to TNG-Dark and Multidark boxes. TNG100-1-Dark and VSMDPL, which have very similar parametrizations, are emphasised with darker colours. The two suites have converged to different solutions.}
\label{fig:multidark_illustris}
\end{figure}

As mentioned above, the Multidark suite and the Illustris-TNG-Dark suite appear to have converged to two different $\langle V_{\rm max}(M_{\rm vir})\rangle$ relations below $M_{\rm vir} \lesssim 10^{11}h^{-1}M_\odot$. We illustrate this in Fig.~\ref{fig:multidark_illustris}.  The orange lines correspond to the mass relations of boxes from Multidark simulations, and the green lines to boxes from TNG-Dark. Linestyle has the same meaning as in Fig.~\ref{fig:mass_trends}. The difference in mass relations emphasises the known fact that convergence within a single simulation suite is not sufficient to establish that a simulation is unbiased.  We discuss this point in Section \ref{sec:intro}. 

In this Section, we focus on isolating the potential causes of this difference without invoking on any specific convergence model. In Section \ref{sec:debias}, we will present a model which predicts that this variation is mostly caused by differences in $\epsilon.$

Our focus in this Section will primarily be on the TNG100-1-Dark and VSMDPL boxes, which have very similar parametrizations. We label the $\langle V_{\rm max}(M_{\rm vir})\rangle$ relations for both simulations in Fig.~\ref{fig:multidark_illustris}, emphasising them with darker line colours. We note that both simulation boxes satisfy {\it internal convergence}, given the consistency of each mass relation with the mass relations of higher resolution boxes of the same suite.

Numerical differences between cosmological simulations come from a finite list of sources: cosmology, sample variance, halo finders, box size, $m_p,$ timestepping, force softening scheme, $\epsilon,$ initial condition generation, code parameters, and code algorithms/versions.  We discuss each potential source below.

{\bf Cosmology}: Cosmology cannot cause the difference between TNG100-1-Dark ($\Omega=0.309$, $h_{100}=0.677$, $\sigma_8=0.8159$) and VSMDPL ($\Omega_m=0.307$, $h_{100}=0.678$, $\sigma_8=0.832$), as the Planck-like parameters they adopt are almost identical. 

{\bf Sample Variance}: Sample variance can never lead to false convergence because such fluctuations would be uncorrelated with simulation suite. Additionally, we find that the sample variance estimated through jackknife resampling for $\langle V_{\rm max}(M_{\rm vir})\rangle$ is small relative to the difference between the suites. 

{\bf Halo Finding}: Halo finder inconsistencies are not a cause of non-convergence, as discussed at length in Appendix \ref{sec:rockstar_versions}. 

{\bf Box Size}: Box size effects are unlikely to be a significant contributor.  Although TNG100-1-Dark ($L=75\,h^{-1} {\rm Mpc}$) and VSMDPL ($L=160\,h^{-1}{\rm Mpc}$) have different box sizes, tests from \citet{Power_Knebe_2006} rule out small-box effects of this magnitude for an $L=75\,h^{-1}{\rm Mpc}$ box.

{\bf Particle Mass}: TNG100-1-Dark ($m_p=6.00\times10^6\,h^{-1}M_\odot$) and VSMDPL ($m_p=6.16\times 10^6\,h^{-1}M_\odot$) have almost identical particle masses, meaning that the source of the difference cannot be related to mass resolution. 

{\bf Timestepping}: Timestepping is very similar between the simulations.  The codes, Gadget-2 and Arepo, use the same timestepping criteria, Eq.~\ref{eq:gadget_timestep}.  Both boxes use nearly identical values of $\eta$: TNG100-1 uses $\eta=0.012$ and VSMDPL uses $\eta=0.01$.

{\bf Force Softening}: Force softening cannot be ruled out as the source of the difference between these two simulations. While both simulations use the same softening scheme, Eq.~\ref{eq:gadget_softening}, TNG100-1-Dark and VSMDPL use softening lengths that differ by a factor of two: $\epsilon/l=0.012$ and $\epsilon/l=0.024$ respectively. In Section \ref{sec:debias} and Appendix \ref{sec:model_predctions_multidark_illustris}, we present a model which suggests that the differences in $\epsilon$ are, in fact, the main culprit.

{\bf Initial Conditions}: Both simulations generate initial conditions in similar ways. Both simulations initialise particle states with the Zel'dovich approximation; this approximation is followed down $z=100$ for VSMDPL \citep{Klypin_et_al_2016} and down to $z=127$ for TNG100-1-Dark \citep{Nelson_et_al_2019}. The impact of starting redshift is well-studied \citep[e.g.][]{Lukic_et_al_2007,Knebe_et_al_2009}, and this small difference in in starting redshift would not alter $z=0$ halo properties at the measured level.

{\bf Code Parameters}: Runtime code parameters (i.e. those defined in configuration files) for the simulations are unlikely to contribute to biases.  After review of the configuration files for both simulations (\citealp{Nelson_et_al_2019}; G. Yepes, personal communication), the only meaningful difference between parametrizations is $\alpha$ (also referred to as $f_{\rm acc},$ and \texttt{ErrTolForceAcc}), which sets the node opening criteria in Gadget's force tree. VSMDPL adopts $\alpha=0.01$, while TNG100-1-Dark adopts a more conservative $\alpha=0.0025.$  Tests from \citet{Power_et_al_2003} indicate that $\alpha=0.01$ can lead to density biases in regions of haloes with $N(<R)\lesssim 100$. However, the biases shown in Fig.~\ref{fig:multidark_illustris} occur at larger $N(<R)$, and the bias discussed in \citet{Power_et_al_2003} is strongly dependent on $m_p$.  The difference in $\alpha$ parametrization is therefore unlikely to contribute to biases that persist across multiple resolutions. While $\alpha$ does not appear to be a primary source of the measured difference, the impact of $\alpha$ deserves further study.

{\bf Code Version}: It is possible that differences in the Gadget gravity-solver contribute to the difference. The Multidark suite was run with LGadget-2, the same optimised version of Gadget-2 which was used to run the Millennium Simulation \citep{Springel_2005,Klypin_et_al_2016}, while IllustrisTNG-Dark was run with Arepo. Arepo's gravity solver is based on Gadget-2, but has implemented various bug fixes and algorithmic improvements over the years \citep{Weinberger_et_al_2019}. We cannot rule out that code changes contribute to the difference between the two suites.\footnote{We note that perhaps the most significant update in Arepo is the removal of a particular optimisation: dynamic force tree updates (\citealp{Weinberger_et_al_2019}; V. Springel, private communication). This optimisation led to force errors which were correlated with timestep sizes, and could bias small-$r$ mass distribution of haloes. Inspection of the LGadget-2 version used to run the Multidark simulations shows that it did not use this optimisation. This leaves only relatively minor code changes as potential culprits.}

Given the above discussion, we identify three potential sources of the differences between these simulations: force softening scale, force accuracy, and code differences. In Section \ref{sec:eps_dependence_mass_trend}, we show that many halo properties have a strong dependence on $\epsilon,$ and in Section \ref{sec:debias}, we present a model for $\epsilon$ biases which predicts that most -- but not all -- of the $V_{\rm max}$ difference between these simulations is caused by differences in $\epsilon.$

\section{The Dependence of Halo Properties on Force Softening Scale}
\label{sec:eps_dependence_mass_trend}

\begin{table}
  \centering
  \caption{Simulation parameters of the resimulated convergence boxes of the Chinchilla-$\epsilon$ resimulation suite. Shared parameters of these simulations are described in Section \ref{sec:eps_dependence_mass_trend}.} 
  \label{tab:chinchilla}
  \begin{tabular}{lll}
  \hline\hline
  Simulations name & $\epsilon$ ($h^{-1}$kpc) & $\epsilon/l$ \\
  \hline
    Chinchilla\_L125\_e1 & 1 & 0.0082\\
    Chinchilla\_L125\_e2 & 2 & 0.016\\
    Chinchilla\_L125\_e5 & 5 & 0.041\\
    Chinchilla\_L125\_e14 & 14 & 0.115\\
  \hline
  \end{tabular}
\end{table}

\begin{figure*}
   \centering
   \subfigure{\includegraphics[width=0.47\textwidth]{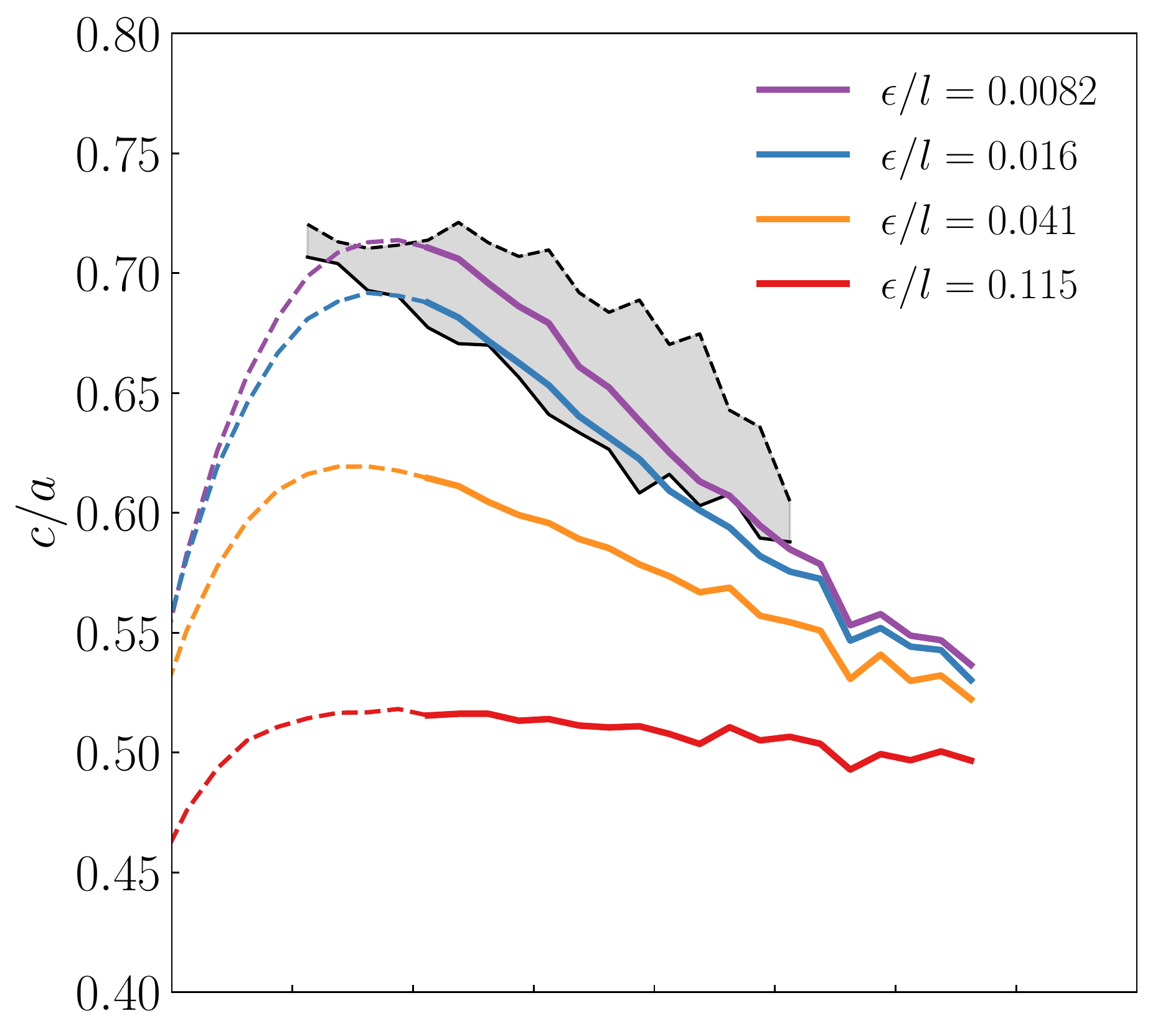}}\ \ \ \ \ \ 
   \subfigure{\includegraphics[width=0.47\textwidth]{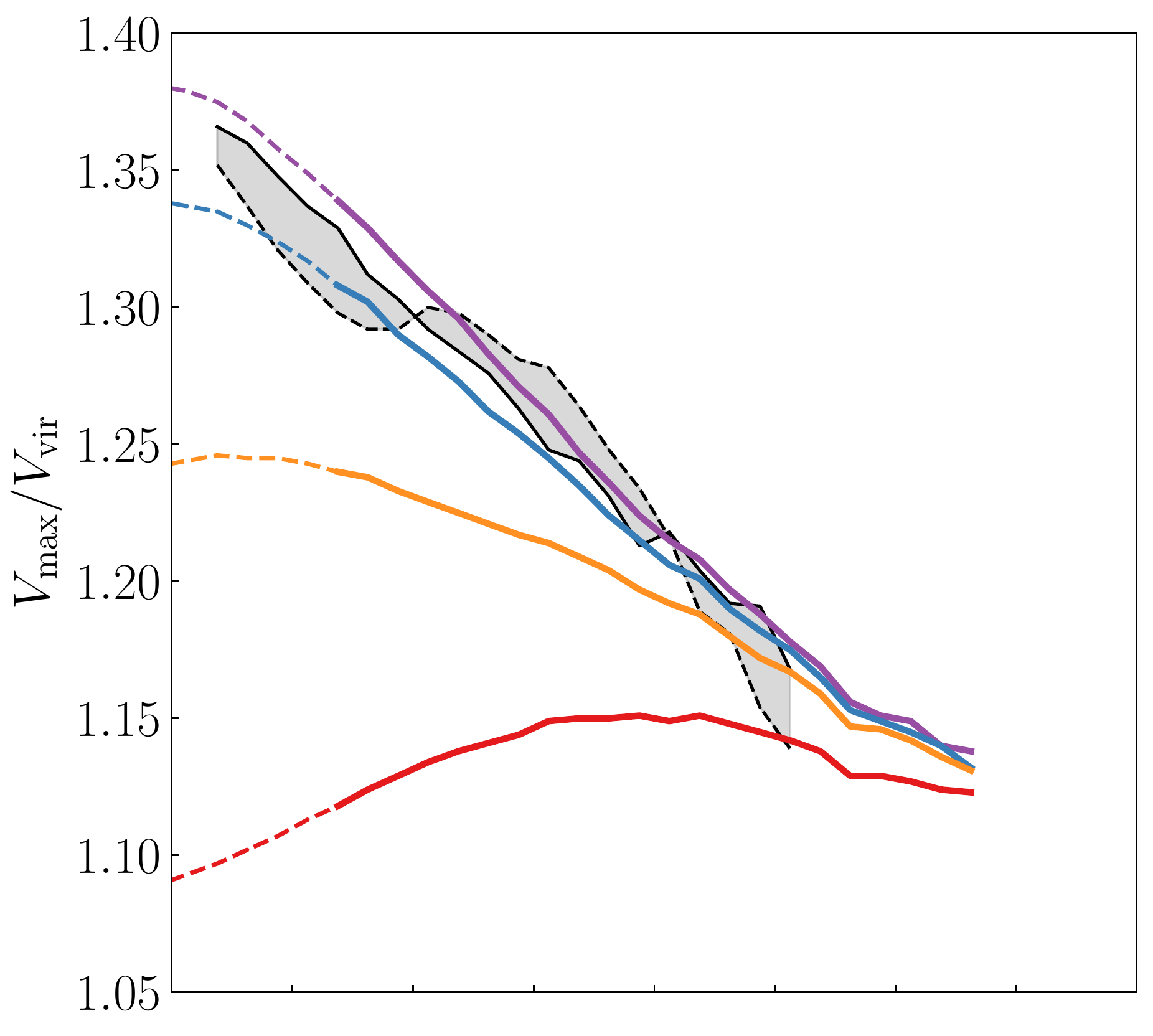}} \\[-2ex]
   \subfigure{\includegraphics[width=0.49\textwidth]{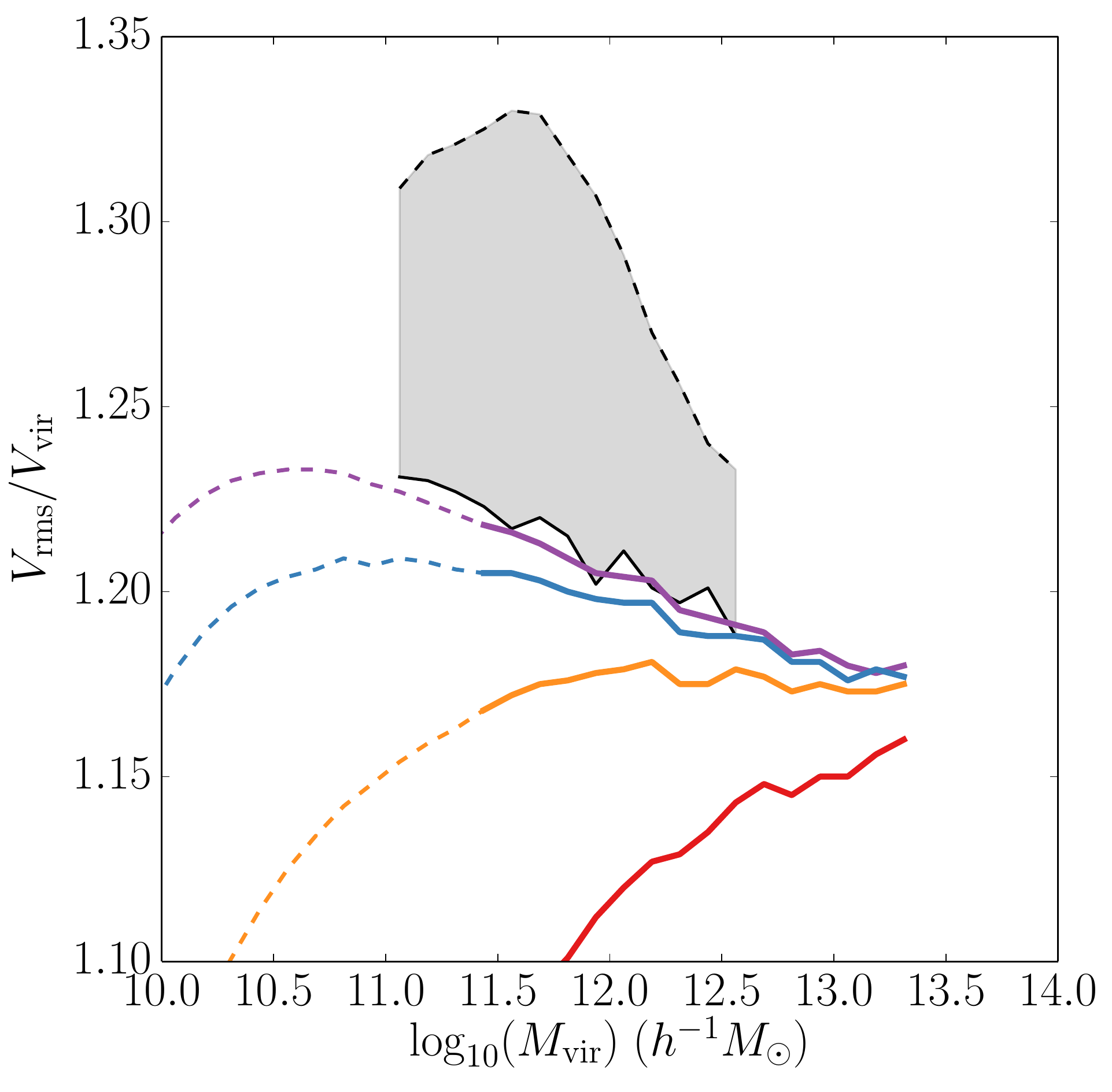}}
   \subfigure{\includegraphics[width=0.49\textwidth]{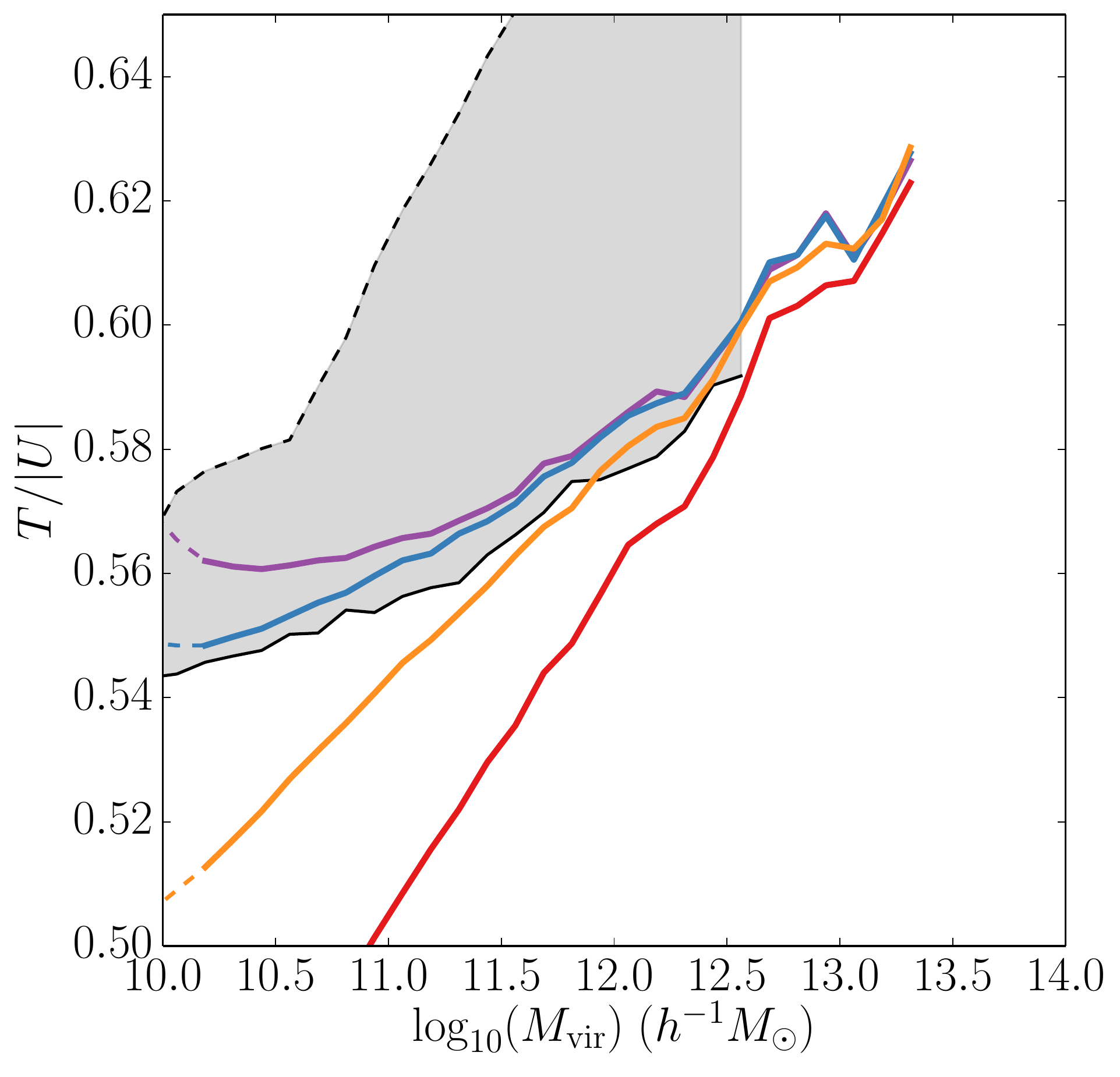}}
\caption{The dependence of various halo properties on force softening scale. Each panel shows the mean mass relation for various halo properties in isolated haloes for four boxes which were resimulated from the same initial conditions with different $\epsilon$. The purple, blue, and yellow curves have $\epsilon$ in the range typically chosen by cosmological simulations. The red curves probe an atypically large value of $\epsilon$. The curves transition from dashed to solid at the cutoffs we visually identified in Appendix \ref{sec:hr_ranges} and represent commonly used resolution ranges in the literature. $c/a,$ $V_{\rm max},$ and $V_{\rm rms}$ are strongly dependent on $\epsilon,$ and $T/|U|$ is dependent on $\epsilon$ for softening scales larger than $\epsilon/l \approx 0.016$. To give a sense of the `practical significance' of these dependencies, we show the impact of baryons in the Illustris-TNG simulations as grey shaded regions. The dashed edges of these regions correspond to mass relations from the baryonic TNG100-2 box and the solid edges correspond to mass relations from non-baryonic TNG100-2-Dark box.}
\label{fig:chinchilla}
\end{figure*}

To investigate the dependence of halo properties on $\epsilon$, we make use of four convergence boxes which were initially run as part of the Chinchilla simulation suite (as seen in, e.g., \citealp{Mao_et_al_2015,Desmond_Wechsler_2015,Lehmann_et_al_2017}).\footnote{Access to these catalogues was generously provided by M. Becker.} These boxes are resimulations of the same set of initial conditions but with different force softening scales (see Table \ref{tab:chinchilla}). They were run with $L=125\,h^{-1}$ Mpc, $N^3=1024^3,$ $\Omega_{M}=0.286,$ $h_{100}=0.7,$ and $m_p=1.44\times10^8\,h^{-1}M_\odot.$ Aside from force softening scale, these boxes are very similar to the Erebos\_CBol\_L125 box (the orange curve in the upper panels of Fig.~\ref{fig:mass_trends}). We refer to this as the Chinchilla-$\epsilon$ suite.

The force softening scales in these boxes span a wide range. In units of the mean interparticle spacing, the smallest force softening scale, in Chinchilla\_L125\_e1, corresponds to $\epsilon/l=0.0082$ a small but not uncommon length which is similar to simulations like Bolshoi or any of the main Chinchilla boxes. The next smallest, Chinchilla\_L125\_e2, corresponds to a fairly typical $\epsilon/l=0.016$ which is similar to SMDPL or Erebos\_CBol\_L63. Next is Chinchilla\_L125\_e5, $\epsilon/l=0.041$ which is close to the upper limit of $\epsilon$ typically found in cosmological simulations and is similar to Erebos\_CBol\_L1000 or the $\nu^2$GC boxes. The last box, Chinchilla\_L125\_e14, has a force softening scale much larger than any box in our simulation suite: $\epsilon/l=0.115.$ This box has force softening and mass resolution similar to high-resolution zoom-in simulations of the early 90's which suffered from `overmerging' (e.g. \citealp{Carlberg_Dubinski_1991}; \citealp{Carlberg_1994}; \citealp{Evrard_et_al_1994}), but this level of force softening has also been recommended by some convergence studies (see Section \ref{sec:timestepping}) to remedy integration errors that occur during large-angle scattering. Timestepping in each simulation is performed via Eq.~\ref{eq:gadget_timestep} with $\eta=0.025,$ meaning that timesteps are not constant between simulations.

We compare the mass-trends for every halo property described in Section \ref{sec:properties} across the boxes in the Chinchilla-$\epsilon$ resimulation suite. Most properties, such as $X_{\rm off}$ or $\lambda_{\rm Bullock}$ show little to no dependence on $\epsilon$ or show agreement for typical values of $\epsilon$ and some mild non-convergence in Chinchilla\_L125\_e14. This is not true for all halo properties.

In Fig.~\ref{fig:chinchilla}, we show the $\langle V_{\rm max}(M_{\rm vir})\rangle,$ $\langle c/a(M_{\rm vir})\rangle,$ $\langle V_{\rm rms}(M_{\rm vir})\rangle,$ and $\langle T/|U|(M_{\rm vir})\rangle$ relations for isolated haloes in each of the Chinchilla resimulation boxes. The curves are solid above the visually-identified `high-resolution' cutoffs (see Appendix \ref{sec:hr_ranges}) and dashed below it. If $N_{\rm vir}$-based convergence limits were sufficient for these halo properties, one would expect that these trends would not depend on numerical parametrization above these cutoffs. We find that all four properties vary with $\epsilon.$ These mass relations change continuously in amplitude and slope across the entire $\epsilon$ range.

To give a sense of the `practical significance' of these trends, we overplot the difference between the DMO TNG100-2-Dark and the baryonic TNG100-2 as a grey shaded region. For $V_{\rm max}$ and $c/a,$ the shift in halo properties due to numerical effects is comparable to or greater than the impact of baryons.

\subsection{Dependence of the Subhalo Mass Function on $\epsilon$}
\label{sec:subhaloes}

\begin{figure*}
   \centering
    \subfigure{\includegraphics[width=0.32\textwidth]{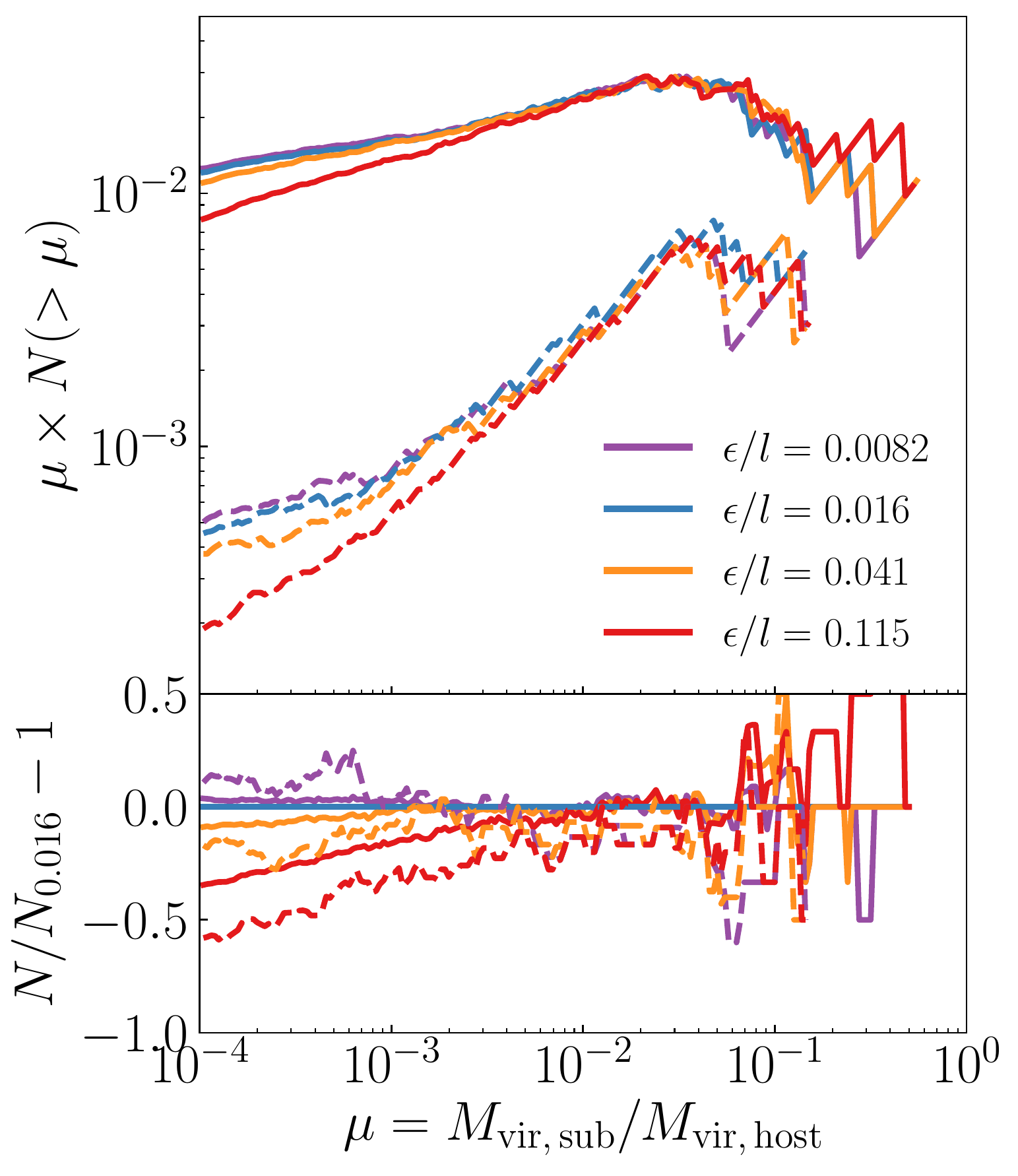}}
   \subfigure{\includegraphics[width=0.32\textwidth]{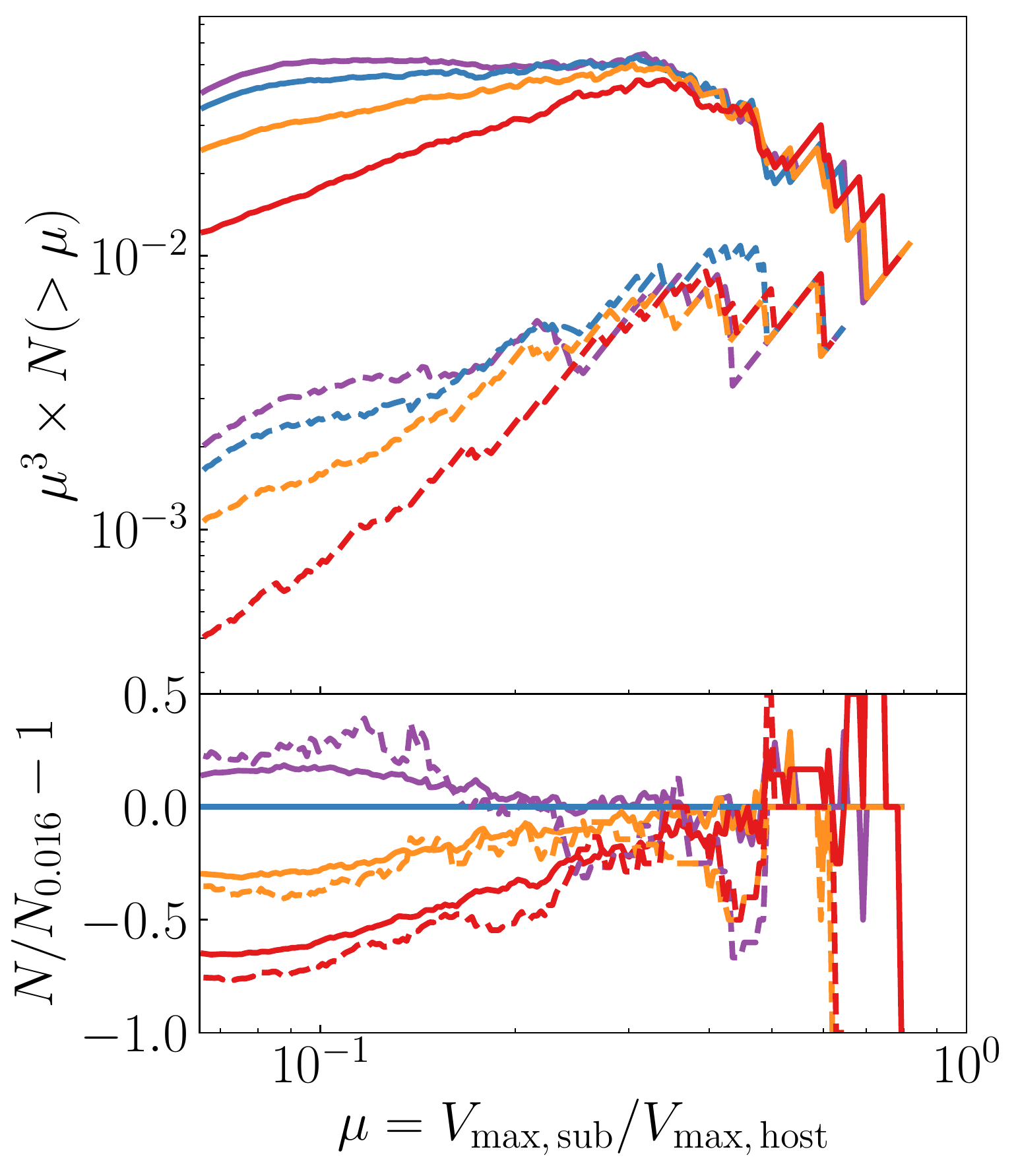}}
   \subfigure{\includegraphics[width=0.32\textwidth]{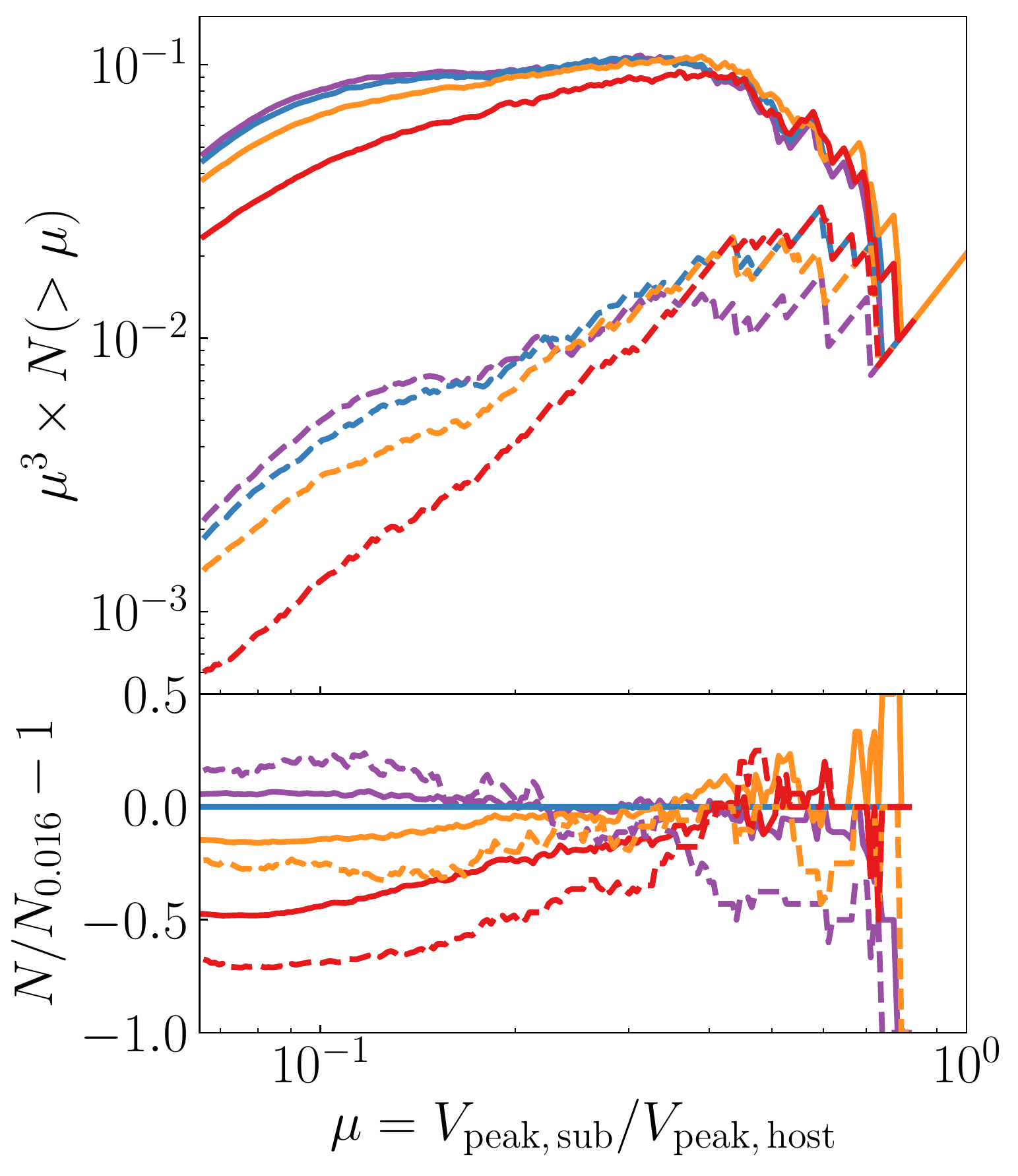}}
\caption{The impact of force softening on the subhalo abundance within the 50 most-massive host haloes from the Chinchilla-$\epsilon$ resimulation suite. These haloes have masses $M_{\rm vir} \approx 10^{14}\,h^{-1}M_\odot$. Left: The dependence of the cumulative subhalo mass function on force softening scale.\ Plotted are the mean mass functions. The solid lines show all subhaloes, and the dashed lines show all subhaloes within $0.2\,R_{\rm vir}$ of their hosts. The fractional deviation from the $\epsilon=0.016$ simulation is shown in the bottom panel. Centre: the same for the mean $V_{\rm max}$ subhalo velocity function. Right: the same for the mean $V_{\rm peak}$ subhalo velocity function. The lowest mass subhaloes shown in each plot have $\approx70$ particles. Note that the red curve corresponds to an atypically large value of $\epsilon.$
The subhalo mass function exhibits only a weak dependence on $\epsilon$ in the outer regimes of the halo. This dependence becomes stronger at small radii, confirming that artificial disruption is stronger in this regime. The subhalo velocity functions depend more strongly on $\epsilon,$ even at large radii. The impact of $\epsilon$ on velocity functions becomes stronger at small radii, although the strength of this radial dependence relative is weak. This implies that most of this effect is from artificial suppression of $V_{\rm max}$ and not from artificial subhalo disruption.
}
\label{fig:subhaloes}
\end{figure*}

In Fig.~\ref{fig:subhaloes} we show the dependence of the subhalo mass and velocity functions on $\epsilon.$ We consider subhaloes around the 50 largest hosts in the Chinchilla-$\epsilon$ suite. This number was chosen so that hosts would have resolution better than $N_{\rm vir}>5\times 10^5,$ ensuring that large subhalo mass ranges can be studied.

The left panel shows the mean subhalo mass functions in terms of $M_{\rm vir}$ for these host haloes, the middle panel shows the mean subhalo $V_{\rm max}$ functions, and the right panel shows the mean subhalo $V_{\rm peak}$ functions.

Both types of subhalo velocity functions show a strong dependence on force softening scale that becomes stronger when considering subhaloes close to the centre of the host. Subhalo mass functions have a weaker dependence on $\epsilon$, although it also becomes stronger for small-$r$ subhaloes, implying that artificial subhalo disruption/stripping becomes stronger at smaller radii. The difference in $\epsilon$-dependence between the mass and velocity functions implies that velocity functions are primarily impacted by artificial suppression of the velocity curve (which does not affect mass functions and which does not have a radial dependence) more than artificial subhalo disruption, but that artificial subhalo disruption likely leads to the radial change in $\epsilon$-dependence.

The locations of even the most massive subhaloes are altered substantially by changes in $\epsilon.$ It is possible that this is due to chaotic errors in halo phase while orbiting their hosts, but given that the tidal disruption rate in the host's central region is dependent on $\epsilon,$ it is also possible that this is caused by an $\epsilon$ dependence in the dynamical fiction experienced by each subhalo. This change in positions makes it impossible to directly measure subhalo disruption using only single-snapshot information. Such analysis would be possible by comparing the trajectories of subhalo progenitors prior to accretion. We defer such analysis of subhalo trajectories to future work.

Note that these tests only study the impact of $\epsilon$ on subhalo abundance. Particle count also substantially impacts the reliability of subhalo velocity functions \citep[e.g.][]{Guo_White_2014,Klypin_et_al_2015} and must be accounted for accordingly.

\subsubsection{Comparison with Previous Work}

The most famous examples of artificial subhalo disruption was the `overmerging problem' expereinced by simulations run prior to the late-90's. Zoom-in simulations at the time \citep[e.g.][]{Carlberg_Dubinski_1991,Carlberg_1994,Evrard_et_al_1994} could simulate a cluster-mass halo with millions of particles, but galaxy-mass subhaloes would rapidly dissolve after accretion. This mismatch with the observed abundance of cluster members -- first noted by \citet{White_et_al_1987,Frenk_et_al_1988} -- would come to be noted as a problem for $\Lambda$CDM simulations. Through a combination of idealised subhalo simulations \citep{Moore_et_al_1996,Klypin_et_al_1999_overmerging} and high-resolution simulations with small-$\epsilon$ \citep{Klypin_et_al_1999_overmerging}, simulators argued that this was because the large softening scales used in these simulations were comparable to or larger than the size of the tidal radius of galaxy-mass subhaloes at pericentre. This would depress the inner density of these subhaloes, making them far easier to disrupt. 

Fig.~\ref{fig:subhaloes} is qualitatively consistent with earlier work on the overmerging problem. The $\epsilon/l=0.115$ simulation (red curves) has $\epsilon=14\,h^{-1}{\rm kpc},$ which as large or larger than many of the aforementioned simulations which experienced overmerging. We find that subhalo mass and velocity functions are suppressed in these simulations, although not as severely as in pre-\citet{Kravtsov_et_al_1998} zoom-in simulations. We find that $\mu\approx 10^{-2}$ ($M_{\rm vir}\approx 10^{12}\,h^{-1}M_\odot$) subhaloes are largely not impacted by such a large force softening and that the amplitude of the mass function is `only' decreased by a factor of $\approx2$ for $\mu\approx10^{-4}$ dwarfs. This is a less severe suppression than was seen in simulations that fell victim to the overmerging problem. It is likely that much of this difference is due to improvements in halo finders, but it is also possible that these earlier simulations suffered from additional significant numerical issues beyond $\epsilon$-induced subhalo disruption. (Although, given the scientific impact of even a 50\% decrease in subhaloes, this observation is, at best, a historical oddity.)

More recently, the idealised tests in \citet{van_den_Bosch_Ogiya_2018} have suggested that force softening may have a larger impact on subhalo disruption than previously thought. They find that simulations of idealised subhaloes experience substantial artificial disruption and that this disruption occurs even at high subhalo resolutions. The rate of tidal stripping is dependent on $\epsilon$ across the range of $\epsilon$ values adopted by the Chinchilla-$\epsilon$ test suite. Our results are not in conflict with these findings, despite the weak dependence of the subhalo mass function on $\epsilon$ for $\epsilon/l\lesssim0.04.$

\citet{van_den_Bosch_Ogiya_2018} found that numerical factors begin to artificially accelerate disruption once haloes have already lost $\gtrsim 90-95$ per cent of their mass due to {\it physical} disruption. Due to the slope of the infalling halo mass function, at any particular snapshot, the majority of subhaloes at a given mass have not yet experienced this level of disruption. Additionally, the effect of artificial disruption due to force softening is strongest in subhaloes on close orbits, with effect becoming particularly strong at $R\approx 0.1\,R_{\rm vir}$, a regime which is not well-probed by the relatively small number of high resolution halos which we have access to. While close-orbit subhaloes make up a small fraction of the host's overall volume (and thus of our sample), the best constraints on the faint end of the satellite luminosity function come from the corresponding satellite population of the Milky Way \citep[e.g.][]{Drlica_Wagner_et_al_2019}. Furthermore, the radial dependence of artificial disruption makes it more difficult to compare observed satellite number density profiles to the predictions of $\Lambda$CDM \citep[e.g.][]{Carlsten_et_al_2020}. These effects are therefore still important for cosmological and astrophysical tests.

\section{Estimating the Impact of Large-$\epsilon$ on $V_{\rm max}$}
\label{sec:debias}

In Section \ref{sec:results}, we showed that the distribution of halo properties measured in different simulations diverge from one another at unexpectedly high particle counts. In Section \ref{sec:eps_dependence_mass_trend}, we showed that varying $\epsilon$ across the range typically used in cosmological simulations has a large impact on many commonly studied halo properties. In this Section, we construct a model that predicts this behaviour for one of the simplest and most fundamental halo properties we have considered: $V_{\rm max}$.

\subsection{Background}

Previous convergence studies have established three primary ways in which the numerical effects in DMO simulations can bias the properties of dark matter haloes:

\begin{itemize}
    \item The suppression of the centripetal force on scales $r \lesssim \epsilon$.
    \item Altered velocity and density structure due to two-body relaxation from repeated minor (`small-angle') collisions between dark matter particles.
    \item Energy non-conservation due to integration errors during major (`wide-angle') collisions.
\end{itemize}

We review these three effects below.

\subsubsection{Centripetal Force Suppression}
\label{sec:background_centripetal_force}

In the large-$\epsilon$ limit, non-Newtonian forces suppress $V(R)$ with increasing $\epsilon$ \citep[e.g.][]{Klypin_et_al_2015,van_den_Bosch_Ogiya_2018,Ludlow_et_al_2019}. This suppression comes from the reduction in centripetal forces for $r\lesssim \epsilon.$

As we show in Appendix \ref{sec:recalibrate_plummer}, the change in $V(R)$ is well-fit by 
\begin{equation}
    \label{eq:v_dev_curve_ein}
    \frac{V(R;h)}{V_{\rm ref}(R)} =1 - {\rm exp}\left(-(Ah/R)^\beta\right).
\end{equation}
Here, $A$ and $\beta$ are parameters of the fit that depend on force softening scheme, and $h$ is the formal resolution of that scheme (see Section \ref{sec:force_softening}).

This fit captures the fact that the reduction in $V(R)$ continues well into the regime where forces are Newtonian. The continued suppression occurs for two reasons. First, $V(R)$ depends on $M(<R)$ and is therefore an integrated quantity.  Second, the decrease in central mass leads to higher total energies for particles outside the non-Newtonian regime, which pushes the particles into orbits with larger radii.

As we discuss in Section \ref{sec:timestepping}, some of Eq.~\ref{eq:v_dev_curve_ein} may be caused by poor timestep resolution under some timestepping schemes.

\subsubsection{Two-Body Relaxation}
\label{sec:background_two_body}

The discretization of dark matter into numerical particles allows particles to collide with one another. Gradually, the cumulative effect of these collisions causes particles to deviate from their original orbits in a process called two-body relaxation (see, e.g., chapter 1.2 of \citealp{Binney_Tremaine_2008}). Through two-body relaxation, the velocity and density structure of matter alters on a position-dependent timescale, $t_{\rm relax}$.\footnote{Despite the similar names, two-body relaxation is unrelated to `dynamical relaxation,' (e.g. chapter 5.5 in \citealp{Mo_vdB_White_2010}). This latter term refers to the processes which allow a collisionless system to adopt an equilibrium state. Unless otherwise specified, the term `relaxation' in this paper refers to two-body relaxation.} CDM is not collisional and real dark matter particles will have orbits which rarely exchange energy with the rest of the halo except through adiabatic contraction  \citep[e.g.][]{Dalal_et_al_2010,Diemer_2017}, meaning that any relaxation of this type is purely numerical.

The effects of two-body relaxation increase as the time since the start of the simulation, $t_{\rm sim}$, increases.  For regions where $t_{\rm sim} \ll t_{\rm relax},$ particles remain collisionless. For regions where $t_{\rm sim} \gtrsim t_{\rm relax},$ particles begin to thermalize. Particles in the high-velocity tails are transported to larger radii, leading to decreasing $V(R)$ \citep[e.g.][]{Power_et_al_2003,Navarro_et_al_2010,Ludlow_et_al_2019}. Studies of globular clusters have long shown that two-body relaxation eventually leads to a `core collapse' when $t_{\rm sim} \gg t_{\rm relax}.$  The collapse results in systems with much higher central densities and higher $V(R)$ \citep[see][for an introduction to the topic]{Lightman_Shapiro_1978}. However, DMO simulations of haloes rarely have $t_{\rm relax}$ small enough for this effect to occur.

We note that $\epsilon$ only has a minor effect on $t_{\rm relax}$ \citep[e.g.][and historical references therein]{Ludlow_et_al_2019}.  Two-body relaxation is predominantly caused by numerous small-angle scatterings instead of rare large-angle scatterings. Consequently, larger force softening scales lead to only modestly longer relaxation timescales, as only a small portion of the Coulomb logarithm is suppressed. Beyond this minor dependence on $\epsilon$, the mean interparticle spacing determines $t_{\rm relax}$.

\subsubsection{Integration Errors}

Although $\epsilon$ has little effect on two-body relaxation, this parameter cannot be set arbitrarily small. $\epsilon,$ combined with $l,$ sets the maximum potential depth of each particle. As $\epsilon$ becomes smaller at a fixed $l,$ rare large-angle collisions can reach higher kinetic energies during pericentre and require finer timesteps to resolve. Thus, small $\epsilon$ both increases the cost of the simulation and increases the risk of energy loss due to integration errors in the event of an insufficiently aggressive timestepping scheme.

The overall impact of integration errors on DMO simulations is a complex topic, and we direct interested readers to our overview in Section \ref{sec:timestepping}.

\subsubsection{Which Numerical Effects Are Likely to Cause the Observed Biases In $V_{\rm max}?$}

The top right panel of Fig.~\ref{fig:chinchilla} illustrates the impact of $\epsilon$ on $V_{\rm max}.$ Larger values of $\epsilon$ lead to lower values of $V_{\rm max}.$ As discussed in Section \ref{sec:background_centripetal_force}, this trend is consistent with large-$\epsilon$ suppression of inner centripetal forces, as quantified by Eq.~\ref{eq:v_dev_curve_ein}.

Section \ref{sec:background_two_body} discusses why two-body relaxation is unlikely to be the primary culprit of the observed biases. The relaxation timescale, $t_{\rm relax}$, has a weak dependence on $\epsilon$ and measurements in \citet{Ludlow_et_al_2019} indicate that for the mass/resolution ranges shown in Fig.~\ref{fig:chinchilla}, relaxation effects will not have a large impact on $V_{\rm max}$. Furthermore, increasing $\epsilon$ leads to longer relaxation timescales, meaning that relaxation effects would suppress $V_{\rm max}$ for the small $\epsilon$ simulations.

The effect shown in Fig.~\ref{fig:chinchilla} could, in principle, be qualitatively consistent with integration errors. For integration errors to cause the observed level of bias, integration errors would need to typically lose energy (as was the case for some Gadget-1 tests in \citealp{Springel_et_al_2001}).  This would further imply that the large-$\epsilon$ simulations are closer to the correct solution. If this were true, it would be an incredibly serious problem for cosmological simulations. Furthermore, the most direct application of the timestepping tests in \citet{Power_et_al_2003} indicates that timesteps in the small-$\epsilon$ Chinchilla-$\epsilon$ boxes (and almost all simulations in Table \ref{tab:simulations}) are too coarse to be converged.  In Section \ref{sec:timestepping}, we address this problem and argue that it is unlikely that simulation timesteps are catastrophically unconverged while acknowledging some open questions related to common timestepping schemes.

Given the above discussion, the $\epsilon$ dependence shown in Fig.~\ref{fig:chinchilla} is likely the result of suppression of centripetal forces in the large-$\epsilon$ simulations. In the following Section, we outline a quantitative model for this effect and show that it predicts the measured $\epsilon$ dependence.

\subsection{A Quantitative Bias Model for $V(R)$}

\begin{figure*}
   \centering
   \subfigure{\includegraphics[width=0.49\textwidth]{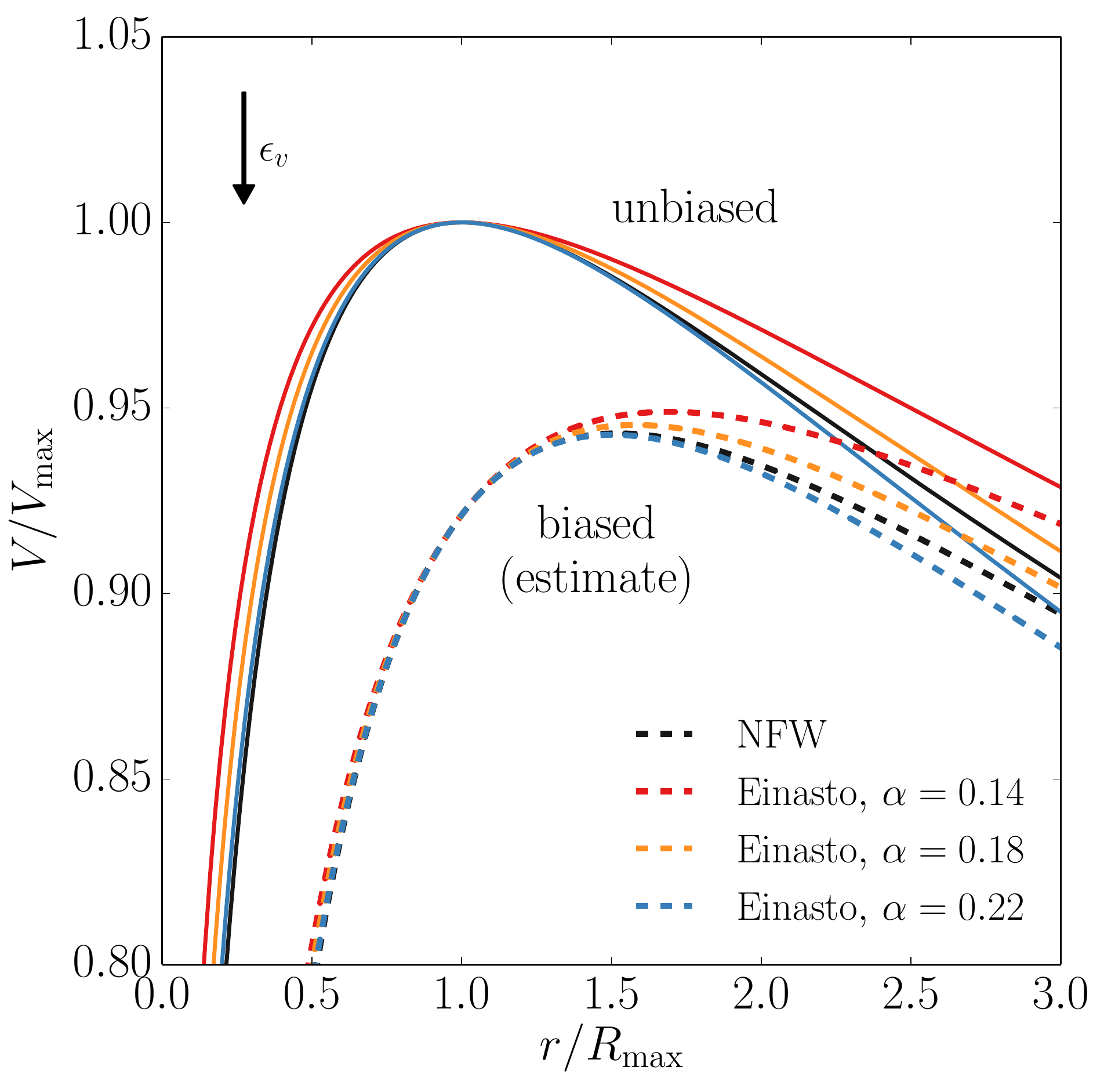}}
   \subfigure{\includegraphics[width=0.485\textwidth]{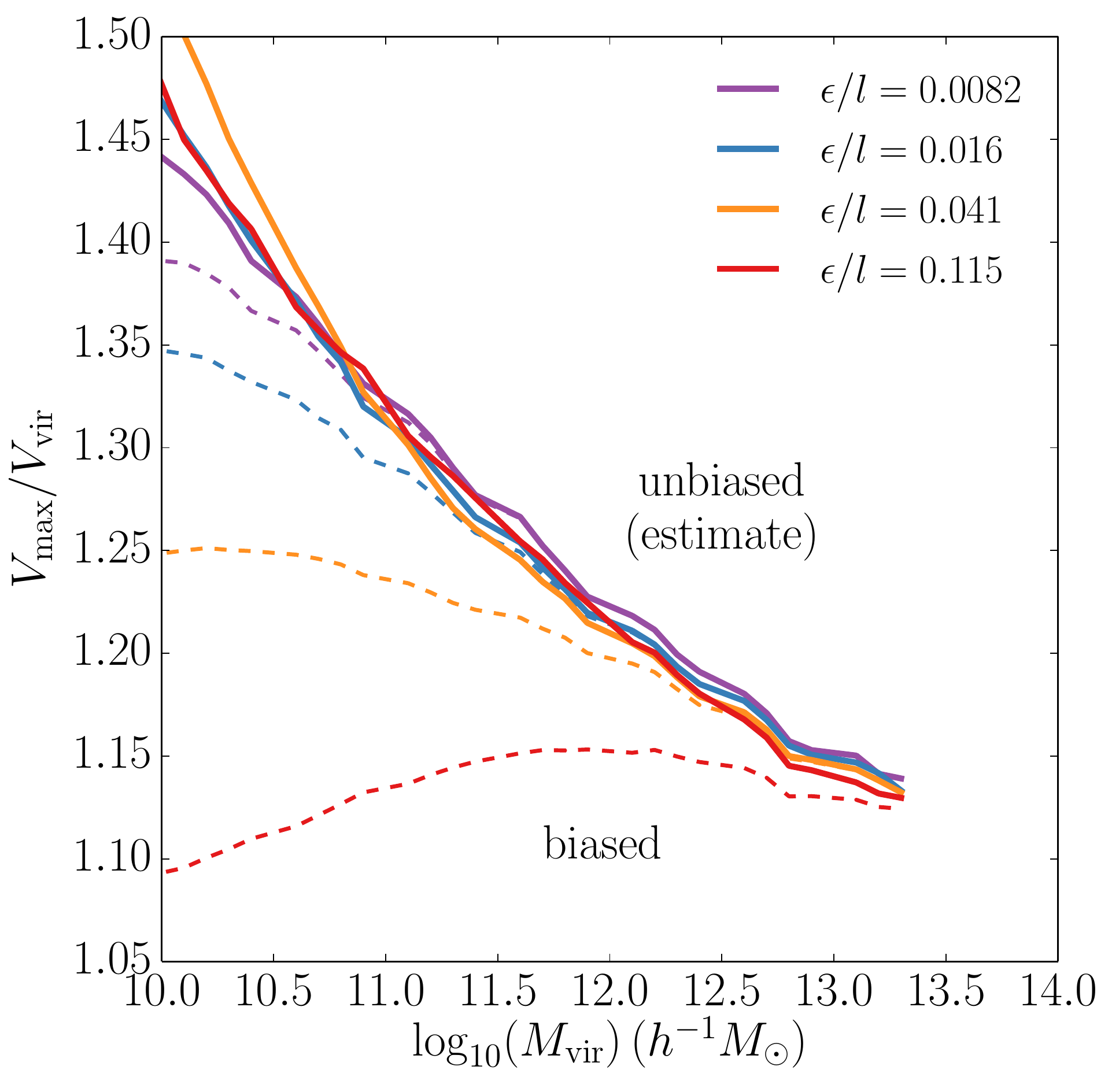}}
\caption{Left: An illustration of the estimated bias due to a large force softening scale on different halo profile shapes. Different colours correspond to different density profile parametrizations. Solid curves show rotation curves unbiased by force softening, and dashed curves show predictions of the biased rotation curves from Eq.~\ref{eq:v_dev_curve_ein} for a Gadget simulation with $\epsilon=0.357\,R_{\rm max}$ ( $h_{\rm Gadget}=R_{\rm max}$). The systemic uncertainty in $V_{\rm max,bias}$ across the profile parameters shown here is $\approx0.007V_{\rm max}$ for the given value of $h_{\rm Gadget}$. Right: The result of applying the bias estimates described in Section \ref{sec:debias} to the top right panel of Fig.~\ref{fig:chinchilla} (note the change of axis range between these Figures). The dashed curves show the same mean $V_{\rm max}$ values measured in each mass bin and the solid curves show estimates of what $\langle V_{\rm max}\rangle$ would be if there were no bias due to force softening. The dependence on $\epsilon$ is almost entirely removed through this bias estimate, indicating that the majority of the $\epsilon$ dependence is due to large-$\epsilon$ biases and not other effects.}
\label{fig:v_bias_curves}
\end{figure*}

When simulators account for large-$\epsilon$ effects, they typically restrict their analysis to haloes where $R>X\epsilon,$ where $X$ is some constant. To give an idea of the typical values of $X$ used, we surveyed several papers which studied the concentration-mass relation to identify values for $X=\langle r_{-2}\rangle(M_{\rm vir,min})/\epsilon$.  We found that this limit ranged from $2.5\leq X \leq6.4$ \citep{Neto_et_al_2007,Duffy_et_al_2008,Gao_et_al_2008,Zhao_et_al_2009,Prada_et_al_2012,Bhattacharya_et_al_2013,Ludlow_et_al_2013,Dutton_Maccio_2014,Klypin_et_al_2016,Poveda_Ruiz_et_al_2016,Child_et_al_2018}. This is broadly consistent with the behaviour of \textsc{Rockstar}, which downweights radii larger than $3\epsilon$.

\citet{Diemer_Kravtsov_2015} performed a detailed review use the results of several zoom-in simulations to conclude that analysis is safe above $>3\epsilon$ for individual haloes, and that analysis of $\langle r_{-2}(M_{\rm vir})\rangle$ should be restricted to masses where $\langle r_{-2}(M_{\rm vir})\rangle\gtrsim 8\epsilon$ to account for scatter in the $\langle c_{\rm vir}(M_{\rm vir})\rangle$ relation.
Below, we take a different approach and use our direct measurements of the impact of $\epsilon$ on rotation curves (Eq.~\ref{eq:v_dev_curve_ein}) to estimate the impact of $\epsilon$ on the distribution of $V_{\rm max}$ in a halo population.

The left panel of Fig.~\ref{fig:v_bias_curves} illustrates the rotation curve bias due to the Gadget force softening scale, as predicted by Eq.~\ref{eq:v_dev_curve_ein} for different halo profile shapes. We reference \citet{Klypin_et_al_2015} for a mathematical summary of NFW rotation curves and \citet{Garrison-Kimmel_et_al_2014b} for a similar summary of Einasto rotation curves.  Einasto profiles require a second parameter beyond $R_s,$ $\alpha$, and provide a more accurate fit than NFW profiles \citep[e.g.][]{Gao_et_al_2008,Springel_et_al_2008}. The solid curves in Figure~\ref{fig:v_bias_curves} show the unbiased rotation curves for an NFW profile in black and Einasto profiles with $\alpha$=0.14, 0.18, and 0.22 in red, yellow, and blue, respectively. The selected $\alpha$ values roughly correspond to the range spanned by $z=0$ haloes \citep[e.g.][]{Child_et_al_2018}. The dashed lines show the biased rotation curves predicted by Eq.~\ref{eq:v_dev_curve_ein} for $h_{\rm Gadget}=R_{\rm max}$ ($\epsilon=0.278\,R_{\rm max}$). The biased maximum velocity, $V_{\rm max,bias}$, ranges from $0.943\,V_{\rm max}$ to $0.949\,V_{\rm max},$ exhibiting a small systematic uncertainty due to halo profile shape, which is $\approx$ 10 percent of $V_{\rm max}-V_{\rm max,bias}.$ This uncertainty consistently stays at or below this level relative to $V_{\rm max}-V_{\rm max,bias}$ regardless of $R_{\rm max}/\epsilon$

By evaluating $\xi_{\rm bias}=V_{\rm max}/V_{\rm max,bias}$ for a range of $\epsilon/R_{\rm max},$ we can empirically construct the invertible function $\xi_{\rm bias}(\epsilon/R_{\rm max})$ for a given halo profile shape. For convenience, we note that for both NFW and Einasto profiles this function is well-fit  by,
\begin{equation}
    \label{eq:debias_solve_1}
    \xi_{\rm bias} = 2 - \left(1 + (A\epsilon/R_{\rm max})^2\right)^\beta.
\end{equation}

We fit this relation for Gadget-like kernels over the range of 0.01 $\epsilon/R_{\rm max}\lesssim h_{\rm Gadget}\lesssim5\,\epsilon/R_{\rm max}$.  Below this range, $V_{\rm max}/V_{\rm max,bias}$ is 1 for all practical purposes. Above this range, Eq.~\ref{eq:v_dev_curve_ein} is poorly constrained. By minimising the least-squared error on $V_{\rm max}/V_{\rm max,bias},$ we find that the parameters $A=6.049$ and $\beta=0.0544$ lead to errors in $V_{\rm max,bias}$ which are $\lesssim 10^{-3} V_{\rm max}$ for NFW profiles and that the parameters $A=5.884$ and $\beta(\alpha)=0.02754\,\ln{(\alpha)} + 0.15566$ lead to errors which are $\lesssim 2\times10^{-3}\,V_{\rm max}$ for Einasto profiles with $\alpha$ ranging from 0.12 to 0.32. However, we use the raw empirical functions in all subsequent analyses, derived from whichever force softening kernel is appropriate.

Note that no function describing $\xi_{\rm bias}(\epsilon/R_{\rm max})$ can be applied on its own to evaluate the bias in $V_{\rm max}$ because these functions depend on the unbiased value of $R_{\rm max},$ which is unknown. Therefore, such a function must be combined with a second, independent equation relating $\xi_{\rm bias}$ to $R_{\rm max}.$

For our application, the systematic errors in $\xi_{\rm bias}$ due to profile shape are small. Therefore, we restrict our analysis to NFW profiles because they depend on only a single parameter.  
With an NFW parametrization, we can directly compute an estimate for $\xi_{\rm bias}$ from the (unknown) unbiased $c_{\rm vir}$ and (known) $V_{\rm max,bias}/V_{\rm vir, bias}$ from the halo catalogue.  The estimate comes in the form of, 
\begin{align}
    \label{eq:debias_solve_2}
    \xi_{\rm bias} &=  0.469 \left(\frac{V_{\rm vir,bias}}{V_{\rm max,bias}}\right)\left(\frac{c_{\rm vir}}{f(c_{\rm vir})}\right)^{1/2} \left(\frac{V_{\rm vir}}{V_{\rm vir,bias}}\right) 
\end{align}
where $V_{\rm vir}/V_{\rm vir, bias}$ is Eq.~\ref{eq:v_dev_curve_ein} evaluated at $R/\epsilon=0.469\,c_{\rm vir}\frac{R_{\rm max}}{\epsilon}$ and $f(x)=\ln{(1+x)} - 1/(1+x)$.  We compute the ratio $V_{\rm vir,bias}/V_{\rm max,bias}$ from halo catalogues, whose measurements are biased due to $\epsilon$. 

For NFW haloes, $R_{\rm max} = 2.164 R_s$.  This identify can transform Eq.~\ref{eq:debias_solve_2} into a function of $\epsilon/R_{\rm max}.$ Therefore, Eq.~\ref{eq:debias_solve_1} and Eq.~\ref{eq:debias_solve_2} are two independent equations for $\xi_{\rm bias}(\epsilon/R_{\rm max}).$ For haloes with $R_{\rm max,bias} < R_{\rm vir,bias}$ (a criterion that holds for virtually all haloes in cosmological simulations), these two relations intersect at exactly one point: a unique solution for $\xi_{\rm bias}=V_{\rm max}/V_{\rm max, bias}$. This statement is only true for single parameter profile models, such as NFW profiles or Einasto profiles with fixed $\alpha.$

With this de-biasing procedure, we can estimate the unbiased $V_{\rm max}/V_{\rm vir}$ for each halo in a given cosmological simulation from the biased measurements of $V_{\rm max,bias}/V_{\rm vir,bias}$.  We can then estimate the mean unbiased $V_{\rm max}/V_{\rm vir}(M_{\rm vir})$ in that simulation. Note that the scatter around $\langle V_{\rm max}(M_{\rm vir}) \rangle$ means that this estimate cannot just be applied to the mean of a particular mass bin, but must first be applied to individual haloes before finding the mean relation as described.

The right panel of Fig.~\ref{fig:v_bias_curves}  shows the result of the `de-biased' estimate of  $V_{\rm max}/V_{\rm vir}$ for the Chinchilla resimulation boxes. From this figure, we see that this procedure completely removes the $\epsilon$ dependence from this sample, \emph{implying that the $\epsilon$ dependence of $V_{\rm max}/V_{\rm vir}$ is almost entirely due to large-$\epsilon.$} 

There is a few-per cent dispersion between curves at moderate-to-high masses. While other numerical effects could cause this dispersion, the level of scatter is consistent with the error level associated with the assumption of an NFW profile in our analysis. As discussed above, assuming a profile shape results in systematic errors in $V_{\rm max}$ on the order of $0.1\,(V_{\rm max}-V_{\rm max,bias}).$ Given that some simulations are estimated to be biased at the 20\% - 30\% level, a 2\%-3\% error is to be expected. The dispersion increases at low particle counts (low halo masses) and small $R_{\rm max}/\epsilon.$ While numerical effects could cause this as well, the dispersion occurs in a regime where corrections are large and Eq.~\ref{eq:v_dev_curve_ein} is poorly constrained.

\begin{figure}
   \centering
   \subfigure{\includegraphics[width=0.49\textwidth]{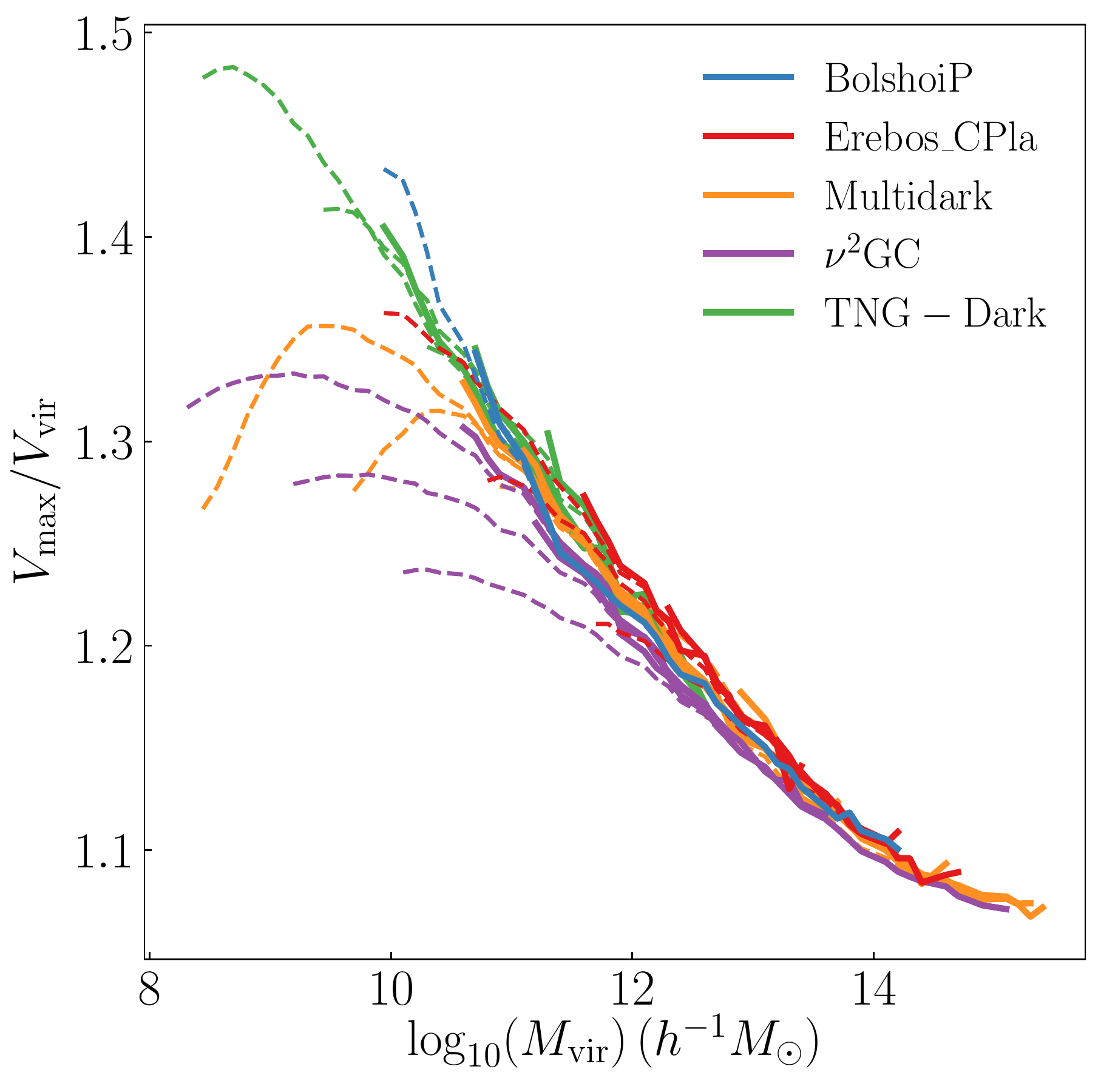}}
\caption{ Fig.~\ref{fig:mass_trends} recreated after correcting for bias using the bias estimates from Section \ref{sec:debias}. As in the right panel of Fig.~\ref{fig:v_bias_curves}, dashed curves show the mean mass trends measured in each simulation and the solid curves show estimates of what these mean trends would have been if not for large-$\epsilon$ biases. To emphasise the mass ranges which are affected by these biases, we only plot solid curves down to mass bins at which they agree with the measured trend to 1 per cent or better. Although there is still some dispersion around a mean relation between simulation suites, all the strong, visually apparent divergences from the mean trend are consistent with being caused by large-$\epsilon.$ Note that there are many science applications where a $V_{\rm max}$ bias of larger magnitudes is perfectly acceptable. This cutoff choice is only meant to mimic the divergences seen by the eye \emph{and does not imply the `usable' mass ranges of these simulations for arbitrary analysis.} Such a mass range must be developed with the tolerances of a given analysis in mind.}
\label{fig:debias_Planck}
\end{figure}

In Fig.~\ref{fig:debias_Planck}, we show the results of applying these bias estimates to various Planck cosmology simulations. In this Figure, dashed lines show the measured mean mass trends in each simulation.  The solid lines show results from our de-biasing procedure, which are estimates of what these trends would have been if not for the large-$\epsilon$ bias.  We cut off the estimated trend when they disagree from the measured trend by more than one per cent. As with the Chinchilla boxes, large-$\epsilon$ biases account for the most visibly-apparent deviations.

We note that there is still a non-trivial amount of scatter between simulation suites about the mean trend (see also, Appendix \ref{sec:model_predctions_multidark_illustris}). While is is possible that this dispersion is also due to numerical factors, another possible explanation is in variations due to slight variations in cosmological parameters. Despite the fact that all are `Planck' cosmology simulations, different suites are either associated with data releases from different years or round their cosmological parameters to a different number of decimal places.

\section{Discussion}

\label{sec:discussion}

\subsection{Timestepping as an Additional Source of Biases}
\label{sec:timestepping}

This Section investigates the impact of timestepping and integration errors on halo circular velocity profiles. Its primary goal is to determine whether integration errors could possibly contribute to the $\epsilon$ dependence seen in Fig.~\ref{fig:chinchilla} (a state of affairs which would be catastrophic to most published cosmological simulations if true). We argue that this is not the case. However, we note that quite a bit of work remains to be done on this topic.

Coarse timesteps have two well-discussed effects on halo profiles \citep[e.g.][]{Power_et_al_2003}. First, particles orbiting a smooth potential can artificially gain or lose energy if their orbits are too poorly resolved in time (e.g. fig.~4 and fig.~6 of \citealp{Springel_2005}). The exact effect on these orbits is dependent on a number of factors including the integration scheme, the local slope of the potential, the ellipticity of the orbits, and the adaptive timestepping scheme \citep{Springel_2001,Springel_2005}.  The second effect occurs with particles orbiting potentials which are noisy due to small force softening scales.  Here, particle-particle scattering can lead to integration errors (e.g. fig.~9 of \citealp{Knebe_et_al_2000}). Numerical integration of this scattering may not conserve energy and can add/remove energy from the affected regions of the halo at a rate which depends on the collision rate, the depth of each particle's potential, and the length of the timesteps relative to the scattering timescale.

We will focus our analysis on the standard Gadget timestepping criteria, Eq.~\ref{eq:gadget_timestep}. Only two of the simulations in Table \ref{tab:simulations} use alternative schemes: Bolshoi and BolshoiP.  For a spherically symmetric NFW potential, Bolshoi and BolshoiP will always have timesteps that are a factor of $\approx10^2-10^3$ smaller than a Gadget simulation run with $\eta=0.025.$  Timestepping errors can be ignored for these two boxes. 

For any spherically symmetric mass distribution, the Gadget timestepping criteria can be conveniently rewritten in terms of the number of timesteps per circular orbit:
\begin{equation}
    \label{eq:gadget_timestep_rewrite}
    \frac{t_{\rm circ}}{\Delta t} = 28.1 \left(\frac{R}{\epsilon}\right)^{1/2}\left(\frac{\eta}{0.025}\right)^{-1/2}.
\end{equation}
We use this relationship to quantify integration errors in Sections \ref{sec:smooth_errors} and \ref{sec:scattering}.

\subsubsection{Integration Errors in Smooth Potentials}
\label{sec:smooth_errors}

Integration errors in smooth potentials, are essentially irrelevant with the conventional Gadget integration settings. Tests in \citet{Power_et_al_2003} show that simulations with constant timestepping converge to $\Delta V/V_{\rm ref} = 0.1$ above radii at which timesteps per circular orbit satisfy, 
\begin{equation}
\label{eq:gadget_tcirc_criterion}
t_{\rm circ}(R)/t_{\rm 200c} > A (\Delta t / t_H)^\alpha,    
\end{equation}
with $A\approx15$ and $\alpha\approx5/6$ for smooth potentials. Empirical criteria are used to determine if the underlying potential of a halo is smooth. Combining the relation in Eq.~\ref{eq:gadget_tcirc_criterion} with Eq.~\ref{eq:gadget_timestep_rewrite} and the $t_{\rm circ}/t_{\rm 200c}$ profile of an NFW halo, we arrive at the requirement,
\begin{equation}
    \label{eq:smooth_errors}
    \frac{R}{\epsilon} \geq 4.96\times 10^{-4} A^{2/\alpha}\left(\frac{\eta}{0.025}\right) \left(\frac{x^2\,f(c_{\rm 200c})}{c_{\rm 200c}^2\,f(x)}\right)^{2 - 2/\alpha}.
\end{equation}
Here, $x=R/R_s,$ $c_{\rm 200c}=R_{\rm 200c}/Rs,$ and $f(x)=\ln{(1+x)} - x/(1+x).$ $R/\epsilon$ has only a weak dependence on $c_{\rm 200c}$ and $x.$ For example, concentrations in the range of $5\leq c_{\rm 200c}\leq15,$ a 10 per cent error in $V_{\rm max}$ due to integration errors in a smooth potential requires a corresponding range of $0.6\leq R_{\rm max}/\epsilon\leq1.2$.

Because of this, integration errors in smooth potentials are typically subdominant or comparable to softening-induced errors in the centripetal force.  As a comparison, Eq.~\ref{eq:v_dev_curve_ein} -- which quantifies the total impact of large-$\epsilon$ on $V(R)$ -- gives $\Delta V/V\approx 0.20-0.30$ at distances where Eq.~\ref{eq:smooth_errors} predicts a fractional error of 0.1. It is possible that even this is an overestimate: concentration- and radius-dependence at the level predicted by Eq.~\ref{eq:smooth_errors} -- while small -- would have been detectable in our tests described in Appendix \ref{sec:recalibrate_plummer}. However, it is possible that the fit values reported in that Appendix have some dependence on Gadget timestepping parameter, $\eta.$ High-precision estimates of $\Delta V/V_{\rm ref}$ likely require measurements at the same $\eta$ as the target simulation. This is a question which deserves further study.

\subsubsection{Integration Errors During Scattering}
\label{sec:scattering}
Excessively small force softening can lead to granularity in the halo potential. In sufficiently granular potentials, integration errors from particle-particle scattering become more severe and require much smaller timesteps to mitigate.\footnote{Note that such scattering is already aphysical regardless of how well resolved it is, and will contribute to two-body relaxation whenever it occurs (see Section \ref{sec:background_two_body}). However, $t_{\rm relax}$ is mainly set by repeated small-angle scattering events rather than infrequent wide-angle scattering due to the form of the gravitational scattering cross-section.} The landmark study on these integration errors is \citet{Power_et_al_2003}.  Empirically, they find that integration errors during scattering occur for $\epsilon < \epsilon_{{\rm opt,P03}}$, where we can express the limit on $\epsilon$ with both
\begin{align}
    \label{eq:eps_opt}
    \epsilon_{\rm opt,P03} &= \frac{2.9\,R_{\rm 200c}}{\sqrt{N_{\rm 200c}}}
\end{align}
and,
\begin{align}
    \label{eq:eps_opt2}
    \epsilon_{\rm opt,P03}/l &= 0.076 \left(\frac{\Omega_{\rm m}}{0.27}\right)^{1/3}\left(\frac{N_{\rm 200c}}{10^3}\right)^{-1/6}.
\end{align}
Here, $l$ is the mean interparticle spacing. Eq.~\ref{eq:eps_opt2} is substantially larger than the $\epsilon/l$ values adopted by virtually all non-zoom in cosmological simulations. Fig.~\ref{fig:chinchilla} shows that any cosmological simulations which abide by such a limit risk substantial biases in halo properties due to softened centripetal forces. Subsequent authors have suggested that Eq.~\ref{eq:eps_opt} is too conservative by a factor of $\lesssim2$ \citep{Zhang_et_al_2019,Ludlow_et_al_2019}.  However, part of the disagreement can be accounted for with a correction of the (now non-standard) Plummer equivalence scale which \citet{Power_et_al_2003} used: $\epsilon=0.5\,h_{\rm Gadget}.$

The most straightforward interpretation of the \citet{Power_et_al_2003} tests is that particle-particle scattering in noisy halo potentials should lead to significant non-convergence in cosmological simulations. Fig.~5 in \citet{Power_et_al_2003} shows that for haloes run at $\epsilon/\epsilon_{{\rm opt,P03},v}\approx 5$ (This would be the case, e.g., for Chinchilla\_L125\_e2 at $N_{\rm 200c}\approx 10^3$),  Eq.~\ref{eq:gadget_tcirc_criterion} is best-fit by $A=11.2$ and $\alpha=0.57$. The small $\alpha$ causes 10 per cent bias to be reached at very large values of $R/\epsilon$ and to become strongly dependent on $c_{\rm 200c}.$ These parameters imply that a thousand particle halo with $c_{\rm 200c}=10$ from a simulation with $\epsilon/l$ similar to Chinchilla\_L125\_e2 would have $>10$ per cent bias out to $V_{\rm vir}!$ Such a strong bias would completely obliterate the inner structure of these haloes.

However, the massive biases predicted by the analysis in the previous paragraph (and comparable predicted biases used to argue for $\epsilon \gtrsim \epsilon_{\rm opt,P03}$) are an artefact of the constant timesteps used in the \citet{Power_et_al_2003}. Under constant timestepping schemes, the size of a timestep relative to the smallest possible collisional timescale, $t_{\rm circ}(m_p(<\epsilon), \epsilon)/\Delta t,$ scales as $\epsilon^{3/2}.$  This dependence on $\epsilon$ means that the resolution of close orbits worsens as $\epsilon$ decreases. However, Eq.~\ref{eq:gadget_timestep_rewrite} shows that with the standard Gadget timestepping criteria, $t_{\rm circ}(m_p(<\epsilon), \epsilon)/\Delta t$ is \emph{independent of $\epsilon$}. Timestepping errors are therefore far less catastrophic with the standard Gadget timestepping criteria.

Other recent convergence studies have investigated the impact of timestepping in the $\epsilon < \epsilon_{\rm opr,P03}$ regime.
\citet{Ludlow_et_al_2019} performed tests on haloes across a wide range of $\epsilon$ values for $\eta=0.025$ and $\eta=0.0025.$ These tests find significant contraction of haloes out to large radii at $\eta=0.025$ for $\epsilon/l \lesssim 0.003,$ but find that haloes in the range of the typical $\epsilon$ of cosmological simulations are relatively unaffected (see \citealp{Ludlow_et_al_2019} fig.~2). 

The non-monotonic behaviour in $\epsilon$ is surprising and deserves further study. The onset of profile contraction occurs at $\epsilon/l$ values that are close to what is needed to avoid large-$\epsilon$ biases in halo properties.  A full characterisation of the profile contraction is therefore of practical relevance. 

One potential explanation for the non-monotonicity is that Eq.~\ref{eq:gadget_timestep_rewrite} ensures that collisions occurring at distances with $\epsilon\ll r_{\rm peri}$ are well-resolved, and the fraction of particle collisions which occur at $\epsilon\approx r_{\rm peri}$ decreases as $\epsilon$ decreases. Although the Gadget timestepping scheme ensures that such collisions are never catastrophically unresolved, modest integration errors are sill possible. \citet{Springel_2005} shows that when using the adaptive timestepping of Gadget-2, small integration errors tend to decrease the energy of the system. Thus, as epsilon decreases, the average energy lost per collision increases as the potential of each particle decreases.  In this case, however, the range of collision parameters that lead to $r_{\rm pericentre}\approx\epsilon$ also decreases until these collisions are so rare that they are not relevant to the internal dynamics of the halo.

\subsection{What is the `optimal' $\epsilon$?}

A number of studies have aimed to identify an optimal choice for $\epsilon$. The \citet{Power_et_al_2003} suggestion for an optimal value, $\epsilon_{\rm opt,P03}$, is shown in Eq.~\ref{eq:eps_opt} and discussed at length in Section \ref{sec:timestepping}. However, cosmological simulations almost always use scales smaller than $\epsilon_{\rm opt,P03}.$ The use of smaller $\epsilon$ values is in part because -- as Fig.~\ref{fig:chinchilla} and Fig.~\ref{fig:subhaloes} show -- haloes simulated at $\epsilon=\epsilon_{\rm opt}$ exhibit large biases at the particle counts that cosmological simulations typically consider. \citet{Klypin_et_al_2015} has also noted this effect in their analysis.

Recent convergence studies \citep{van_den_Bosch_Ogiya_2018,Ludlow_et_al_2019} have argued for an alternative optimal choice in $\epsilon$:
\begin{equation}
    \label{eq:eps_opt_VO18}
    \epsilon_{\rm opt,VdB,O,18}/l = 0.017.
\end{equation}
The level of bias implied by Fig.~\ref{fig:chinchilla} and Fig.~\ref{fig:subhaloes} at $\epsilon_{\rm opt,VdB,O,18}$ would be acceptable for many applications, but is not zero. These Figures do not conclusively establish convergence in $\epsilon,$ but, the model presented in Section \ref{sec:debias} would predict that simulations with
\begin{equation}
    \label{eq:eps_opt_vmax}
    \epsilon_{\rm opt,Vmax}/l\approx0.008
\end{equation}
would exhibit bias in $V_{\rm max}$ which is smaller than sample variance for simulations with comparable resolution and box sizes to the Chinchilla-$\epsilon$.

However, we caution against uncritically accepting Eq.~\ref{eq:eps_opt_vmax} as a blanket prescription for $\epsilon$ for several reasons:
\begin{itemize}
    \item Different particles in the same simulation may have different optimal force softening scales \citep[e.g.][]{Dehnen_2001,van_Kampen_2000,Power_et_al_2003}. Due to the myriad of numerical effects associated with $\epsilon,$ an $\epsilon/l$ which is too large for one system may be too small for another. However, understanding this trade-off requires robust models for the impact of $\epsilon$ on halo properties. As we have argued in Sections \ref{sec:eps_dependence_mass_trend} and \ref{sec:debias}, there remains much work to be done on this front.
    \item The level of acceptable bias in a measurement is highly dependent on the science goals. While striving for zero numerical bias (a formally impossible goal) is the safest generic option, all analyses can tolerate at least some deviation from the true predictions of $\Lambda$CDM.
    \item This recommendation is based solely on reducing bias in $V_{\rm max}.$ Halo properties which depend on the mass distribution at radii smaller than $R_{\rm max}$ will require smaller $\epsilon.$
    \item Our simulation suites did not explicitly establish a range of converged $\epsilon$ and this recommendation is thus model-dependent.
    \item Poorly-explored timestepping effects can cause significant halo contraction for $\epsilon$ values somewhat smaller than Eq.~\ref{eq:eps_opt_vmax} for standard timestepping schemes \citep{Ludlow_et_al_2019}.
\end{itemize}
All four considerations must be accounted for before applying Eq.~\ref{eq:eps_opt_vmax} or any other $\epsilon_{\rm opt}$ prescription.

\section{Conclusions}
\label{sec:conclusion}

In this paper, we study the impact of DMO simulation parameters on halo properties. We provide several tools to help analysts avoid and quantify these numerical biases. We do this by comparing a number of publicly available cosmological simulation suites against one another and by measuring the dependence of halo properties on both particle mass and on several secondary simulation parameters. The most important of these is the `force softening scale', which controls the distance scale at which the gravitational field of dark matter particles becomes non-Newtonian. We also consider the impact of coarse timestep sizes.

\begin{itemize}
    \item We report the $N_{\rm vir}$ cutoffs where the mean value of various halo properties diverge from the converged values of higher resolution simulations (Section \ref{sec:typical}). We report these cutoffs for a large collection of publicly available simulations.
    \item There are many halo properties (e.g. $x_{\rm off},$ $a_{1/2}$) where these cutoffs are consistent between simulations. For these properties, most analyses can simply use a set of conservative `convergence limits' at modest values of $N_{\rm vir}$ (Table \ref{tab:iso_cut}).
    \item For similarly high levels of agreement, other commonly used properties (e.g. $V_{\rm max}$, $c/a$) behave differently between simulations. High levels of agreement can require $N_{\rm vir}$ as large as $\approx10^5-10^6$ (Section \ref{sec:limit_variation} and Fig.~\ref{fig:mass_trends}).
    \item This disagreement is partially because some simulation suites have internally converged to different solutions. We demonstrate this for Multidark and IllustrisTNG-Dark (Section \ref{sec:multidark_illustris}). We argue that this disagreement is mostly caused by differences in force softening.
    \item We show that many halo properties (e.g., $V_{\rm max},$ $c/a$, and subhalo abundances) exhibit a strong dependence on force softening (Section \ref{sec:eps_dependence_mass_trend}). The biases associated with this dependence can be comparable to the impact of baryons on these properties.
    \item We develop a model which estimates the bias in $V_{\rm max}$ due to large force softening scales (Section \ref{sec:debias}). This model predicts the measured dependence of $V_{\rm max}$ on force softening and most of the dispersion in simulated $\langle V_{\rm max}(M_{\rm vir})\rangle$ relations.
    \item We review previous studies on timestep size and conclude that commonly used timestepping schemes are unlikely to significantly bias halo properties (Section \ref{sec:timestepping}).  However, we outline several open questions in this topic.
\end{itemize}

We emphasise that {\em all analyses can accommodate some level of numerical bias}. This paper does not assert what those levels are. There is nothing incorrect about studying haloes below the most conservative convergence limits, however such analyses should incorporate some estimate of the associated systematic uncertainty. The results of this paper will help analysts to identify the regimes where this is necessary and to estimate the resultant biases.

\section*{Acknowledgements}

We would like to thank Matthew Becker for generously sharing halo catalogues from the Chinchilla and Chinchilla-$\epsilon$ simulation suites. We would like to thank Gustavo Yepes and Stefan Gottloeber for allowing us to use halo catalogues from the ESMDPL simulation. We would also like to thank Aaron Ludlow, Frank van den Bosch, and Lehman Garrison for sharing electronic versions of data and/or unpublished data from \citet{Ludlow_et_al_2019}, \citet{van_den_Bosch_Ogiya_2018}, and \citet{Joyce_et_al_2020}, respectively and Benedikt Diemer, Tomo Ishiyama, Anatoly Klypin, and Gustavo Yepes for sharing detailed configuration information from their simulations and halo catalogues with us.

We would like to thank Andrey Kravtsov, Nick Gnedin, Benedikt Diemer, Peter Behroozi, Anatoly Klypin, Tomo Ishiyama, Aaron Ludlow, Frank van den Bosch, Matthew Becker, Gustavo Yepes, Andrew Hearin, Yao-Yuan Mao, Risa Wechsler, Chistine Simpson, Gus Evrard, and Neal Dalal for useful discussion which helped improve the quality of this work.

Many catalogues in this paper were accessed through the CosmoSim database. The CosmoSim database is a service by the Leibniz-Institute for Astrophysics Potsdam (AIP). The MultiDark database was developed in cooperation with the Spanish MultiDark Consolider Project CSD2009-00064.

The authors gratefully acknowledge the Gauss Centre for Supercomputing e.V. (www.gauss-centre.eu) and the Partnership for Advanced Supercomputing in Europe (PRACE, www.prace-ri.eu) for funding the MultiDark simulation project by providing computing time on the GCS Supercomputer SuperMUC at Leibniz Supercomputing Centre (LRZ, www.lrz.de). The Bolshoi simulations have been performed within the Bolshoi project of the University of California High-Performance AstroComputing Center (UC-HiPACC) and were run at the NASA Ames Research Center.

PM would like to thank The Grainger Foundation for the James Cronin Fellowship, which has supported his research at the University of Chicago. PM was also supported during other portions of this project by the Kavli Institute for Cosmological Physics at the University of Chicago through grants PHY-1125897, AST-1714658, and an endowment from the Kavli Foundation and its founder, Fred Kavli. CA was supported by the Leinweber Center for Theoretical Physics and the LSA Collegiate Fellowship at the University of Michigan.

The analysis in this paper was facilitated by the use of the \texttt{NumPy} \citep{NumPy}, \texttt{SciPy} \citep{SciPy}, and \texttt{matplotlib} \citep{Matplotlib} libraries.

\bibliographystyle{mnras}
\bibliography{main} 

\appendix

\section{Calibrating a Shared Scale for Different Force Softening Schemes}
\label{sec:recalibrate_plummer}

\begin{figure*}
   \centering
   \subfigure{\includegraphics[width=0.49\textwidth]{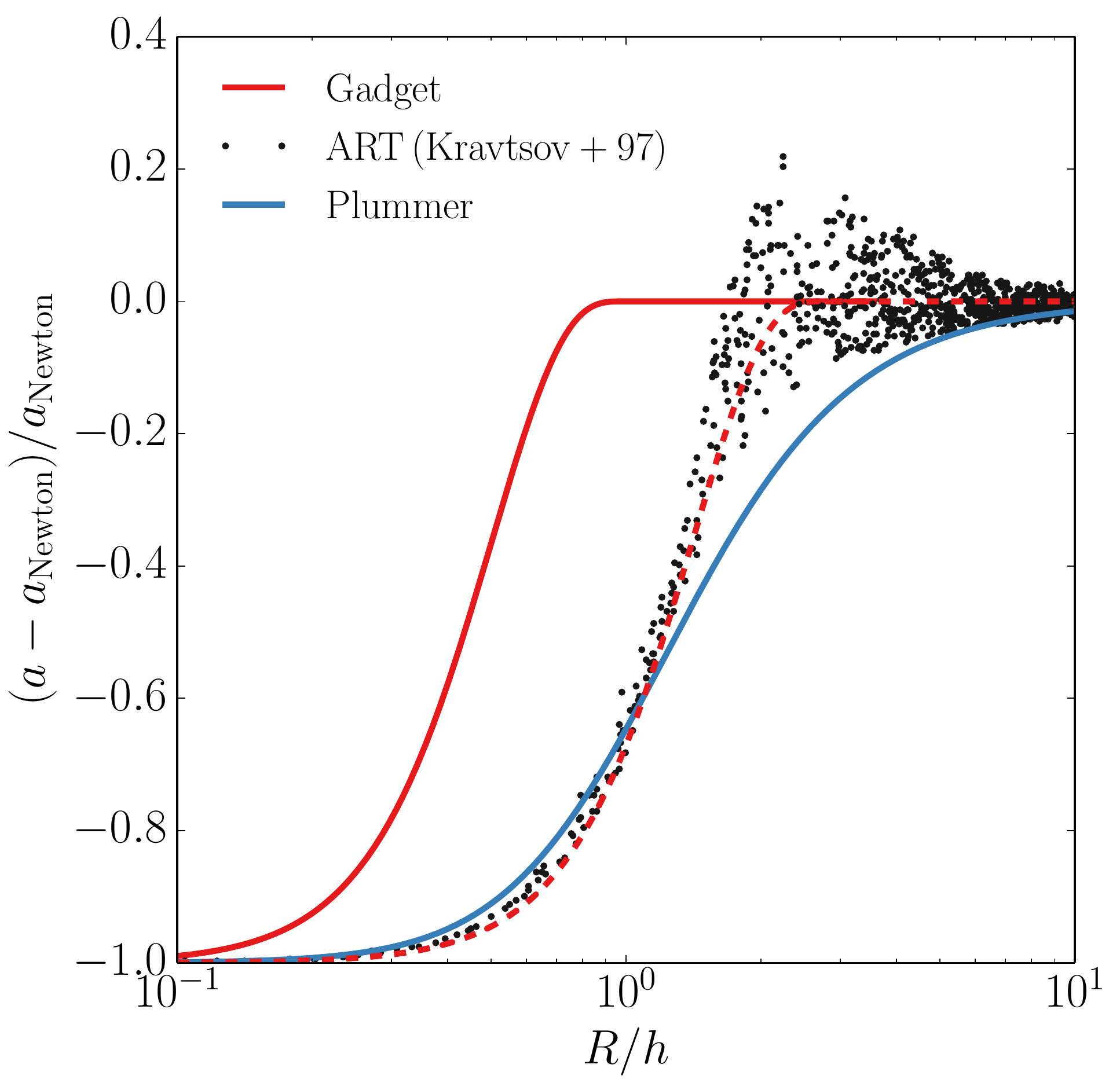}}
   \subfigure{\includegraphics[width=0.475\textwidth]{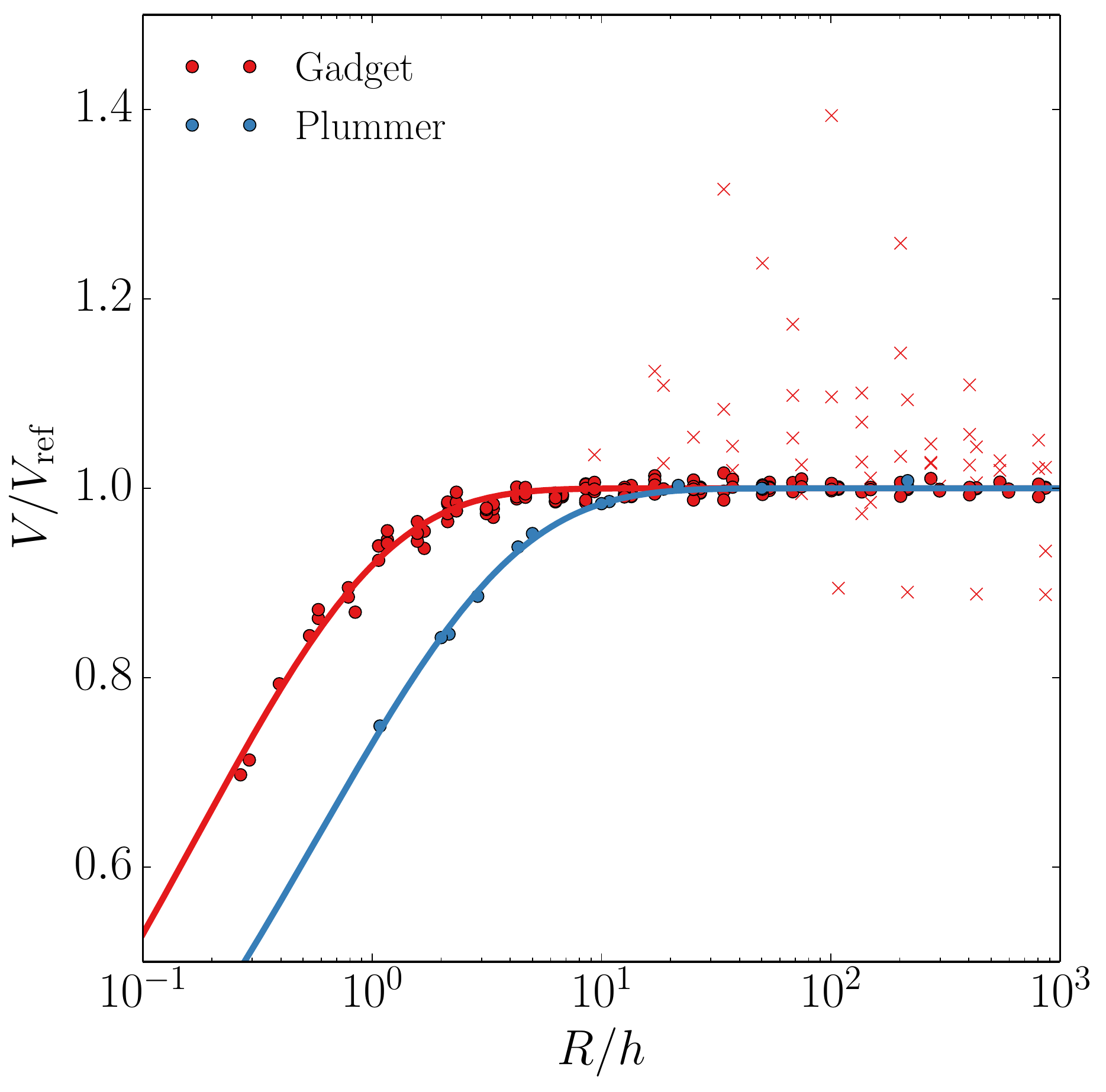}}
\caption{ Left: The gravitational acceleration under various force softening schemes at some distance, $r$, from a particle. This acceleration is shown relative to the gravitational acceleration due to a Newtonian point source. Distances are normalised by $h,$ the scheme-specific formal resolution described in Section \ref{sec:force_softening}. The points for ART are taken from \citet{Kravtsov_et_al_1997}. The dashed red line shows the deviation for the Gadget force kernel, scaled by $h=0.357\,h_{\rm Gadget}.$ This plot illustrates the known fact that the traditional ``Plummer-equivalent'' conversion between formal resolution parameters leads to similar deviations for $r<\epsilon_\phi,$ but highly discrepant deviations at larger radii. Right: The impact of different force softening schemes on halo circular velocity profiles. The points in this plot show the measured biases in circular velocity profiles as a function of the formal resolution, $h,$ for different force softening schemes. Gadget measurements are from \citet{Ludlow_et_al_2019}, and Plummer measurements are from \citet{Klypin_et_al_2015} and \citet{van_den_Bosch_Ogiya_2018}. Points shown as red `x's correspond to measurements from \citet{Ludlow_et_al_2019} where deviations from the reference rotation curve were caused by integration errors. Curves show the results of fits against Eq.~\ref{eq:v_dev_curve_ein}. These fits form the basis for our conversion of formal resolutions onto a shared scale. See Appendix \ref{sec:recalibrate_plummer} for discussion.}
\label{fig:eps_kernels}
\end{figure*}

This work considers the results of simulations run with a variety of codes. Some of these simulations use different schemes for softening forces. Given the importance of force softening to the analysis in this paper, this Appendix contains a detailed analysis of the impact that force softening has on circular velocity profiles. The main result of this Appendix is Eq.~\ref{eq:v_dev_curve_ein_app}, which quantifies the impact of different force softening schemes on $V(R).$

We use this fit in two ways in this work. First, we use the best-fitting parameters to construct an alternative to the standard `Plummer-equivalent' conversion between different force softening schemes. Second, we use this fit as a core component of our model for $V_{\rm max}$ biases in Section \ref{sec:debias}.

We compare the results of \citet{Klypin_et_al_2015}, \citet{van_den_Bosch_Ogiya_2018} and \citet{Ludlow_et_al_2019}, which measured circular velocity profiles for haloes simulated with varying $h$ for Plummer and Gadget kernels. \citet{Klypin_et_al_2015} and \citet{van_den_Bosch_Ogiya_2018} considered idealised isolated NFW haloes, while \citet{Ludlow_et_al_2019} studied stacked mass profiles from a series of small cosmological boxes.

We first consider the profiles in \citet{Ludlow_et_al_2019}. These tests were performed with a `standard' Gadget timestepping parameter of $\eta=0.025$ and with a higher resolution $\eta=0.0025.$ The $\eta=0.025$ boxes were run with formal resolutions of $h_{\rm Gadget}(z=0)=\{2^{-6},\,2^{-5},\,...,\,2^4\}\times h_{\rm Gadget,fid}$ for $h_{\rm Gadget,fid}=$0.6642 $h^{-1}$ kpc and the $\eta=0.0025$ boxes were run with $h_{\rm Gadget}(z=0)=\{2^{-6},\,2^{-5},\,...,\,2^9\}\times h_{\rm Gadget,fid}.$ These profiles were stacked in mass bins centred on $M_{\rm 200c} = \{10^9,\,10^{10},\,10^{11},\,10^{12}\}h^{-1}M_\odot,$ and widths of 0.3 dex, corresponding to median $N_{\rm 200c}$ values of $\{6.7\times10^2,\,6.3\times10^3,\,6.5\times10^4,\,6.4\times10^5\},$ respectively.

This range of parameters means that the \citet{Ludlow_et_al_2019} measurements can probe the impact of $h_{\rm Gadget}$ across a wide range of halo radii, particle counts, and concentrations. The $\eta=0.0025$ simulations allow the impact of numerical scattering due to coarse timesteps to be separated from timestep-independent effects like two-body relaxation effects and overly-large $h.$

The variation in profiles between $\eta=0.0025$ boxes is at the per cent level and does not show strong dependence on $h$ for the small $h$ scales probed by these boxes, so we take the $h_{\rm Gadget}=h_{\rm Gadget,fid},$ $\eta=0.0025$ box as our `reference' simulation. Our results are nearly identical if smaller values of $h$ are used.

For each mass bin and $h_{\rm Gadget}$ value in the $\eta=0.025$ boxes, we measure the value of $V_{\rm circ}(R)/V_{\rm circ,ref}(R)$ for $R=\{2^{-4},\,2^{-3},...,2^2\}\times R_{\rm max,ref},$ where $V(R)$ is the circular velocity at radius $R$, $R_{\rm max}$ is the radius at which the circular velocity profile reaches its maximum value, and quantities subscripted with `ref' are measured in the reference simulation. We discard $V_{\rm circ}(R)/V_{\rm circ,ref}$ values at radii smaller than the convergence radii advocated for by \citet{Ludlow_et_al_2019}, although we find that our fits are strongly insensitive to this minimum radius. We also remove values which deviate by more than two per cent from values measured in $\eta=0.0025$ simulations with identical $h_{\rm Gadget}.$ While deviations in these regimes are relevant to convergence studies, they are caused by two-body scattering and time integration errors and not by errors due to large $h_{\rm Gadget}.$

We find that $V(R;\,h)/V_{\rm ref}$ does not depend on particle count or halo concentration and that the ratio can be reparametrized as $V(R;\,h)/V_{\rm ref}=V(R/h)/V_{\rm ref}$ without loss of accuracy \citep[see also, the first three panels of fig.~5 in][]{Ludlow_et_al_2019}. After experimenting with multiple functional forms, we fit these measurements against
\begin{equation}
    \label{eq:v_dev_curve_ein_app}
    \frac{V(R;h)}{V_{\rm ref}(R)} =1 - {\rm exp}\left(-(Ah/R)^\beta\right).
\end{equation}
Here, $A$ and $\beta$ are free parameters. We perform our fit using non-linear least squares minimisation, because manual inspection of the likelihood posterior confirms that it is unimodal and approximately Gaussian near the minimum.

\begin{table}
  \centering
  \caption{The best-fitting parameters for Eq.~\ref{eq:v_dev_curve_ein_app} for different force softening schemes. Gadget velocity deviations are measured at $\eta=0.025$.
  }
  \label{tab:v_dev_fit}
  \begin{tabular}{llll}
  \hline\hline
  Fit type & Scheme & $A$ & $\beta$ \\
  \hline
  Free $\beta$ & Gadget & $0.172 \pm 0.006$ & $-0.522 \pm 0.010$ \\
  & Plummer & $0.580 \pm 0.026$ & $-0.497\pm 0.016$ \\
  \hline
  Fixed $\beta=-0.522$ & Plummer & $0.616\pm0.011$ & \\
  \hline
  Fixed $\beta=-0.497$ & Gadget & $0.160\pm0.002$ & \\
  \end{tabular}
\end{table}
We show this fit in the right panel of Fig.~\ref{fig:eps_kernels} and give its best-fitting parameters in Table \ref{tab:v_dev_fit}. $V/V_{\rm ref}$ measurements removed prior to fitting due to timestepping dependence are shown as `x's. As an internal consistency check, we find that this fit predicts deviations equal to $0.1\,V(R;\,h)$ at $0.76\,h$, which is consistent with fig.~5 in \citet{Ludlow_et_al_2019}.

As mentioned above, $V(R;h)/V_{\rm ref}(R)$ does not depend on particle count, or concentration. This means that we can safely compare these fits against tests performed on narrower radius, particle count, and concentration ranges.
We combine the $R=R_{\rm max}$ measurements from \citet{Klypin_et_al_2015} and the $R=R_s/2$ measurements from \citet{van_den_Bosch_Ogiya_2018} for our Plummer kernel data set. Our results are unchanged if we restrict ourselves to the results of either paper.Both studies analyse idealised NFW profiles instead of cosmological boxes, so we use NFW profiles as our reference $V_{\rm ref}(R)$ curves. The timestepping schemes used in both papers are substantially more aggressive than an $\eta=0.025$ Gadget simulation, so we do not need to remove any simulations due to integration errors, as was done for the \citet{Ludlow_et_al_2019} data set. However, we do remove the $h=10^{-4}R_{\rm vir}$ simulation from \citet{van_den_Bosch_Ogiya_2018} before fitting because that halo is undergoing thermalisation at $R=R_s/2.$

We show this fit in the right panel of Fig.~\ref{fig:eps_kernels} and give its best-fitting parameters in Table \ref{tab:v_dev_fit}.

Because $A$ and $\beta$ are slightly covariant, comparison between the $A$ values of different fits can only be performed at a constant $\beta.$
If $\beta$ is fixed to $-0.522$ for the Plummer fit, $A_{\rm Plummer}=0.616\pm0.011,$ indicating that $\epsilon_{\rm Gadget}=A_{\rm Plummer}/A_{\rm Gadget} = 0.279\pm0.006.$ Fixing $\beta=-0.497$ for the Gadget fit results in $\epsilon_{\rm Gadget}=0.277\pm0.006.$ Because Gadget-like softening kernels are more common in modern simulations than Plummer kernels, we choose to normalise the relation to preserve the commonly-used conversion between $h_{\rm Gadget}$ and $\epsilon:$
\begin{align}
    \label{eq:appendix_plummer_equivalent}
    \epsilon=1.284\,h_{\rm Plummer}=h_{\rm ART}=0.357\,h_{\rm Gadget}.
\end{align}
Note that without comparable ART-based tests, we have arbitrarily chosen to take the convention from \citet{Klypin_et_al_2016} that $h_{\rm ART}=0.357\,h_{\rm Gadget}$. This leads to comparable mean deviations from Newtonian gravity to those caused by the Gadget kernel at all radii. No analysis in this paper relies on this portion of the convention.

We have performed this fit with several other functional forms in the place of Eq.~\ref{eq:v_dev_curve_ein_app} and found results which are similar. For example, when using $V(R;\,h)/V_{\rm ref}(R)=(1 + (Ah/R)^2)^\beta$ -- a form similar to the one used in \citet{Klypin_et_al_2015} -- we find that $\epsilon$ ranges from $1.29\,h_{\rm Plummer}$ to $1.28\,h_{\rm Gadget}.$

While Eq.~\ref{eq:appendix_plummer_equivalent} is most appropriate when estimating the effects of reduced centripetal forces on halo profiles, force softening also impacts halo profiles through two-body scattering and time integration errors. In regimes where these effects dominate, the depths and shapes of the kernel potentials may be more important than the long-distance deviations from Newtonian gravity. If so, these two body-scattering effects would be be best analysed through $\epsilon_\phi.$ To prevent readers from needing to frequently convert between $\epsilon$ conventions, we have converted all values used in this paper to $\epsilon,$ except in cases of specifying an algorithm which depends on $\epsilon_\phi.$

We note that Eq.~\ref{eq:v_dev_curve_ein_app} appears to `predict' that $\epsilon$ can be made arbitrarily small without error. This is only true over the $R/\epsilon$ range fitted here and only when timesteps are very fine. Coarse timesteps lead to very real errors at small $\epsilon$ (see Section \ref{sec:timestepping}), the `convergence radius' which we use to select our fitting ranges has a weak dependence on $\epsilon$ \citep{Ludlow_et_al_2019}, and fig.~6 of \citet{van_den_Bosch_Ogiya_2018} shows that aggresively small softening scales ($\epsilon \lesssim 10^{-4}R_{\rm vir}$) can accelerate the impact of two-body scattering. Similar effects can be seen in fig.~13 of \citet{Klypin_et_al_2015}. Large-$\epsilon$ effects are only a portion of the story.

\section{The Impact of \textsc{Rockstar} Versions on Halo Properties}
\label{sec:rockstar_versions}
The simulations we consider in this paper use a number of different versions of the \textsc{Rockstar} halo finding software. \textsc{Rockstar} has undergone a number of bug fixes since its original release, and halo catalogues generated with different versions can have significantly different property distributions. To understand the impact of different software versions, we obtained the approximate \textsc{Rockstar} download times and configuration files for every simulation suite considered in this paper to identify the corresponding software version (B. Diemer; A. Klypin; M. Becker; T. Ishiyama; P. Behroozi, personal communication). 

We then isolated the source of version-dependent results.  First, we regenerated halo catalogues for the CBol\_L125 simulation using the different versions -- matching the exact commit hash if known -- as well as the relevant parameters in each respective configuration file, and we cross-matched these catalogues against one another. Second, we performed an extensive review of the \textsc{Rockstar} and consistent-trees version control commit histories\footnote{available at \texttt{https://bitbucket.org/gfcstanford/rockstar} and \texttt{https://bitbucket.org/pbehroozi/consistent-trees}}. By combining these two analyses, we determined that there were two sets of variables which gave version-dependent results and that all other variables were consistent between versions. These variables are (1) the axis ratios calculated within $R_{\rm 500c}$ and (2) properties that depend on internal energy calculations. The \textsc{Rockstar} changelogs document both of these issues, meaning that, fortunately, our cross-matching of catalogues did not reveal any new significant inconsistencies.

Incorrect axis ratio measurements at $R_{\rm 500c}$ affect Erebos\_CBol and Erebos\_CPla, which used \textsc{Rockstar} catalogues generated with code downloaded prior to October 22$^{\rm nd}$, 2013. There is no method for correcting this issue, but as discussed below the convergence properties of these inner axis ratios are largely similar to the conventional larger axis ratios.  We therefore do not analyse this property. \textsc{Rockstar} catalogues generated with code downloaded prior to May 15$^{\rm th},$ 2014 estimate internal energies which are too large by a factor of two. This can be corrected by replacing variables, $X,$ with updated versions, $X'.$ In the cases of the virial ratio and the Peebles spin parameter, the replacement variables would appear as
\begin{align}\label{eq:internal_e_correction}
    T/|U|' &= 2\,T/|U| \\
    \lambda_{\rm Peebles}' &= \lambda_{\rm Peebles}\, \frac{\sqrt{1-T/|U|'}}{\sqrt{2-T/|U|'}}
\end{align}
We apply the corrections of Eq.~\ref{eq:internal_e_correction} to catalogues for the Erebos\_CBol, Erebos\_CPla, Bolshoi, and BolshoiP suites. We also applied these corrections to Chinchilla\_L250 and Chinchilla\_L400, but Chinchilla\_L125 did not require these corrections.

Another potential source of variation amongst \textsc{Rockstar} catalogues is the choice of primary mass definition, which changes the values of other reported halo properties (see section 4 and appendix A of \citet{mansfield_kravtsov_2019} for a full discussion). However, we confirmed that all of the halo catalogues that we consider in our analyses used $M_{\rm vir}$ as the primary mass definition.  This particular source of variation does not impact our results.

\section{Finding Empirical Convergence Limits}
\label{sec:procedure_app}

This Appendix serves as an expanded and more technical version of Section \ref{sec:procedure}

\subsection{Separation into Subgroups}
\label{sec:subgroups}
Because many halo properties depend on cosmology (especially properties which depend on accretion histories), we separate simulations by cosmology to avoid misinterpreting these cosmological dependencies as non-convergence. We analyse the WMAP suites Bolshoi, Chinchilla, and Erebos\_CBol as a group and the Planck suites $\nu^2$GC, BolshoiP, Multidark, and Erebos\_CPla as a group. The exact parameters used still vary from suite-to-suite, mostly due to the year of each mission which these simulations attempt to match. This is most apparent when comparing the Chinchilla suite to other WMAP simulations like Bolshoi or Erebos\_CBol (see Table \ref{tab:simulations}).

We tested the impact of these small cosmology differences by repeating our analysis with groups based on the exact cosmological parameters and did not find a meaningful difference in our results. Because this split significantly reduces the number of simulations which have higher-resolution boxes available for comparison, we do not use this approach in the rest of this paper. 

We also separate haloes by subhalo and isolated halo status (see Section \ref{sec:properties}). This is important both because subhaloes and isolated haloes may have difference convergence properties and because numerical parametrization can lead to changes in the artificial subhalo disruption rate \citep[e.g.][see also Section \ref{sec:subhaloes}]{van_den_bosch_et_al_2018,van_den_Bosch_Ogiya_2018}. Artificial disruption would lead to isolated haloes being over-represented at a constant mass, and in cases where host haloes and subhaloes follow different mass relations this would propagate to a change in the global mass relation.

Using $R_{\rm vir}$ to define subhalo status (as we do here) is suboptimal. There is a large population of `splashback subhaloes' which are qualitatively indistinguishable from other subhaloes but whose orbits have apocentres outside the arbitrarily-defined virial radius \citep{Balogh_et_al_2000,Mamon_et_al_2004,Gill_et_al_2005,Ludlow_et_al_2009,Bahe_et_al_2013,Wetzel_et_al_2014,Xie_et_al_2015}. \citet{mansfield_kravtsov_2019} showed that this population of misidentified subhaloes is responsible for the entire high-concentration tail of the `isolated' halo population, thus opening the possibility that numerical subhalo disruption could affect our convergence limits. Although many schemes for identifying splashback subhaloes exist \citep[see][for review]{mansfield_kravtsov_2019}, we do not use them here: they rely on merger tree information and/or raw particle data, which are not available for all the simulations considered here, and these methods have non-trivial convergence properties themselves \citep{Mansfield_et_al_2017} which would be a larger complicating issue than subhalo contamination.

\subsection{Defining High-Resolution Particle Ranges}
\label{sec:hr_ranges}
For each simulation, $s,$ we measure $\langle X(M_{\rm vir})\rangle_s$ within logarithmic 0.125 dex mass bins. We restrict analysis to mass bins containing at least 100 haloes.

We identify non-convergence by identifying where simulations deviate from the mass-relation implied by the high-resolution regimes of other simulations. We identify a such a corresponding high-resolution cutoff, $N_{\rm HR},$ by eye such that no simulations in our sample deviate from others in their subgroup when mass relations are constructed for haloes with $N_{\rm vir}>N_{\rm HR}$.

This cutoff is chosen separately for each halo property and each analysis subgroup, although we use the same cutoffs for both Planck and WMAP cosmology. These cutoffs are given in the online supplement. Our tests indicate that our results are not sensitive to the exact $N_{\rm HR}$ choices used.

The cutoff chosen for the $\langle X_{\rm off}(M_{\rm vir})\rangle$ relation is shown as the transition of from solid to dashed lines in the left panel of Fig.~\ref{fig:procedure}.

As discussed in Section \ref{sec:limit_variation} and Appendix \ref{sec:fitting}, several simulations diverge significantly from other simulations at aberrant high particle counts for various halo properties. These simulations are not included in our determination of $N_{\rm HR}$ and are discussed extensively throughout section \ref{sec:results}.

\subsection{Fitting Mean Relations}
\label{sec:fitting}

For each halo property, $X,$ simulation, $s,$ and mass bin, $i,$ we measure both $\langle X(M_{\rm vir})\rangle_{s,i},$ and the uncertainty due to sample variance, $\sigma_{X,s,i}$ as estimated by jackknife resampling.

We fit these points with a mass-dependent Gaussian distribution which has a centroid given by the $d$-degree polynomial, $X_d:$
\begin{align}
    X_d(M_{\rm vir}) &= \epsilon + \sum_{i=0}^d p_i \, \log_{10}(M_{\rm vir}/M_0)^i \\
    \epsilon &\sim {\rm Norm(0,\, \sigma_0)}.
\end{align}
Here, $p_i$ are polynomial coefficients, $\sigma_0$ is the intrinsic scatter in the distribution, and $M_0=10^{12.5}\,h^{-1}M_\odot.$ We include intrinsic scatter in the fit because sample variance alone -- as measured by $\sigma_{X,s,i}$ -- is insufficient to explain the full scatter in $\langle X\rangle_{s,i}$ at a given $M_{\rm vir}.$ This is likely due to slight differences in otherwise-similar cosmologies or subtle numerical parameter differences which do not result in major non-convergence.

Because the posterior distributions for the corresponding likelihood functions are generally smooth and unimodal on inspection, we fit $(p_i,\sigma_0)$ by maximising the log-likelihood function.

We use a similar procedure to fit mass functions, $\phi(M_X).$ We fit $\log_{10}(\phi(M_X))$ against
\begin{align}
    X'_d(M_{\rm vir}) &= \epsilon + \sum_{i=0}^d p_i \, \log_{10}(M_X/M_0)^i \\
    \epsilon &\sim {\rm Norm(0,\, \sigma_0)}.
\end{align}
For velocity functions, $\phi(V_X),$ we fit $\log_{10}(\phi(V_X))$ against:
\begin{align}
    X''_d(M_{\rm vir}) &= \epsilon + \sum_{i=0}^d p_i \, \log_{10}(V_X/V_0)^i \\
    \epsilon &\sim {\rm Norm(0,\, \sigma_0)}.
\end{align}
for $V_0 =$ 100 km/s.

We give the best-fitting parameters for each halo property in the online supplement.\footnote{\texttt{https://github.com/phil-mansfield/halo\_convergence}}

For many halo properties, a subset of simulations diverge significantly from other simulations within the same suite. We remove all $\nu^2$GC boxes, all TNG boxes, VSMDPL, HMDPL, and Chinchilla\_L250 prior to fitting the $\langle c/a(M_{\rm vir})\rangle$ relation, Chinchilla\_L250 before fitting the $c_{\rm vir}$ relation, all $\nu^2$GC boxes, VSMDPL, SMDPL, and Chinchilla\_L250 prior to fitting the $V_{\rm max}$ relation, and Chinchilla\_L250 prior to fitting the $V_{\rm peak}$ relation.

Three sets of outlier removals require special comment. We found that the $\langle c_{\rm vir}(M_{\rm vir})\rangle$ relation was well fit by a power law for each simulation suite individually, but that amplitude and power law index of these relations were noticeably different for each suite. As such, we fit each suite independently with the additional removal of Chinchilla\_L250.

As is discussed in section \ref{sec:multidark_illustris}, Illustris-TNG and the high resolution MDPL simulations appear to `converge' to different $\langle V_{\rm max}(M_{\rm vir})\rangle$ relations. Because $\nu^2$GC-H2 and $\nu^2$GC-H1 give aberrant results, the only Planck-cosmology simulations in Table \ref{tab:simulations} which probe halo masses below $M_{\rm vir} \lesssim 10^{11}\,h^{-1}M_\odot$ fall into one of these two suites. To avoid a fit which `splits the difference' between the two, we perform 
two fits removing ESMDPL, VSMDPL, and SMDPL from one 
fit and all the TNG-Dark boxes from a second fit. We analyse both fits.

This removal of outlier simulations serves to emphasise that these fits cannot be interpreted as approximating the `correct' converged solutions for these mass relations, but as approximating the high-resolution solutions for a particular subset of simulations: we explicitly do not claim that any individual simulation considered in this paper is converged or correct (or that the inverse is true). As such, we do not provide any of the fits produced from this part of this analysis to prevent their potential misuse.

\subsection{Measuring Significance}
\label{sec:significance}

Traditional convergence tests are either performed by eye or by measuring the mass at which halo properties deviate from a reference relation by more than some fixed level of acceptable bias. We do not take these approaches for three reasons: first, even the mass relations of converged simulations can deviate from the high resolution relation due to Poissonian noise, sample variance, and uncertainties in the underlying fit. This makes percentage cuts sub-optimal. Second, visual identification is time-consuming, especially given the number of simulations, sub-groups, and halo properties considered in this paper. Third, we noted unintentional researcher confirmation bias in our own tests of visually identified convergence limits. For these reasons, we have opted to use a different statistical test.

Instead, we construct a null hypothesis, $H_0(\delta, M_{\rm vir}),$ which states that `$\langle X(M_{\rm vir})\rangle_s$ deviates from $\langle X(M_{\rm vir})\rangle_{\rm HR}$ by less than $\delta\langle X(M_{\rm vir})\rangle_{\rm HR}$.' For each $M_{\rm vir}$ bin, we measure the probability of measuring a deviation, $\Delta_s(M_{\rm vir})=|\delta\langle X(M_{\rm vir})\rangle_s - \langle X(M_{\rm vir})\rangle_{\rm HR}|$ at least as large as the measured $\Delta$ if $H_0(\delta,M_{\rm vir})$ were true. To do this, we use the frequentist $z$- test. For each simulation, $s,$ and mass bin, $M_{\rm vir},$
\begin{equation}
    z_s(M_{\rm vir})= \frac{\Delta - \delta\langle X(M_{\rm vir})\rangle_s}{\sqrt{\sigma_0^2 + \sigma_s(M_{\rm vir})}}.
\end{equation}
Here, $\sigma_s(M_{\rm vir})$ is the sample variance in the mass bin $M_{\rm vir}$ for simulation $s$, and $\sigma_0$ comes from the fit in Appendix \ref{sec:fitting}. We then compute the {\em upper-tailed} $p$-value associated with $z_s(M_{\rm vir}).$ The lowest mass bin with $p \geq 0.05$ is the convergence limit for that simulation and property, $X$. To deal with cases where the unconverged behaviour of $\langle X(M_{\rm vir})\rangle_s$ is non-trivial and potentially crosses $\langle X(M_{\rm vir})\rangle_{\rm HR}$ multiple times, we ignore mass ranges where $p \geq 0.05$ for fewer than three consecutive mass bins.

We show this test in the right panel of Fig.~\ref{fig:procedure}. Mass relations are colour-coded by the value of $p$ as a function of mass.

To confirm this procedure, we visually identified convergence limits for every simulation and halo property in each sub-group without knowledge of the bins that our statistical method selected. Visual identifications were generally within 0.125-0.25 dex of the statistical measurements described above. Qualitatively, no major results in this paper change if these visual cutoffs are used. However, as noted above, the particle cutoffs for some halo properties showed somewhat smaller dispersions when visual cutoffs were used. Inspection of individual cases caused us to interpret this as confirmation bias.

There are some simulations which are converged across the entire mass range used to fit the high-resolution relation. This is the case for the highest resolution fits in Fig.~\ref{fig:procedure}. Rather than extrapolate our fits, the convergence limits of these simulations are left as upper limits. This is almost always the case for the highest resolution simulation in a sub-group. In rarer cases, there are simulations which are unconverged across the entire high-resolution mass range. The convergence limits for these boxes are left as lower limits.

\section{The Impact of $\epsilon$ On Multidark and Illustris-TNG $V_{\rm max}$ Distributions}
\label{sec:model_predctions_multidark_illustris}

\begin{figure}
    \centering
    \includegraphics[width=0.49\textwidth]{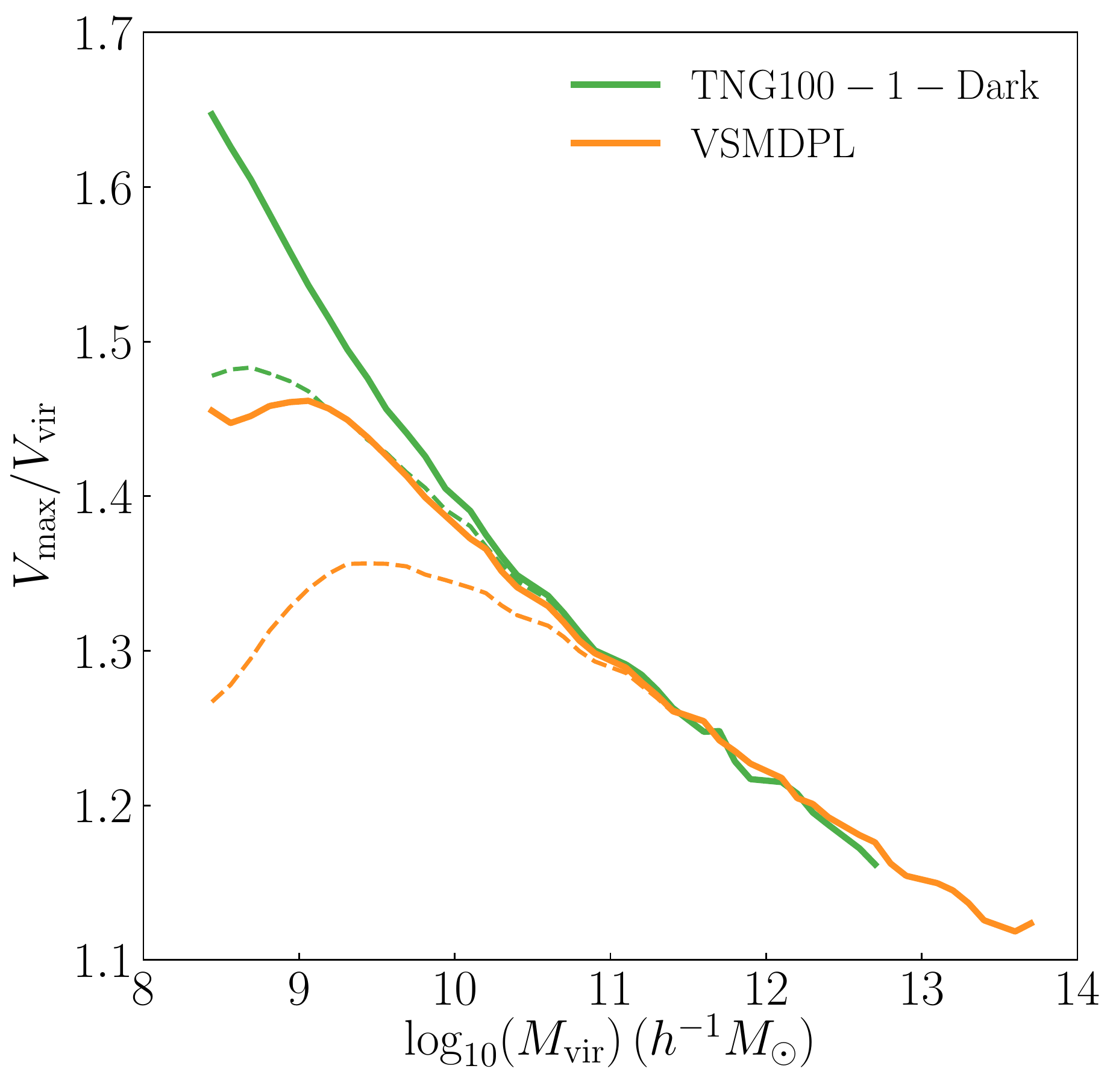}
    \caption{The results of applying the bias estimates described in Section \ref{sec:debias} to TNG100-1-Dark and VSMDPL. As in Fig.~\ref{fig:v_bias_curves}, the dashed curves show the mean $V_{\rm max}$ values measured in each mass bin and the solid curves show estimates for what $\langle V_{\rm max}\rangle$ would be if there were no bias due to force softening. The difference between the simulations is mostly -- but not entirely -- accounted for by large-$\epsilon$ biases.}
    \label{fig:multidark_illustris_app}
\end{figure}

In Section \ref{sec:multidark_illustris}, we discussed the fact that the Multidark and IllustrisTNG-Dark simulation suites appear to have converged to two separate $\langle V_{\rm max}(M_{\rm vir})\rangle$ relations. We argue that this difference can be well understood by only considering the near-identical TNG100-1-Dark and VSDMPL boxes. We further argue that the cause of the difference is likely to be either differences in force softening, differences in force accuracy, or updates to the Gadget force solver that were implemented in Arepo.

In Section \ref{sec:debias}, we present a model which predicts the impact of large-$\epsilon$ biases on the $V_{\rm max}$ distribution in simulations. We show the result of applying this model to VSMDPL and TNG-100-1-Dark in Fig.~\ref{fig:multidark_illustris_app}. This Figure shows that large-$\epsilon$ biases account for most of the difference between the two simulations. The estimated de-biased $\langle V_{\rm max}(M_{\rm vir})\rangle$ relations agree to lower $M_{\rm vir}$ and the divergence at $M_{\rm vir} = 10^{9.5}\,h^{-1}M_\odot$ decreases by a factor of 60\% - 70\%.

As we showed in the similar Fig.~\ref{fig:v_bias_curves}, this procedure can entirely remove the $\epsilon$-dependence in $\langle V_{\rm max}(M_{\rm vir})\rangle$ if all other factors are held constant. This means that the remaining difference between the debiased curves in Fig.~\ref{fig:multidark_illustris_app} is likely the cause of additional numerical factors. Given the discussion in Section \ref{sec:multidark_illustris}, we suggest that future work consider the impact of force accuracy and code differences between LGadget-2 and Arepo.

\bsp	
\label{lastpage}
\end{document}